%% file: A-main-thesis-file.tex
\documentclass[12pt]{ull_thesis22}
\usepackage[vlined,linesnumbered,ruled,boxed]{algorithm2e}
 \usepackage{amsmath}
 \usepackage{amsfonts}
 \usepackage{amssymb,xcolor}

 \usepackage{layout}
 \usepackage{mydef}
 \usepackage{url}
\usepackage{bibentry}

\usepackage{tabularx} 
\usepackage{algorithmic}

\nobibliography*

 \usepackage{wrapfig,color,caption,subcaption,balance,multirow}
 \usepackage[vlined,linesnumbered,ruled,boxed]{algorithm2e}
 \usepackage{indentfirst}

\usepackage{etoolbox}
\makeatletter
\patchcmd{\@chapter}{\addtocontents{lof}{\protect\addvspace{10\p@}}}{}{}{}
\patchcmd{\@chapter}{\addtocontents{lot}{\protect\addvspace{10\p@}}}{}{}{}
\makeatother

\makeatletter
\AtBeginEnvironment{thebibliography}{%
   \clubpenalty10000
   \@clubpenalty \clubpenalty
   \widowpenalty10000}
\makeatother

\newcommand{\etal}{{\it et~al.}}
\newcommand{\comments}[1]{}
\newcommand{\mtx}[1]{\ensuremath{\mathbf{#1}}}

\usepackage{adjustbox,tabu,tabularx}

\newcommand{\etc}{{\it etc.}}
\newcommand{\name}{Edge-MultiAI\xspace}

\makeatletter
\def\thickhline{%
  \noalign{\ifnum0=`}\fi\hrule \@height \thickarrayrulewidth \futurelet
   \reserved@a\@xthickhline}
\def\@xthickhline{\ifx\reserved@a\thickhline
               \vskip\doublerulesep
               \vskip-\thickarrayrulewidth
             \fi
      \ifnum0=`{\fi}}
\makeatother

\newlength{\boxfigwidth}

\usepackage{titlesec}

\newcommand*{\secref}[1]{%
  \begingroup
    \renewcommand*{\hspace}[1]{}%
    \ref{#1}%
  \endgroup
}

 \titleformat{\chapter}[block]
{\singlespacing\bfseries\filcenter}{\chaptertitlename\ \thechapter: \ }{0pt}{}
\titlespacing*{\chapter}
{0pt}{-0.5\baselineskip}{0pt}  

\titleformat{\section}
{\bfseries}{\thesection\ }{1em}{}
\titlespacing*{\section}{0pt}{*0}{0pt}

 \titleformat{\subsection}[runin]
{\bfseries}{\thesubsection}{1em}{}
\titlespacing{\subsection}{0pt}{*0}{0pt}

\titleformat{\subsubsection}[runin]
{\bfseries}{\thesubsubsection}{1em}{}[ ] 
\titlespacing{\subsubsection}
{\parindent}{*0}{0pt}

 \raggedright
\usepackage{graphicx,multirow}
\usepackage[vlined,linesnumbered,ruled,boxed]{algorithm2e}

 \usepackage{layout}
 \usepackage{url}
\usepackage{wrapfig}
\usepackage{bibentry}

\usepackage{tabularx} 
\usepackage{algorithmic}
\usepackage{amssymb}
\usepackage{bbold}

\nobibliography*

\usepackage{wrapfig,color,caption,subcaption,balance,multirow}

\usepackage{adjustbox,tabu}

\usepackage[vlined,linesnumbered,ruled,boxed]{algorithm2e}
 \usepackage{amsmath}

 \usepackage{caption}
  \usepackage{subcaption}

\usepackage{bibentry}

\makeatletter
\def\thickhline{%
  \noalign{\ifnum0=`}\fi\hrule \@height \thickarrayrulewidth \futurelet
   \reserved@a\@xthickhline}
\def\@xthickhline{\ifx\reserved@a\thickhline
               \vskip\doublerulesep
               \vskip-\thickarrayrulewidth
             \fi
      \ifnum0=`{\fi}}
\makeatother

\newlength{\thickarrayrulewidth}
\nobibliography*

 \usepackage{graphicx,multirow}

\usepackage{diagbox}

\usepackage{cleveref}
\newcommand\hl{\bgroup\markoverwith
  {\textcolor{yellow}{\rule[-.5ex]{2pt}{2.5ex}}}\ULon}

\def\BibTeX{{\rm B\kern-.05em{\sc i\kern-.025em b}\kern-.08em
    T\kern-.1667em\lower.7ex\hbox{E}\kern-.125emX}}

\begin{document}

    \title{AI-Driven Confidential Computing across Edge-to-Cloud Continuum}
    
    
     
    \author{Sm Zobaed}
    \convocationdate{Fall}
    \gradyear{2022}
    \degree{Doctor of Philosophy}
    \major{Computer Science}
    \supervisor{Mohsen Amini Salehi}
   \ranksupervisor{Associate Professor of Computer Science \\ The Center for Advanced Computer Studies}
    \deanofgraduateschool{Mary Farmer-Kaiser}
    \firstcommitteemember{Raju Gottumukkala}
  \rankfirstcommitteemember{Director of Research, Informatics Research Institute 
  }
    \secondcommitteemember{Sheng Chen}
  \ranksecondcommitteemember{Associate Professor of Computer Science \\ The Center for Advanced Computer Studies}
    \thirdcommitteemember{Li Chen}       
  \rankthirdcommitteemember{Assistant Professor of Computer Science \\ The Center for Advanced Computer Studies}
 \fourthcommitteemember{Xiali Hei}     
   \rankfourthcommitteemember{Assistant Professor of Computer Science \\ The Center for Advanced Computer Studies}
 \fileforabstract{B-abstract}
  \filefordedication{B-dedicatory}
    \fileforacknowledgement{B-acknowledgement}





\prefatorypages

    \include{C1-Intro_2}
    \include{C2-RW}
    \include{C2-2}

\include{C2-3}

    \include{C2-4}
    \include{C3-clus}

    \include{C3-4-static}

    \include{C3-4-static-1}

    \include{C3-4-static-2}
    \include{C3-5-dyna}
    \include{C3-6-secu}
    \include{C3-7-perf}
    \include{C3-7-1}
    \include{C3-8-sum}
    \include{C4-saed}

    \include{C4-3-archi}

    \include{C4-4-plug}
    \include{C4-5-perf}

    \include{C4-6-sum}

    \include{C5-Memo}
    \include{C5-3-archi}

    \include{C5-4-perf}

    \include{C5-5-sum}
    \include{C6-Conc}
    \include{C7-Future}

\bibliographystyle{IEEEtran}
\bibliography{myBib}


\closingpages
    \include{F-biosketch}

\end{document}

%% file: c1-Intro_2.tex
\chapter{Introduction}
\section{Motivation: Data Confidentiality in the Current Age }

More than half of the world's population is now connected to the internet thanks to the proliferation of information and communication technologies that have shaped today's digital world. The expeditious growth of digitalization has been producing a massive volume of data in various forms. 
It is estimated that every day 2.5 exabytes of data are being generated in which, over 80\% of the data is in unstructured (\eg audio, streaming, text) form~\cite{clustcrypt}. 
Data can range widely from a person's first and last name to sensitive (a.k.a. confidential) information such as biometric information, law-enforcement records, healthcare reports, and so on. Such confidential data must always be safeguarded to prevent unauthorized access. As an example, most of the current smartphones are featured with biometric-based security protocol and so, they retain biometric data for unlocking the device after ensuring proper authorization. If this biometric information is compromised as a result of a data breach, it could assist criminals in stealing identities, forging documents, and committing crimes.

\section{Essence of Maintaining Data Confidentiality }
Maintaining data confidentiality while data is stored either on-premises denoted as (\emph{data at rest}) is a widely known problem with numerous established encryption solutions~\cite{seth2022integrating,S3BD,sun2014privacy,zhu2020privacy}. On another front, solutions like transport layer security (TLS) protocol are globally adopted to tackle the challenge of maintaining data confidentiality during node-to-node transmission (a.k.a. \emph{data at transit}). Another state of data that needs to be protected is known as \emph{data in use} that refers to preserving data confidentiality and security while it is being accessed and processed by users. 
Although adopting an encryption technique can provide security assurance while data is stored
or transmitted, it does not guarantee data privacy when the data is being used in memory. Compared to other two states, data is most vulnerable during computing (\ie when it is in use).
The degree of data vulnerability is further elevated when owners of confidential data are either individuals or institutions that rely on cloud services for their storage demands. 

Cloud providers (\eg AWS, Azure, Google cloud) have come forward offering various services for large-scale data storing and processing, however, confidential data owners are hesitant to adopt cloud services due to the valid data-privacy concerns~\cite{S3BD,al2019cybersecurity} on the cloud data centers. In fact,
 cloud adoption increases the risks associated with ubiquitous access to the data. In fact, they provide larger attack surface that can be exploited by intruders. That is why clouds have been the target platform for numerous recent privacy violations incidents ~\cite{cloudacci,zobaedbig}.  
In one notable incident, confidential information of over three billion Yahoo users were exposed~\cite{yahooacci}. In another incident, information of over $14$ million Verizon customer accounts were exposed from the company's cloud system~\cite{verizonacci}.

Considering these incidents, currently, a large spectrum of applications ranging from personalized healthcare, search, archives, and finance to social network (\eg Twitter, Facebook) and IoT industries are under similar cloud-based data breaching threats~\cite{cloudacci}. Even if cloud providers can offer strict security control against external threats, subscribers dealing with sensitive content are still concerned, hence, cannot fully embrace cloud services due to potential of insider attacks. 
As such, securing confidential data processing both within and across a wide range of systems-- from user devices to clouds and even multi-cloud environments-- that is not fully controlled by the data owner is the pressing need of the IT industry globally. 

\section{Confidential Computing }
\subsection{\textit{Basic Definition }}~\\

There are numerous solutions to ensure data confidentiality for data at rest and in transit. However, preserving  confidentiality of data \emph{in use} remains an open problem that needs further attention. 
In this regard, the idea of \emph{confidential computing} has emerged over the recent years 
that has given birth to  hardware-enforced \emph{trusted execution environment (TEE)} systems for secure computing (\eg data processing) without compromising data privacy. 
TEE allows user-level code to allocate private regions of memory, called \emph{enclaves} to confidentially process data without trusting operating system or hypervisors~\cite{shepherd2016secure,ning2018preliminary}. Hence, it
prevents unauthorized access or modification of applications and data while they are ``in use". By that means, confidential computing enhances the data security assurances.
Recently, the use cases of confidential computing are getting popular both in industry and academia, and the total market of the concept is expected to grow at least $26\times$ over the next five years~\cite{sl2021}.

\subsection{\textit{Confidential Computing across Edge-to-Cloud Continuum }}~\\
Adoption of cloud services is virtually unavoidable to successfully store and process large volume of data; nevertheless, due to simultaneous threats arriving from both within and outside the cloud systems, confidential data owner cannot put their faith in the cloud and liberally utilize its services. Accordingly,
\emph{the goal of confidential computing on the cloud is defined as to provide the users with the secure access to third-party cloud computing services in a public domain}. However, apart from the security aspect, due to their centralized nature, clouds also suffer from high communication latency that can be detrimental for many of the IoT-based solutions that have latency constraints~\cite{deng2020edge,zobaed2021saed,hussain2020analyzing}. That is the reason for the emergence of a new computing paradigm over the past few years that goes beyond conventional cloud systems and encompasses a continuum of computing tiers---from the device tier to edge, fog, and the cloud \cite{deng2020edge,sahan2019edge,zobaed2021saed,hussain2020analyzing}.

\begin{figure}[!h]
    \centering
    \includegraphics[width=\textwidth]{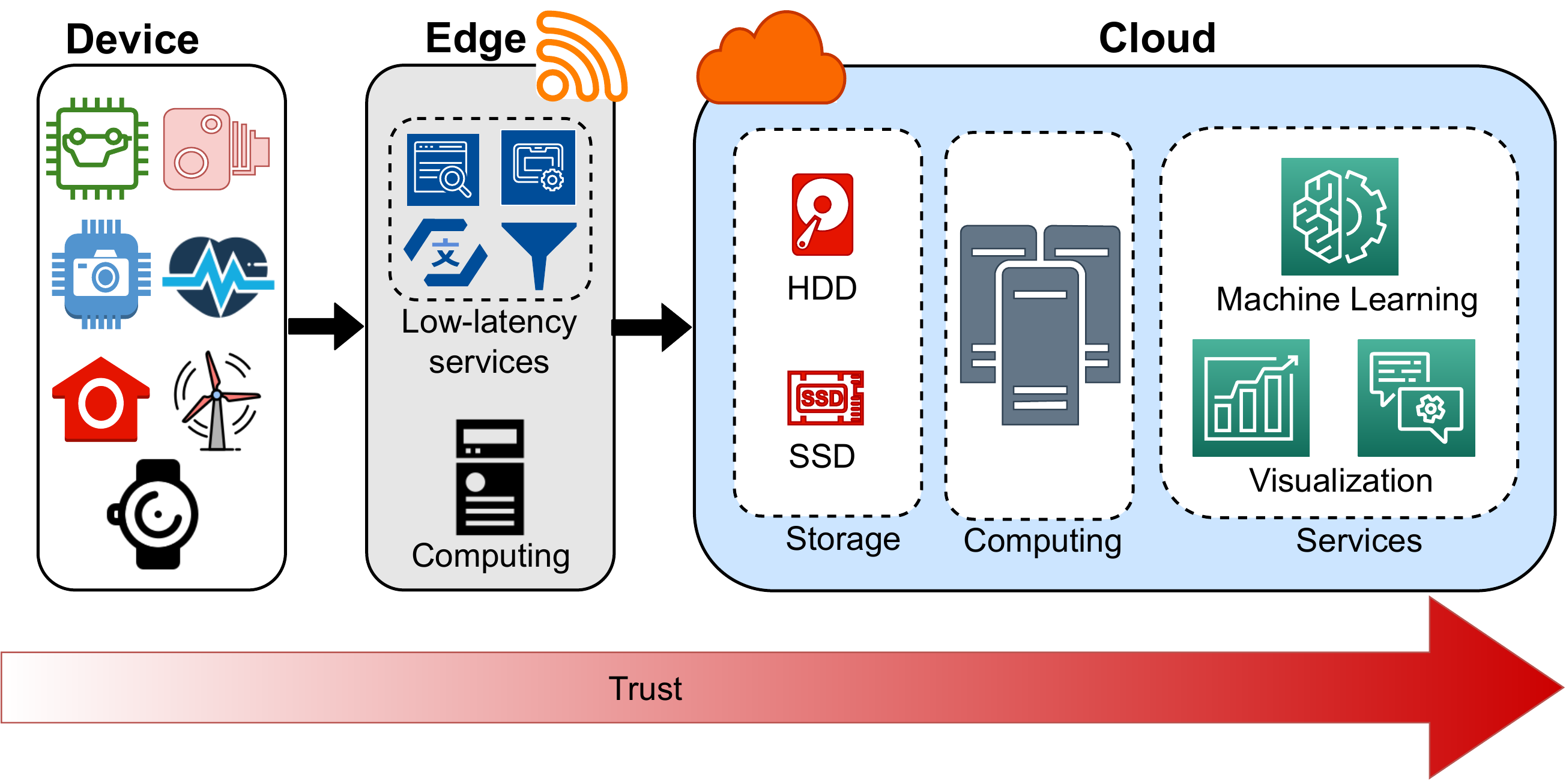}
    \caption{High-level workflow diagram of performing confidential computing on an edge-cloud system. The bottom arrow indicates the degree of trust across the continuum.}
    \label{fig:C1intro1}
\end{figure}

The device-to-cloud continuum increases the vulnerability surface beyond the cloud, hence, confidential computing solutions have to be expanded across the entire continuum to enable integrity of the IoT-based systems. 
Figure~\ref{fig:C1intro1}, represents a computing continuum  with applications span across the user-device to edge and cloud. The data generated by the user is first pre-processed on the device-tier (\eg IoT devices); Then, it is processed by the services on the edge and cloud tiers, depending on the on the low-latency and resources demands.

In our vision, confidential computing across edge-to-cloud continuum is defined as \emph{protecting the integrity and confidentiality of the users' data while are in use by the applications span across the continuum.} One challenge in providing confidential computing to across the continuum is that both the device tier (\eg UAV~\cite{almalki2021green} and smartglasses~\cite{felare2022}) and the edge tier (\eg smartphones, companion devices), often, are resource- and energy-limited and fall short in executing trusted applications needed for confidential computing. 
Trustworthiness throughout the continuum is another challenge that must be overcome. This is due to the fact that as soon as data is transited away from the user's end, the vulnerability surface expands (Figure \ref{fig:C1intro1}), and as a result, the degree of trust falls as data is sent to edge and cloud tiers. In addition, various encryption techniques such as client side encryption~\cite{amazonencry22} are adopted to encrypt data at user-premise, thereby, ensure data confidentiality while utilizing any cloud services. This is because, in this case, clouds providers are not capable of decrypting the data. The inability to decrypt  data, however, prevents accessing the data.
It is these challenges that we aim at addressing in this dissertation. 
Specifically, this dissertation investigates ways to enable confidential computing across edge-cloud while considering (a) the trustworthiness level of each tier in the continuum; and (b) the low-latency constraints of the applications. 

\subsection{\textit{Confidential Computing of Unstructured data }}~\\
An organization with a massive volume of confidential unstructured text-based data  desires a trusted application that is executed on confidential computing platform to provide secure semantic searchability over the  data in latency-sensitive manner. One instance of such organization is a law enforcement agency with encrypted crime report data, with officers who would require to search over the reports using their handheld devices while at the office or on the move in low-latency. 
In the context of confidential unstructured data processing, various searchable encryption systems (\eg~\cite{pham2019survey, zhang2020lightweight,zhang2020se,salehi2017reseed}) have been developed to enable secure search ability over the encrypted data. Upon using encrypted data, such systems build an encrypted index, which is then traversed against a search query at the search time to discover relevant documents.
\begin{figure}[!h]
    \centering
    \includegraphics[width=0.8\textwidth]{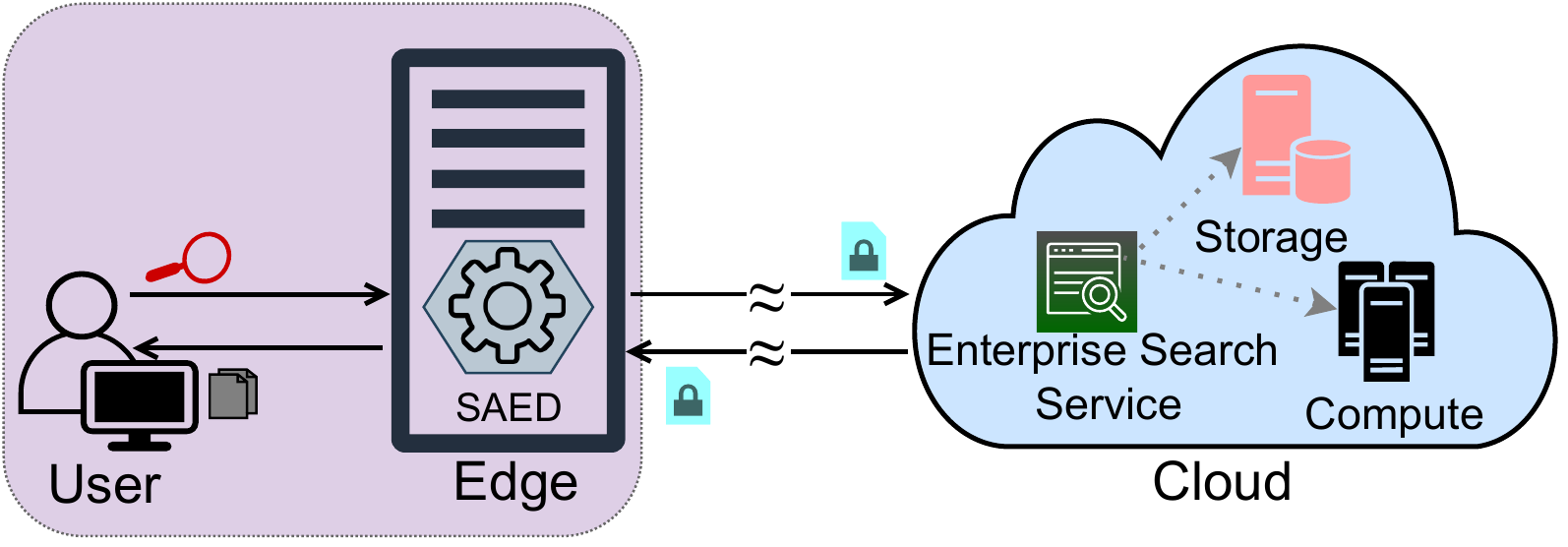}
    \caption{A high-level diagram of user-edge-cloud based three-tier architecture to facilitate smart and confidential enterprise search service.}
    \label{fig:C1intro1.1}
\end{figure}

Searching exhaustively over the whole index for a given search query prohibits the low-latency constraint of the search operation. 
Therefore, index partitioning (a.k.a. clustering) is required to prune the search space so that search can be performed over a pruned index with minimal overhead. 
Clustering is one of the crucial data analytics methods that are commonly used to group datapoints based on their shared attributes. Therefore, upon applying clustering on the search index, we can prune it into multiple subsets that can improve search time overhead in orders of magnitudes~\cite{clustcrypt,S3BD}.


It is possible that the user (\eg law enforcement officers) do not remember the specific keywords that are included in the documents they are looking for. Hence, they need to retrieve documents \emph{semantically} and contextually related to their given search query.
As example, if a officer searches for ``robbery'', he/she  can also be interested in finding documents about ``mugging'', ``theft'', or ``break in". 
In addition, since the officer performs the searches on their limited resourceful handheld devices when he/she is on the move, the solution should incur a minimal processing overhead and scale well to massive amount of unstructured text data.
To this end, a robust and secure enterprise search service in the form of a trusted application is the need of the hour to search semantically over the encrypted confidential  data. In Figure~\ref{fig:C1intro1.1}, a high-level architecture of secure enterprise search service is depicted. Such service can provide the secure search intelligence utilizing the on-premises edge resources. The high-end storage and compute resources on the cloud tier are utilized by the existing search systems to exhaustively carry out pattern matching on the entire dataset.

\section{Research Problems and Objectives}
With the aim of facilitating confidential computing across edge-to-cloud continuum, in this dissertation, we address the following research problems:

\begin{enumerate}
\item How to develop a trusted application to optimally, scalably, and securely cluster keywords in an encrypted unstructured dataset? 
\item How to cluster the data when there is dynamism in the dataset meaning that the contents are being added to or removed from? 
\item How to enable secure semantic search over encrypted data with minimum overhead?
\item How to develop a trustworthy robust encrypted enterprise search service?
\item How to manage NN models of trustworthy DL applications to stimulate their concurrent executions without compromising their inference accuracy?

\end{enumerate}

\section{Contributions}

 In light of the research topics outlined in the preceding section, this dissertation makes the following significant \textbf{contributions}:

\begin{enumerate}
    \item  Proposing two trusted applications to enable confidential clustering of encrypted unstructured data in the cloud: (1) ClustCrypt- cloud-only architecture and (2) ClusPr- edge-cloud architecture. 
    While ClustCrypt can estimate the suitable number of clusters (\textit{K}) and then cluster encrypted static data only, by incorporating edge, ClusPr can go beyond by clustering data that contain dynamism. ClusPr against other schemes in the literature, on three different test datasets demonstrates between 30\% to 60\% improvement on the cluster coherency. Moreover, we notice that employing ClusPr within a privacy-preserving enterprise search system can reduce the search time by up to 78\%, while improving the search accuracy by up to 35\%.

    \item Proposing an open-source search mechanism (titled as \emph{SAED}) that overcomes the privacy problem by separating the intelligence aspect of the search from its pattern matching aspect. In SAED, the search intelligence is provided by an on-premises edge tier and the shared cloud tier only serves as an exhaustive pattern matching search utility. Leveraging the edge tier, SAED offers personalized semantic searchability on existing cloud-based enterprise search services with low-latency constraint while maintaining data privacy. Evaluation under real settings and verified by human users demonstrate that SAED can improve the relevancy of the retrieved results by on $\approx 75\%$ for encrypted generic datasets with negligible search time overhead.

    
   \item Proposing an NN model management framework, called \emph{Edge-MultiAI} that facilitates continuous execution of confidential DL applications on the trustworthy edge server to avoid the risk of cloud execution. This is because, NN models of the trusted applications cannot be outsourced to the public clouds. With the help of approximate computing, Edge-MultiAI efficiently utilizes the edge memory such that the multi-tenancy degree is maximized without any major compromise on the inference operations. Edge-MultiAI dynamically loads the high-precision NN model for the requester application, while loading low-precision ones for others. The framework proposes \emph{iWS-BFE} policy along with three other baseline heuristic policies within Edge-MultiAI to choose the suitable model for the application performing inference, and to decide how to allocate memory for it. Experiment reveals that  Edge-MultiAI can stimulate the degree of multi-tenancy on the edge by at least $2\times$ without any major loss on the inference accuracy.
    
\end{enumerate}

\section{Dissertation Organization }

 \begin{figure}[!h]
    \centering
    \includegraphics[width=.8\textwidth]{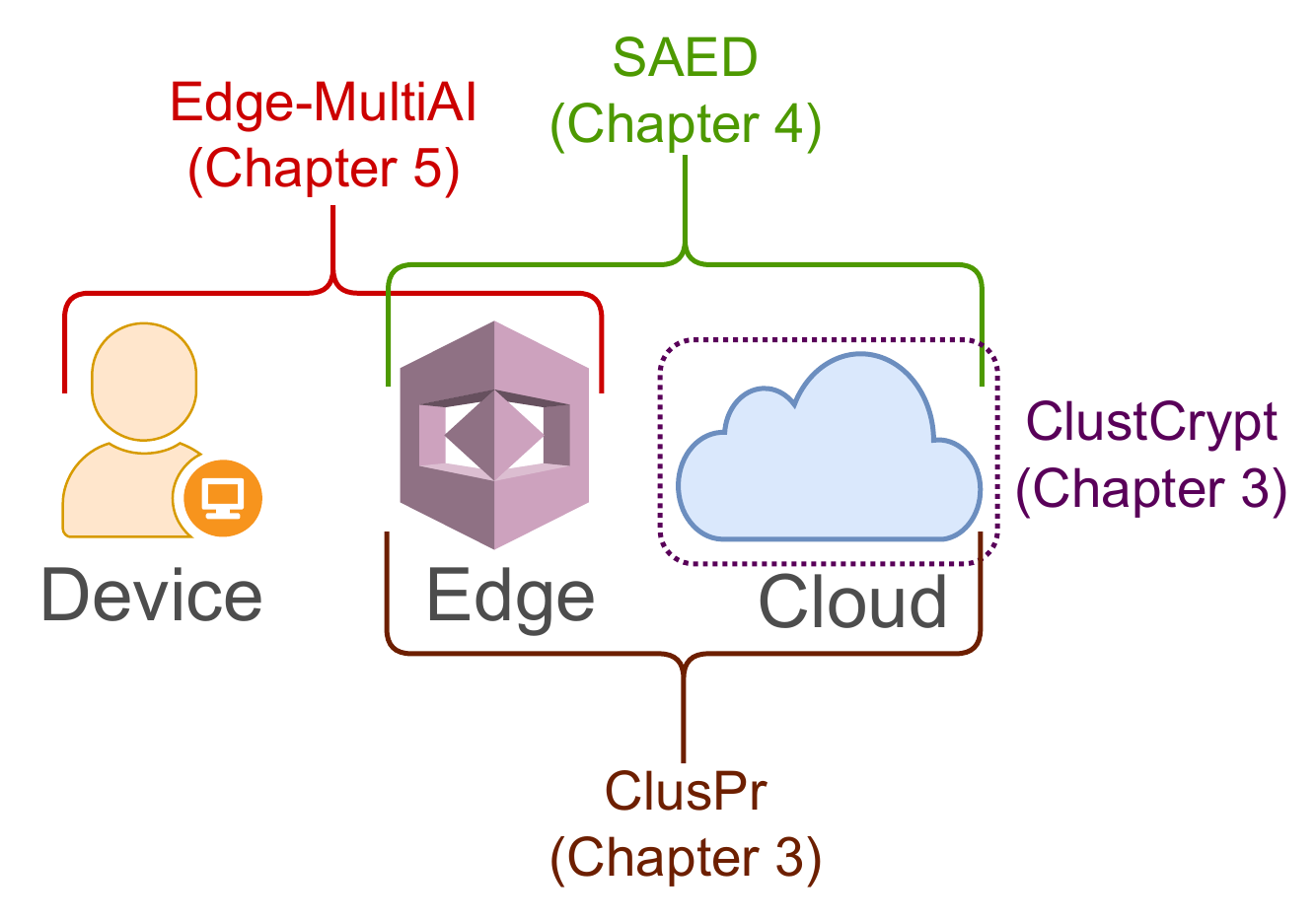}
    \caption{Interrelationship between chapters and related contribution.}
    \label{fig:c1intro_org}
\end{figure}

Figure~\ref{fig:c1intro_org} depicts the relationships between chapters and the contribution to which they are related to.
The core chapters of this dissertation are derived from several research papers published during the course of the  Ph.D. candidacy. 
\begin{itemize}
\item Chapter~\ref{chap:bg} provides background for: emergence of edge-cloud continuum, trustworthy compute tiers, enterprise search service, confidential machine learning, and explores the related research works.

\begin{itemize}
     \item \textbf{Sm Zobaed}, Mohsen Amini Salehi,
     \textit{Big Data in the Cloud} published in Encyclopedia of Big data, Springer, ISBN: 978-3-319-32009-0.
     \item \textbf{Sm Zobaed}, Md Enamul Haque, Md Fazle Rabby, Mohsen Amini Salehi,
     \textit{Senspick: Sense Picking for Word Sense Disambiguation}, Published in proceedings of the 15th IEEE International Conference on Semantic Computing (ICSC'21), Online, 2021.
          \item \textbf{Sm Zobaed}, Mohsen Amini Salehi,
     \textit{A Survey on Confidential Computing over Edge-to-Cloud Continuum}, Preparing to be submitted.
\end{itemize}

\item Chapter~\ref{chap:clus} explores the Benefits of clustering privacy-preserving text-based big data. The semantics of the data is lost after the encryption. However, the data can be clustered topically 
by utilizing the statistical characteristics of the data. This Chapter discusses our proposed approach of clustering encrypted static and dynamic data. In addition, the Chapter compares the clusters obtained by proposed approach and others in the measure of popular cluster goodness metrics (\eg Silhouette Coefficient, Davis-Boudin index). Finally, it presents a set of experiments carried out in a realistic environment to show the effectiveness of the clustering. 
\begin{itemize}
\item \textbf{Sm Zobaed}, Sahan Ahmad, Raju Gottumukkala, Mohsen Amini Salehi,
\textit{Clustcrypt: Privacy-preserving clustering of unstructured big data in the cloud}, Published in proceedings of the 21st IEEE International Conference on High Performance Computing and Communications (HPCC'19), China, 2019.  (Full code in Github repository: https://github.com/hpcclab/ClustCrypt).
\item Sahan Ahmad, \textbf{Sm Zobaed}, Raju Gottumukkala, Mohsen Amini Salehi,
\textit{Edge Computing for User-Centric Secure Search on Cloud-Based Encrypted Big Data}, Published in proceedings of the 21st IEEE International Conference on High Performance Computing and Communications (HPCC'19), China, 2019.
\item \textbf{Sm Zobaed}, Mohsen Amini Salehi,
\textit{Privacy-Preserving Clustering of Unstructured Big Data for Cloud-Based Enterprise Search Solutions}, Published in Journal of Concurrency and Computation: Practice and Experience (CCPE),Volume 34, Issue 22, 2022. (Full code in Github repository: https://github.com/zobaed11/Jorunal-Version).
\end{itemize}

\item  Chapter~\ref{chap:saed} studies the significance of secure and personalized semantic search over the encrypted data. This Chapter explains the workflow and mechanisms of the proposed secure search service architecture. Finally, it presents a set of experiments carried out in AWS Kendra service environment to show the effectiveness of the search relevancy.
\begin{itemize}
\item \textbf{Sm Zobaed}, Mohsen Amini Salehi, Rajkumar Buyya,
\textit{SAED: Edge-Based Intelligence for Privacy-Preserving Enterprise Search on the Cloud}, Published in proceedings of the 21st ACM/IEEE International Conference on Cluster Cloud and Grid Computing (CCGrid ’21), Australia, 2021. 
(Full code in Github repository: https://github.com/hpcclab/SAED-Security-At-Edge)
\end{itemize}

\item Chapter~\ref{chap:concurrent} explores multi-tenant execution of latency-sensitive DL applications on edge server. This Chapter explains the architectural overview of \name and the heuristics within \name for managing models of the multi-tenant DL applications.   
\begin{itemize}
\item \textbf{Sm Zobaed}, Ali Mokhtari, Jaya Prakash Champati†, Mathieu Kourouma, Mohsen Amini Salehi,
\textit{Edge-MultiAI: Multi-Tenancy of Latency-Sensitive
Deep Learning Applications on Edge}, Accepted in proceedings of the 15th ACM/IEEE International Conference on Utility and Cloud Computing (UCC'22), USA, 2022. 
(Full code in Github repository: https://github.com/hpcclab/SAED-Security-At-Edge)
\end{itemize}

\item Chapter~\ref{chap:conc} concludes the dissertation with a discussion of our major findings
and explores further research topics and directions that emerged during the course of this research but have not discussed in this thesis.

\end{itemize}

%% file: C2-RW.tex
\chapter{Background and Literature Study} \label{chap:bg}
This chapter provides background and a survey of other research works undertaken in the fields most related to the confidential computing across edge-to-cloud.
\section{Background} 

\subsection{\textit{Trusted Execution Environment }}~\\
A trusted execution environment (TEE) is a tamper-resistant processing environment that are leveraged to run trustworthy applications, such as biometric authentication, privacy-preserving search over encrypted data \etc with hardware-enforced isolation via a trusted hardware (\ie secure processor). TEEs have their own memory regions where trusted applications  (TAs) reside with complete isolation aiming to prevent unauthorised
accesses from generic (a.k.a. untrusted) space, manipulation of software adversaries (\eg malware, hacked OS) or even hardware adversaries (\eg channel attack) who have physical access to the platform.
TEE is considered as the kernel of confidential computing and so, recent advancement in TEE technology has brought solution ranges from  microcontrollers to large servers. The widely adopted TEE technologies are Intel SGX, AMD SEV, and ARM TrustZone. Intel SGX and AMD SEV provide TEE support for serverside and personal computers, while ARM TrustZone-based TEEs are designed for resource constraint devices (\eg edge devices, smartphones and Raspberry Pis).

\begin{figure} [H]
\centering
	\includegraphics[width=.7\linewidth]{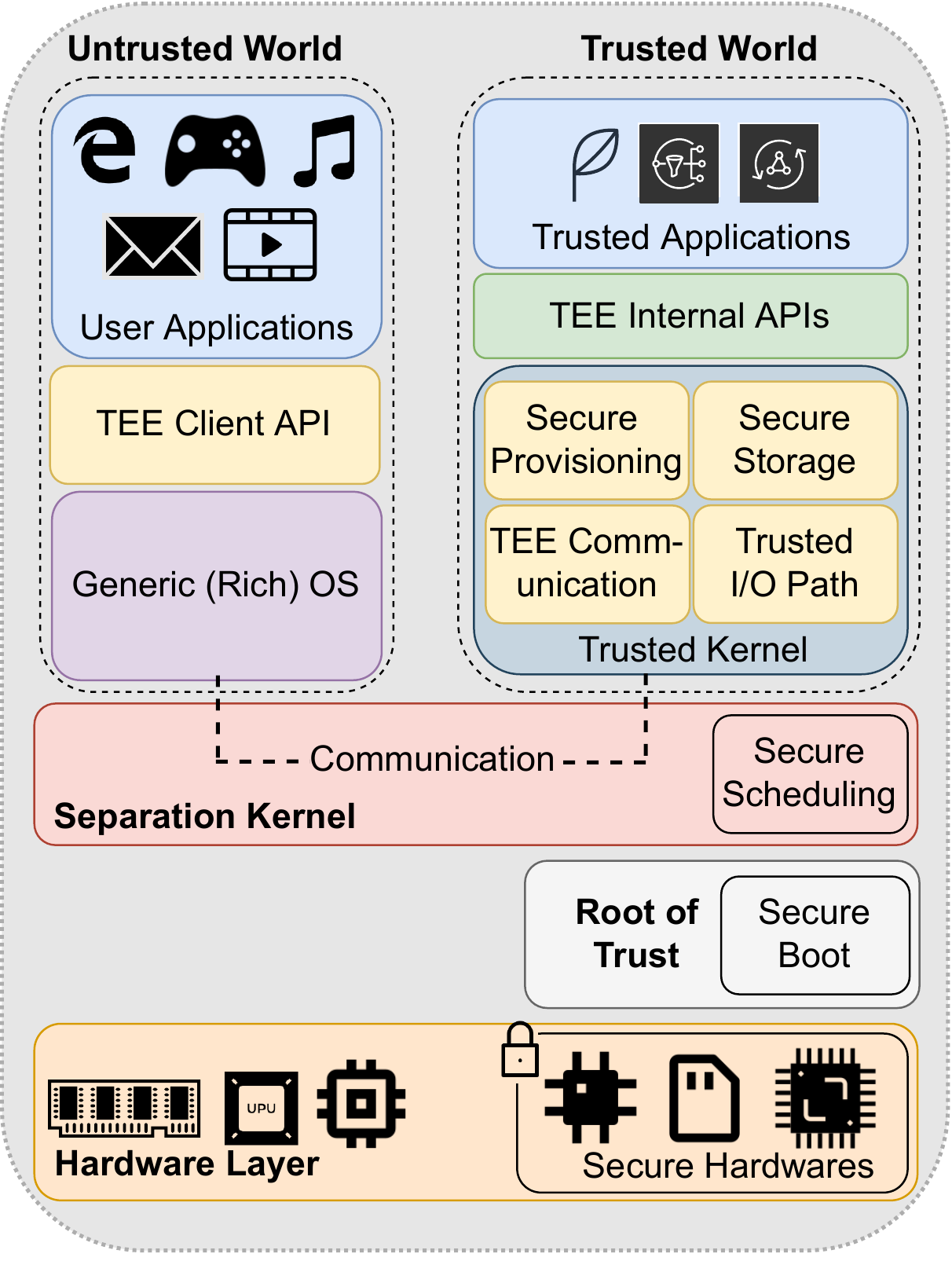} 
	\caption{High-level architectural overview of TEE building
blocks.} 
	\label{fig:tee}
\end{figure}

In Figure~\ref{fig:tee}, we represent a high-level architectural overview of TEE components. Generally, a TEE maintains two separate spaces for all trusted and generic applications, namely trusted and untrusted world. The trusted world contains a trusted OS or, kernel that communicates with TAs using the TEE Internal API, whereas generic applications from the untrusted world communicate with the trusted world via the TEE Client API. In addition, a TEE can offer secure storage utilizing the sealing abstraction (\eg GPTEE, SGX);  a trusted user-interface API for establishing secure paths between TAs and output display; a secure provisioning API for initiating TEE network connections using POSIX-style sockets.

\subsection{\textit{Trustworthy Infrastructure for Confidential Computing }}~\\

\begin{figure} [H]
\centering
	\includegraphics[width=.8\linewidth]{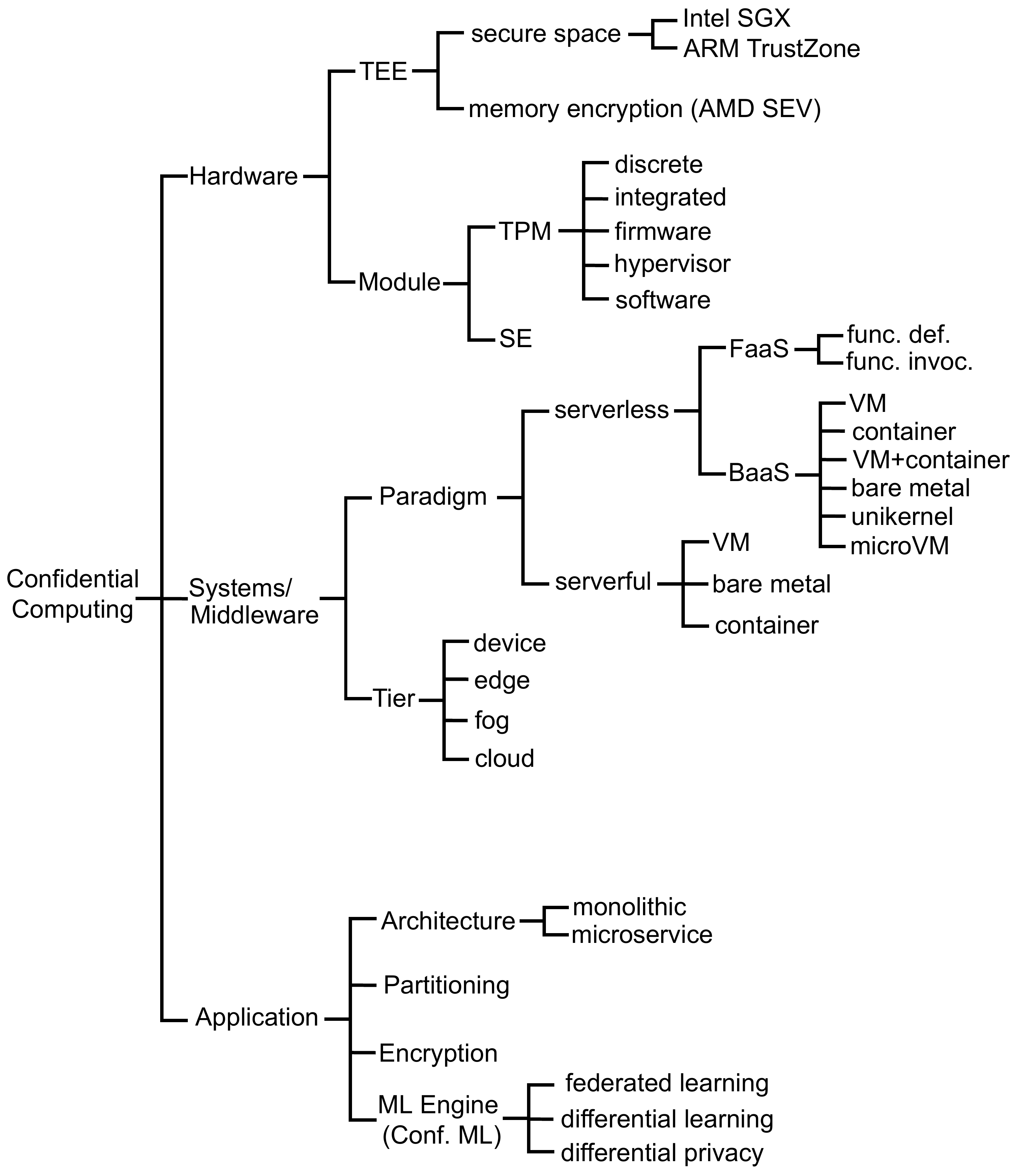} 
	\caption{A taxonomy of the scopes of confidential computing.} 
	\label{fig:tax_cc}
\end{figure}

Towards deploying confidential computing pipeline, trust should be ensured in hardware, middleware (OS), and application layer. Breaching confidentiality while execution can be occurred due to tempering any of the layers.
Figure~\ref{fig:tax_cc} represents a taxonomy of the scopes implementing confidential computing in high-level.
Dealing with big data size confidential data, confidential computing on the cloud tier is crucial where the chance of breaching always remains peak. Although confidential computing provides isolated execution environment, related hardwares, OS, and applications should be attested locally or remotely via third party (\ie trusted authority) prior to any executions. In~\cite{valadares2018achieving}, Valadares \etal provided different attestation mechanisms for preventing hardware attacks (\eg side -channel) on Intel SGX-enhanced edge-IoT systems.

\subsection{\textit{Cloud-based Enterprise Search Services over Unstructured Text Data }}~\\
Providing access and search ability over big data is essential and data without these abilities is not much of use. However, organizations that deploy cloud services for their big data are concerned about data exposure (\cite{clustcrypt, S3BD}). Hence, accessing the data without exposure is required. 
\emph{Enterprise search} cloud services are becoming increasingly popular to enable searching over and providing legitimate access to organizational big data (\cite{enter:secu}). 
Enterprise search services often maintain a dynamic index structure based on timely crawling in organizational documents. Then, the user's query is searched against the index structure and the result-set, referencing the relevant documents, is displayed to the legitimate user.

 \noindent { \textbf{Amazon cloud}} has provided a semantic enterprise search service named Kendra by leveraging machine learning and natural language processing methods. Amazon argues that their clients, such as Woodside, 3M, and Sage have improved the accuracy and speed of searching and accessing their organizational documents, in compared to other existing solutions(\cite{enter:example}). 
Semantic searchability comes with the cost of compromising the users' data privacy \cite{S3BD,sahan2019edge,clustcrypt}. This is, in fact, the trapdoor that particularly internal attackers can misuse to breach the confidentiality or even the integrity of the users' data. \textit{It is this type of attack model that we try to make the cloud-based enterprise search services resistant against.} We note that, for encrypted datasets, the current enterprise search services cannot offer anything beyond na\"ive string matching. 

We note that currently Amazon Kendra does not support enterprise search service over datasets encrypted by the user's key (aka user-side encryption). This leaves the organizational data privacy concern an open question in the cloud era. To address this concern, multiple solutions are provided to enable semantic search over user-side encrypted big data (\cite{S3BD},\cite{clustcrypt},\cite{sahan2019edge}). These solutions aim at performing real-time search operation without compromising data privacy.

Even for plain-text datasets, our investigations revealed that Kendra covers only ontological semantics in the search and it falls short in providing context-aware and personalized semantics. 
For instance, we tested Kendra to verify the ability of capturing context-aware semantics by feeding \texttt{soccer} as a query and in the result set, there were documents about \texttt{rugby}~\cite{zobaed2021saed}. In another test, \texttt{river bank} query returned documents about \texttt{commercial bank} that indicates the lack of context-awareness in the search. 

\noindent{\textbf{Microsoft Azure Cognitive Services}} provide different APIs for performing various useful NLP tasks including sentiment analysis, conversational AI, and translator on Azure cloud. Such services give the scopes of using both customizable and pretrained models to deploy anywhere either on demand or spot instance basis~\cite{azure}.  



\subsection{\textit{Emergence of Edge-to-Cloud Continuum }}~\\
 
 The edge computing paradigm~\cite{ning2018preliminary,yu2017survey,gong2020intelligent} becomes widely adopted because of the latency-sensitive feature that ensures secure real-time data processing. However, because of the constrained processing power, edge nodes are limited to process small volume (\ie light-weight) of data. Therefore, the edge paradigm is not effective processing massive volume of sensitive data in standalone manner and application developers and data owner adopt to cloud. Although a large body of research regarding performance improvement in terms of real-time processing, scalability, and output accuracy have been performed on edge-to-cloud continuum, comparatively less attentions are paid to confidential data processing ability~\cite{ning2018preliminary}.
In addition, edge computing is capable only for processing light-weight data and hence, from big data aspect, no alternative exists except processing on the cloud.

Generally, edges are dispersedly distributed and have a large attack surface. As a result, there is high chance that off-premises edge can be compromised. The recent move of the hardware vendors who design dedicated hardware-assisted TEE compatible to the both cloud and edge computing infrastructures.

\subsection{\textit{Machine/Deep Learning for Unstructured Data-driven Applications }}~\\
Due to the volume and complexity of the data, conventional data analytics tools (such as frameworks) are unable to handle unstructured data in an efficient manner. We need to employ a variety of computer vision- and natural language modeling (NLP)-based solutions that are founded on machine learning and deep learning architecture so that we can carry out data analytics on unstructured data. Recent advances in vision and natural language processing algorithms, such as convolutional neural networks, autoencoders, generative adversarial networks, long short-term memories (LSTM), transformers, and multi-headed attention mechanisms, have made it possible to deal with unstructured text data in an effective manner.

There have been several advancements made in cloud-based, AI-powered, and specific use-case driven data analytics tools as a result of the availability of artificial intelligence services from major cloud service providers such as AWS and Azure.
It is necessary to train a model by providing it with a curated dataset in order to construct machine learning and deep learning-enhanced applications for unstructured data. This is done so that the model can comprehend the underlying intricate pattern, relation, or advanced features.
It has been established, after validating the validity of the model, that the model is prepared to carry out the activity that has been stated. Following this, the model will move on to the inference phase, where it will undertake predictive analysis based on live data in order to produce results that may be acted upon.

 \subsection{\textit{Edge Multi-Tenancy for Latency-Sensitive Processing }}~\\
 An edge server is an indispensable part of an IoT-based edge-cloud system that has to continuously execute multiple (a.k.a. multi-tenant) smart (\eg deep learning) applications with low-latency and
high accuracy. However, due to memory limitation, executing latency-sensitive multi-tenant applications on an edge server can cause memory contention problem that decreases execution rate. This is because, DL applications utilize bulky Neural Network (NN) models at their kernel to infer on the inputs received from the sensors. The NN models have to be kept in memory to enable low-latency (a.k.a. \emph{warm-start}~\cite{dang2017dtrust}) inference operations. Otherwise, because the NN model size is often huge, loading it into the memory in an on-demand manner (a.k.a. \emph{cold-start}) is counterproductive and affects the latency constraint of the DL applications. As the edge servers naturally have a limited memory size (\eg 4 GB in the case of Jetson Nano \cite{nvidia}), multi-tenant execution of DL applications on them leads to a memory contention challenge across the processes~\cite{deng2020edge,gholami2021survey}.
To this point, in a multi-tenant execution environment, it is crucial to dynamically load a suitable model in memory from the set of models available to the application  such that  it neither interrupts the execution of
other applications, nor causes a cold-start inference for them.

%% file: C2-2.tex
\section{ Prior Literature for Confidential Computing for Unstructured Data}



\subsection{\textit{Privacy-preserving Unstructured Data Clustering Schemes }}~\\
\label{sec:back-privacy-clust}
Clustering is essential for various Natural Language Processing (NLP) tasks, particularly a pre-requisite for most of the advance search systems. Once the data is encrypted, only statistical characteristics of data remains. Therefore, secure data clustering is performed based on considering only the statistical properties of the cipher-texts of the document set.   

A large body of research has been undertaken to enable processing of the encrypted data (ciphertext). Zhou~\etal~proposed a linear transformation-based solution for matching queries against encrypted data while ensuring data privacy on the cloud without any intervention of the data owner~\cite{zhou2017efficient}. However, linear transformation methods support secure \textit{K}-nearest neighbor (KNN)-based query matching approaches but not the clustering. This is because clustering is not invariant to linearly transformed data. 
The optimal linear transformation has a prerequisite of knowing the true cluster means, which is not possible to obtain before generating the cluster~\cite{tarpey2007linear}.
In addition, we assume that the data are tokenized and encrypted before transferring to the cloud. Therefore, unlike~\cite{zhou2017efficient}, where the entirety of encrypted data is queried using time-consuming cryptographic calculations, we use the statistical properties of the data without revealing any meaningful part of it to the cloud. Sun \etal~proposed a searchable encryption method by forming a tree index structure that operates based on the cosine similarity and $TF \times IDF$~\cite{sun2013privacy,sun2014privacy} measures. However, the solution is not scalable for big data, because the search index can become large to the extent that it impacts timeliness of the search operation. We believe that our proposed clustering approach can be a complement to \cite{sun2013privacy,sun2014privacy} where the central index is partitioned topically into multiple small size index structures that can improve the search time and efficiency.

Homomorphic encryption has become a popular method to perform computation over the encrypted data. Several variations of the homomorphic encryption such as
fully or partially Homomorphic encryption~\cite{homomorphic:modern,xing2017mutual} have been proposed to enable privacy-preserving data processing on the cloud.
Zhu~\etal~\cite{zhu2020privacy}
proposed a secure aggregation and division
protocol based on homomorphic encryption to securely compute
clusters without tampering with the privacy of individual peers in a peer-to-peer
system. However, their clustering technique does not consider data dynamism. Pang and Wang proposed a homomorphic scheme that provides security to outsourced data uploaded from multiple parties in a twin-cloud system~\cite{pang2020privacy} that is assumed to be a semi-honest environment, whereas, we assume cloud to be untrusted in terms of storing/processing sensitive data~\cite{li2018privacy}.

Wang \etal~proposed \textit{HK}-Means++ that combines~\textit{K}-Means clustering with finding the suitable cluster numbers~\cite{wang2019research}. In addition, the work leverages homomorphic encryption scheme to solve the encrypted data manipulation, distance, and convergence calculation. Although our
work is comparable to \textit{HK}-Means++, it can only cluster static datasets. Moreover, the experiments were performed only on one dataset and it is not clear how the method performs on other datasets.
We note that the current implementations of the homomorphic encryption technique imply a high computational overhead~\cite{homomorphic:slow} which affects the real-time response of a search system, particularly, for big datasets~\cite{zhu2020privacy}.    

Vaidya and Clifton~\cite{privvad} proposed a solution to cluster encrypted datasets in which different data attributes are stored in distinct storage systems. Then, the clustering was carried out in each one of the data storage systems individually. However, this solution is time consuming and cannot serve the real-time constraint we consider in this work. 

Very few research have been undertaken in the context of privacy-preserving big data processing in real-time. S3BD, proposed by Woodworth~\etal~\cite{S3BD} is one of them. S3BD is a cloud-based secure semantic search system that performs searching over big data using cloud services without exposing any data to cloud providers. To maintain the constraint of real-time search on big data, S3BD proactively prunes the search space to a subset of the whole dataset. For the sake of pruning, they proposed a method to cluster the encrypted big data. Once the clustering is done, an \emph{abstract} (a representative set) of each cluster is maintained on the client-end to navigate the search operation to appropriate clusters at the search time. 


\subsection{\textit{Searchable Encryption and Encrypted Index }}~\\ 
Several research works have been undertaken recently to initiate different types of search over encrypted data in the cloud. Most of the searchable encryption based solutions generate cipher-text of the search query and search over encrypted text in a na\'ive straightforward way. Particularly, each word in a given document is
encrypted independently and later, the document set is sequentially scanned while searching for getting match with the queried cipher-text (encrypted query)~\cite{song2000practical,boneh2004public}. These solutions are generally chosen
as they require no storage overhead on the server but they are commonly slower~\cite{S3BD,song2000practical}.

\begin{figure} [H]
\centering
	\includegraphics[width=.8\linewidth]{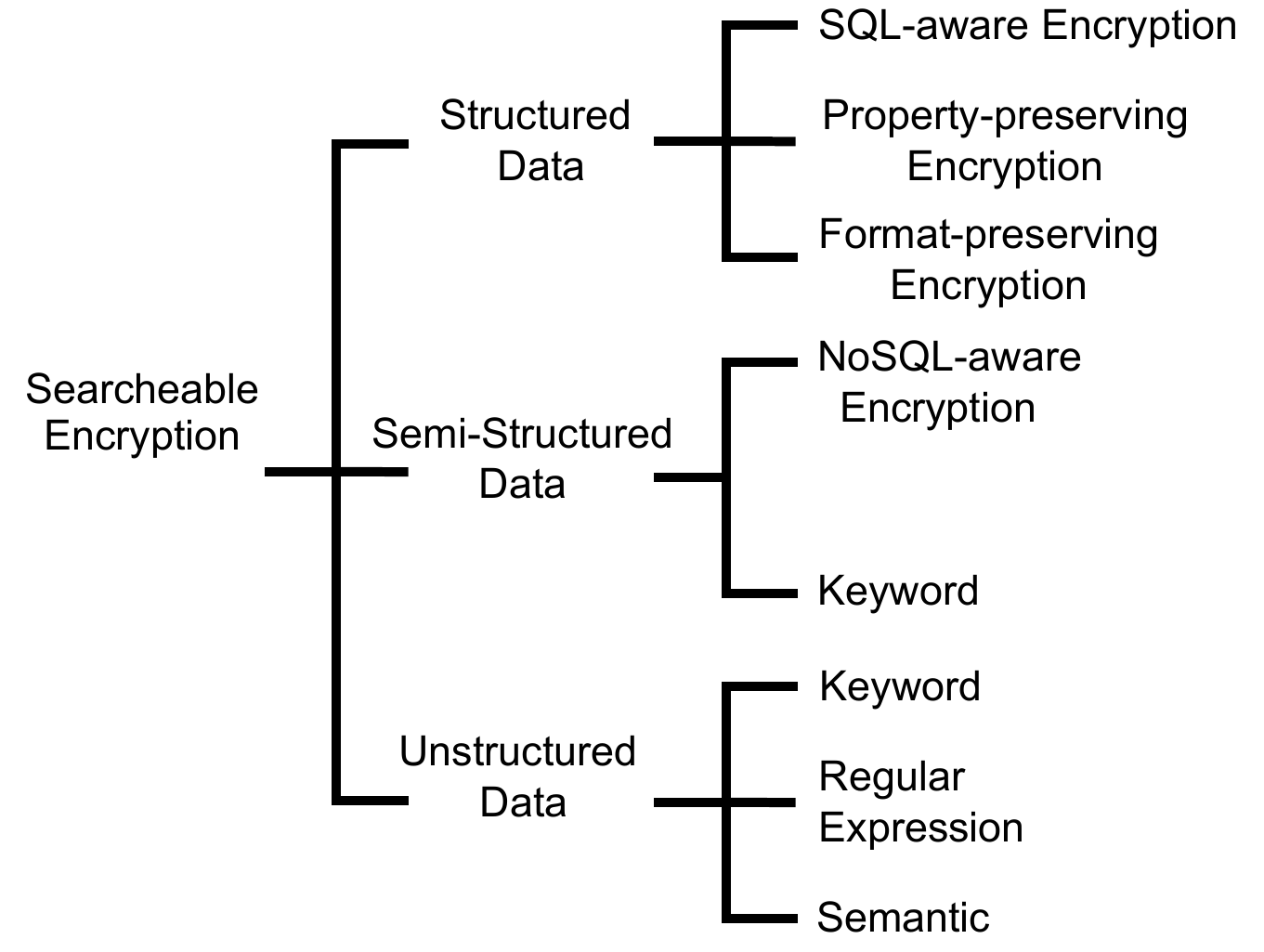} 
	\caption{Taxonomy of different types of search over encrypted big data in the cloud.} 
	\label{sencryption}
\end{figure}

Figure~\ref{sencryption} provides a high-level taxonomy of research works on the search over encrypted data in the cloud.
Privacy-Preserving query over encrypted graph-structured data (\cite{Cao:2011:PQO:2014697.2014770}), cryptDB (\cite{popa2012cryptdb}), and dragonfruit (\cite{rozier2013dragonfruit}) are the instances of search over encrypted structured data. 
SecureNoSQL (\cite{ahmadian2017securenosql}), SemiLD (\cite{kettouch2019semild}, and XSnippets (\cite{naseriparsa2019xsnippets}) are the instances of search over encrypted semi-structured data. 
REseED (\cite{salehi2017reseed}), SSE (\cite{curtmola2006searchable}) S3C (\cite{S3C}) are tools developed for  regular expression (Regex), keyword, and semantic searching respectively over unstructured big data in the cloud.

Some of the searchable encryption based solutions maintains central index structure to store store selected data from each document for the sake of making the search
operation relatively quicker and well adapted to big data aspects~\cite{S3C,S3BD,curtmola2006searchable,clustcrypt,zobaed2020privacy}.

\subsection{\textit{Privacy-Preserving Cloud-based Search Systems }}~\\
In addition to plain-text data, searching is performed on privacy-preserving data ensuring negligible chances of data leakage. Therefore, various searchable encryption-based solutions are adopted to facilitate search over such data.
Few works at the time of writing have combined the ideas of semantic searching and searchable encryption. Works that attempt to provide a semantic search often only consider word similarity instead of true semantics.  
Li \etal~\cite{li} proposed a system which could handle minor user typos through a fuzzy keyword search. Moataz~\etal~\cite{moataz} use various stemming approaches on terms in the index and query to provide more general matching. Sun~\etal~\cite{sun} present a system that used an indexing approach over encrypted file metadata and data mining techniques to capture the semantics of queries. This approach, however, builds a semantic network only using the documents that are given to the set and only considers words that are likely to co-occur as semantically related, leaving out many possible synonyms or categorically related terms.  
S3BD~\cite{S3BD}, a secure semantic search system that could search semantically over encrypted confidential big data. 

They expand their search query by incorporating semantic data extracted blindly from an ontological network.
They do not consider context-aware query expansion that created confusion for the search system while processing ambiguous or multi-context keywords in a query. To perform query processing in client devices, they end up requiring additional computational overhead in the client tier.

%% file: C2-3.tex
\subsubsection{\textbf{Semantic Representation of Search Query Keywords. }}
Query expansion is a process to seek keywords that are semantically related to a given query and fill the lexical gap between the user queries and the searchable documents. One of the widely-used methods of query expansion is Pseudo-Relevance Feedback (PRF)~\cite{liang2012exploiting,wang2019query} that extends an unsuccessful query with various related keywords and then re-ranks the search results to increase the likelihood of retrieving relevant documents. Although the PRF-based approach generally improves the retrieval effectiveness, it is sensitive to the quality of the original search results.

Latent semantic analysis~\cite{deerwester1990indexing}, latent dirichlet analysis~\cite{albishre2017effective}, and neural-based linguistic models~\cite{wang2019query,diaz2016query} are some of the query expansion methods that can obtain the semantic representation of a given query.    
In these methods, vectors are commonly referred to as \emph{word embeddings} that represent words into a low-dimensional semantic space, where the vicinity of words demonstrates the syntactic or semantic similarity between them~\cite{mikolov2013efficient}. 
However, pre-trained word embedding models, such as Word2vec \cite{mikolov2013efficient}, always generate the same vector representation for an input word, regardless of the context in which the word has appeared in. Hence, if any ambiguous keyword(s) present in a query, the underlying topic of the query could not be detected. 

WordNet \cite{miller1995wordnet} is one of the widely-used and lexically-rich resources in English that is utilized to infer the sense of ambiguous words in a given corpus. 
In WordNet, words containing similar meanings are grouped into synonym sets, whereby each set has a semantic and conceptual relationship with the other sets. Song \etal~\cite{song2007integration} and Nakade~\etal~\cite{nakade2018preliminary} evaluate the  effectiveness of utilizing WordNet for query expansion in National Institute of Standards and Technology (NIST) and Twitter datasets. They identify important key-phrases of the query and use WordNet to obtain the relevant synonym sets. Later, they utilize the synonym sets to construct the expanded query.
Nevertheless, in most of the prior research on query expansion using WordNet (\eg~\cite{leung2013collective}), the elements of the expanded query set are considered uniformly that undermines the relevancy and ranking of the result set. 


\subsection{\textit{Edge Computing for Privacy-preserving Unstructured Data Processing }}~\\

To facilitate secure personalized search, most of the enterprise search services rely on the computational capability of the client devices. Therefore, it imposes a significant overhead on the user devices (\ie thin client) to perform a secure query processing or to encrypt/decrypt user documents.  
To this regard, on-premises edge computing has potential to perform personalized search based on the historical search data stored in the client devices and also perform encryption/decryption on demand. To this context, it is ensured that the on-premises edge is fully trusted and offer uninterrupted confidential computing environment. Prior work S3BD~\cite{S3BD} imposes overhead to the client device while performing secure search over encrypted big data. On-premises edge computing is an appropriate approach for such system. By extending their two-tiered architecture with an on-premises/trusted edge can reduce a significant overhead from the client devices.

%% file: C2-4.tex
\section{Prior Literature on Multi-Tenant AI-based Executions on Edge }
\subsection{\textit{Edge AI }}~\\
Numerous research have been undertaken to explore the applications, scopes, and benefits of edge-based AI for the seamless execution of latency-sensitive smart applications~\cite{chen2019deep,murshed2021machine,deng2020edge,zhou2019edge}.
Murshed~\etal discussed different DNN-based practical applications such as video analytics and image
recognition for enabling edge AI~\cite{murshed2021machine}. 
Zhou~\etal  surveyed on various training and inference techniques for NN models on edge devices~\cite{zhou2019edge}. Chen and Ran discussed different techniques that can help to accelerate the DL training and inference on the edge-based systems~\cite{chen2019deep}. Han~\etal explored the ways to accelerate the training convergence for the edge-based architectures~\cite{deng2020edge}.
Wang~\etal surveyed the development of DL applications on edge from the latency and bandwidth perspectives~\cite{wang2020convergence}.
Zhou \etal~\cite{zhou2019edge} claimed that although higher edge intelligence reduces data offloading and improves the privacy, the latency and energy consumption overhead can increase.

\subsection{\textit{Multi-tenant Execution on Edge }}~\\
Prior studies investigated AI multi-tenancy on the edge servers. Mao~\etal proposed a mobile computing framework, MoDNN, to execute DL applications simultaneously on resource-constrained devices~\cite{mao2017modnn}. MoDNN can
partition pre-trained DNN models across several mobile devices to accelerate tensor processing with reduced device-level computing cost and memory usage while achieving $2.17\times$---$4.28\times$ speedup.

Multi-tenant execution across edge servers can lead to undesirable latency in application execution.  Ko~\etal proposed DisCo, a multi-tenant DL application execution offloading framework that enables execution of both the compute- and data-intensive parts of applications either on the device or on the edge \cite{ko2017disco}. Hadidi~\etal discussed that complex DNN models are sensitive to data loss as they depend more on the nuances in the data~\cite{hadidi2018distributed}. They mentioned losing one layer of the Inception V3 model can deteriorate the accuracy by more than $50\%$. 
They utilized distributed DNN models on IoT systems to reduce the processing and the memory footprints. 

The aforementioned research works addressed the problem of accelerating multi-tenant applications without considering the memory constraint of the edge servers. The only exception, to the best of our knowledge is \cite{xie2017towards}, in which the authors explored the executing the obstacle detection application in an autonomous vehicle with ultra low-latency constraint upon compromising with other executing applications. They proposed a reinforcement learning-based technique to scavenge memory from a non-priority application, hence, executing the obstacle detection application immediately and avoid accidents. Although their technique is effective to serve the latency-sensitive task, multi-tenant executions is out of their scope~\cite{xie2017towards}. 
In contrast to these works, we investigate the problem of memory management to increase the degree of multi-tenancy and the number of warm-start inferences, thereby, improving the practical usability of IoT-based systems.

\subsection{\textit{DNN Model Compression }}~\\
Model compression techniques allow for running a model on different resource-constrained devices. There are mainly two techniques to reduce the complexity of a given DNN model: making use of a fewer bit widths (a.k.a. \emph{quantization}) and using fewer weights (a.k.a. \emph{pruning}). These techniques have been considered individually and together to serve the purpose of model compression.  

\noindent{\textbf{Quantization}} reduces the computational resource demand at the expense of a diminutive loss in accuracy. By default, the model weights are float32 type variables which means 4 bytes are associated with each model weight with a significant amount of memory requirements. Model weights can be reduced from 32 bits to 8 bits (or even shorter~\cite{gholami2021survey}) to accelerate inference operation.

\noindent{\textbf{Pruning}} technique is applied 
to reduce the memory consumption of the model to accelerate the inference operations. An effective pruning technique removes redundant connections and/or reduces the width of a layer while ensuring a slight impact on the inference accuracy. Therefore, the pruned models are retrained to compensates the loss in accuracy. 
Failure of selecting proper pruning candidates affects inference tasks and make the pruned model futile. Some studies have also been conducted on the selection of appropriate pruning candidates.


For compatibility with the IoT devices, Yao~\etal proposed DeepIoT~\cite{yao2017deepiot}, a reinforcement learning pruning technique for DNN models in the IoT devices. However, during pruning the model parameters, 
they only considered the execution time speed-up, hence, the technique inevitably exhibits inferior inference accuracy performance.
As noted above, aggressive pruning often
substantially degrades the inference accuracy. Training and
inference with high pruning with negligible impact on the accuracy is still an open research problem~\cite{gholami2021survey}.

\subsection{\textit{Warm-Start vs Cold-Start DL Inference }}~\\

Provided the increasing complexity of DNN models, loading even compressed models to the edge memory is a burden. The problem is further complicated in scenarios where the edge server has to continuous maintain multiple applications in its memory (\ie multi-tenancy) which is cost-prohibitive.
Nonetheless, cold-start inferences should be avoided as they bring about a remarkable inference latency (see \textit{loading time} in Table \ref{tab:model_des} for more details). Some research works have been accomplished to avoid cold-start inferences. For instance, to support latency-sensitive applications, in~\cite{ma2019moving}, the authors proposed cold-start of a DNN model in the background while the user is browsing a specific web page. By utilizing system resources, their technique tracks the user's browsing activity and loads the task-specific model in parallel during browsing activity to avoid the cold-start.

\section{ Summary and Positioning of this Dissertation }


Prior literatures neither provided a confidential computing-enabled system design for confidential unstructured data processing, nor low-latency constraint and multi-tenacy requirements on the resource limited edge computing systems. For the confidential computing, we propose to logically partition the system to perform intelligence within the on-premise edge tier and use the cloud tier to perform simple and large scale processing. There has been much research accomplished in the fields of confidential computing, clustering, searchable encryption, and enterprise searching. However, there has been little done in the intersection of these fields.

It is this intersection that we position our contributions. In this regard, we develop  (1) Data clustering  for confidential static and dynamic unstructured data that can be used in delay sensitive systems such as cloud-based search systems (2) Enabling trusted enterprise semantic searching over encrypted confidential data across edge-to-cloud (3) Stimulating the ability of edge systems to execute multi-tenant DL applications with low-latency without the help of unstructured cloud systems.



%% file: C3-Clus.tex
\chapter{Privacy-Preserving Clustering for Unstructured Cloud Data  }
\label{chap:clus}

\section{Overview }
 Our preliminary research depicted in the previous chapter has confirmed that clustering is possible in encrypted data. Topic-based clustering can improve the performance of various NLP tasks, particularly, in the context of secure search system. However, forming topic-based clusters with encrypted text data is a challenge.
To overcome this challenge, clustering is achieved based on \emph{statistical semantics}. The idea is to locate tokens that are semantically close to each other in the same cluster. To achieve this, we first need to know the number of clusters ($k$) that should be created to cover topics exist in token of a given dataset. Then, we find the central tokens for each cluster and assign the rest of tokens to the most topically related clusters. 
We develop the proposed clustering solutions, namely ClustCrypt~\cite{clustcrypt} and ClusPr~\cite{zobaed2020privacy}. We replace the existing clustering policy of S3BD with the proposed schemes.

This chapter presents data-characteristics specific different clustering schemes and the architectural overview of the context where the proposed clustering schemes can be deployed.   
Note that, we consider that the frequency and co-occurrences of all tokens in the dataset are available in the proposed clustering works. 

\section{Problem Statement }
 The prior clustering schemes of S3BD and other works require to specify number of clusters to initiate partitioning that is detrimental to optimal clustering of tokens (keywords) in the most appropriate cluster. We cannot predetermine the same number of clusters and cluster size regardless of any sized datasets. In addition, if the data contains dynamism, the clustering scheme needs to accommodate new tokens added to the dataset. On the contrary, the clusters can be shrunk due to the deletion of some documents from the document set.
Therefore, in this chapter, we investigate how to optimally and scalably cluster keywords in an encrypted unstructured dataset. The outcome of this research enhances clustering of encrypted keywords by estimating the appropriate ($k$) and distributing keywords across the clusters.
We highlight the importance of  
 probabilistical semantic similarity among the encrypted tokens for clustering to measure the tendency for each token to be separate from others.

\section{Positioning of the Proposed Clustering Works }
Our proposed works are motivated from Woodworth~\etal method for topic-based clustering on encrypted tokens (aka keywords) over the central index using \textit{K}-means method~\cite{S3BD}. The cluster-wise token distribution function was determined based on the statistical data of each encrypted token. The authors used a predefined \textit{K} value. Such \textit{K} value is inefficient, because the appropriate number of clusters could be varied based on the dataset characteristics. Moreover, as the authors only considered static/unchanged data, the proposed scheme is not capable of processing dynamic data.  On the other hand, proposed solutions provides a heuristic to approximate the suitable number of clusters and then, clustering the data while maintaining the data privacy on the cloud. For a dynamic dataset, where documents are added or removed over time, because of the re-clustering operation, clusters are shrunk or expanded to reflect the dynamism of the dataset.  

Note that, ClustCrypt can effectively cluster the encrypted static data only and hence, it is not capable to manage the cluster set if dynamism exists in the data. Our next solution ClusPr can work with static, semi-dynamic, and also fully dynamic encrypted unstructured data. Particularly, we propose three different clustering schemes namely  
S-ClusPr, SD-ClusPr, and FD-ClusPr for compatibility with respect to static, semi-dynamic, and fully-dynamic datasets.
Table~\ref{tab:summary} summarizes the notable related studies in the literature and positions the contribution of the proposed clustering works with respect to them. 

\begin{table}[h]
\centering
\resizebox{\linewidth}{!}{

\begin{tabular}{|l|c|l|l|c|c|c|c|}
\hline
\textbf{Research Works}         & \textbf{\begin{tabular}[c]{@{}c@{}}Estimating\\ \#Clusters\end{tabular}} & \textbf{\begin{tabular}[c]{@{}c@{}}Encryption\\ Approach\end{tabular}} & \textbf{\begin{tabular}[c]{@{}c@{}}Cloud's\\ Trustworthiness\end{tabular}} & \textbf{\begin{tabular}[c]{@{}c@{}}Using Edge \\Computing\end{tabular}} & \textbf{\begin{tabular}[c]{@{}c@{}}Real-time \\ Support\end{tabular}} & \textbf{\begin{tabular}[c]{@{}c@{}}Dynamic Data \\ Clustering\end{tabular}} & \textbf{\begin{tabular}[c]{@{}c@{}}Multiple\\ Data Owners\end{tabular}} \\ \hline
 Wang \etal~\cite{wang2019research}      & No                    & Homomorphic         & Semi-honest                                                                  & No                                                                      & No                                                                                       & No                                                                          & No                                                                      \\ \hline
Valdiya \& Clifton~\cite{privvad}      & No                    & Homomorphic         & Semi-honest                                                                  & No                                                                      & Yes                                                                                 & No                                                                          & Yes                                                                     \\ \hline
Pang \& Wang~\cite{pang2020privacy}     & No                    & Homomorphic         & Semi-honest                                                                  & No                                                                      & Yes                                                                                 & No                                                                          & Yes                                                                     \\ \hline
 Sun \etal~\cite{sun2013privacy,sun2014privacy} & No                    & User-side           & Honest-but-curious                                                           & No                                                                      & Yes                                                                                 & No                                                                          & No                                                                      \\ \hline
 Zhu \etal~\cite{zhou2017efficient}                  & No                    & Homomorphic         & Honest                                                                       & No                                                                      & No                                                                                       & No                                                                          & Yes                                                                     \\ \hline
 Woodworth \etal~\cite{S3BD}                 & No                    & User-side           & Honest-but-curious                                                           & No                                                                      & Yes                                                                                      & No                                                                          & Yes                                                                     \\ \hline

 ClustCrypt (Proposed) \etal~\cite{clustcrypt}                 & Yes                    & User-side           & Honest-but-curious                                                           & No                                                                      & Yes                                                                                      & No                                                                          & Yes                                                                     \\ \hline

ClusPr (proposed) \etal~\cite{zobaed2020privacy}              & Yes                   & User-side           & Honest-but-curious                                                           & Yes                                                                     & Yes                                                                                      & Yes                                                                         & Yes                                                                     \\ \hline
\end{tabular}

}
\caption{Summary of the existing privacy-preserving clustering approaches and positioning our proposed works (ClustCrypt and ClusPr) with respect to them.}
\label{tab:summary}
\end{table}

\section{Architecture to Facilitate Clustering in Secure Search System }
\label{sec:clus-archi}
\subsection{\textit{Architecture: ClustCrypt }}
\begin{figure} [H]
\centering
	\includegraphics[width=.9\linewidth]{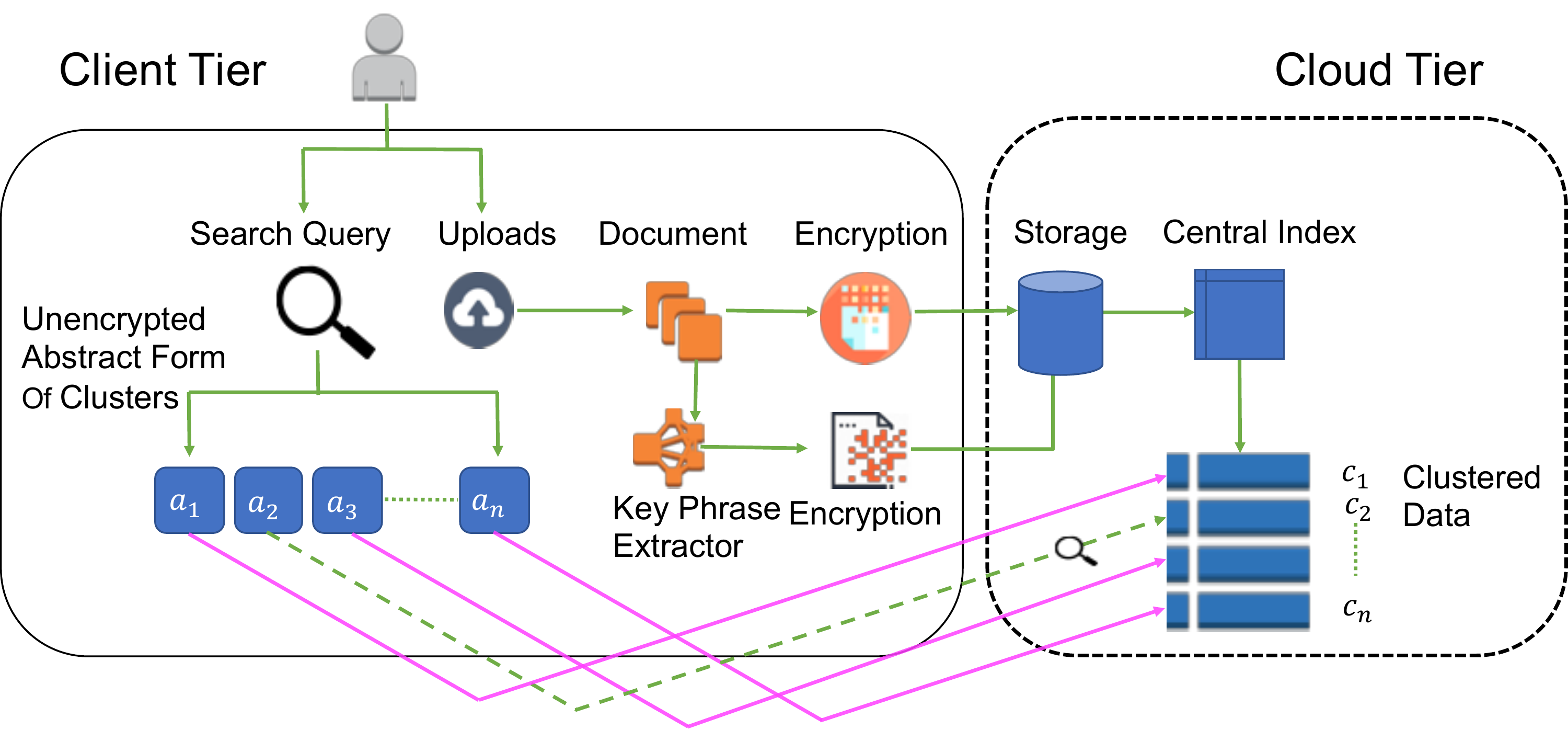} 
	\caption{High-level two-tiered (client-cloud) Search System Architecture Integrating ClustCrypt Approach.} 
	\label{fig:archi-clustcrypt}
\end{figure}

~\\
Although we implemented ClustCrypt in the context of S3BD, the approach is generic and can be deployed in other systems that require clustering of encrypted data (\eg~\cite{paladhi2019enhanced, samantaray2019efficient}). 
Figure~\ref{fig:archi-clustcrypt} presents where ClustCrypt is positioned within the S3BD system. We can see that S3BD is composed of a \emph{client tier} and a \emph{cloud tier}~\cite{S3BD}. The client tier is considered trusted and it provides \emph{upload} and \emph{search} functionalities for the users. The cloud tier is considered honest but curious, therefore, all the documents and their indexed tokens are stored in encrypted form. To enable real-time searching, the encrypted indexed tokens have to be clustered. 
The illustrated system consists of ``Client tier'' for the user (who can be the data owner as well) and ``Cloud Tier'' where the index and clusters reside. 
Users are able to upload documents to cloud or input search queries to look for documents that are semantically relevant to the queries.
In this setup, if a user wants to upload documents, first, the keywords or tokens are extracted from the original documents, then the documents and tokens are both encrypted and sent to the cloud tier. RSA deterministic encryption technique \cite{RSA} is used to encrypt documents and extracted tokens. Individual data users (\eg law-enforcement agent) who want to perform search share the same RSA key. 

The Cloud Tier maintains a \emph{central index} structure with a key-value pairs. Each key-value pair represents, respectively, an encrypted token, and the list of documents (locations) where the token appears in, plus the frequency of the token in each one of those documents. Homomorphic encryption~\cite{homomorphic:modern} can be used to encrypt the token frequency information. However, due to the slow down imposed by processing homomorphically encrypted data~\cite{homomorphic:slow} and to practically maintain the real-time search quality, currently, the frequency information is stored in unencrypted form. Upon issuing a search query by a user, the search keywords are encrypted and searched against the central index in the Cloud Tier to retrieve the relevant documents. Upon receiving the list of matching documents, the user can download and decrypt them utilizing his/her private key.

Clusters $c_1...c_n$ are constructed based on the tokens of the index structure and to mitigate exhaustively searching the whole index structure for every single search query. The clusters are topic-based and they are constructed so that the union of the \textit{k} clusters is equivalent to the index structure. For a given search query, instead of searching the whole index, the search space is pruned and gets limited to only those clusters that are topically related to the search query. The pruning is achieved based on a set of Abstract structures (denoted $a_1...a_n$) that are sampled from each one of the clusters and reside either on the Client tier or possibly on a trusted edge server. In our prior research, we proposed to formulate user-centric Abstract for personalized search~\cite{sahan2019edge}. 
Details of the ways sampling can be accomplished are mentioned in~\cite{S3BD}. Upon issuing a  search query, the most similar abstracts to the search query are chosen and then, their corresponding clusters are searched.

\subsection{\textit{Architecture: ClusPr }}
\begin{figure} [H]
\centering
	\includegraphics[width=.9\linewidth]{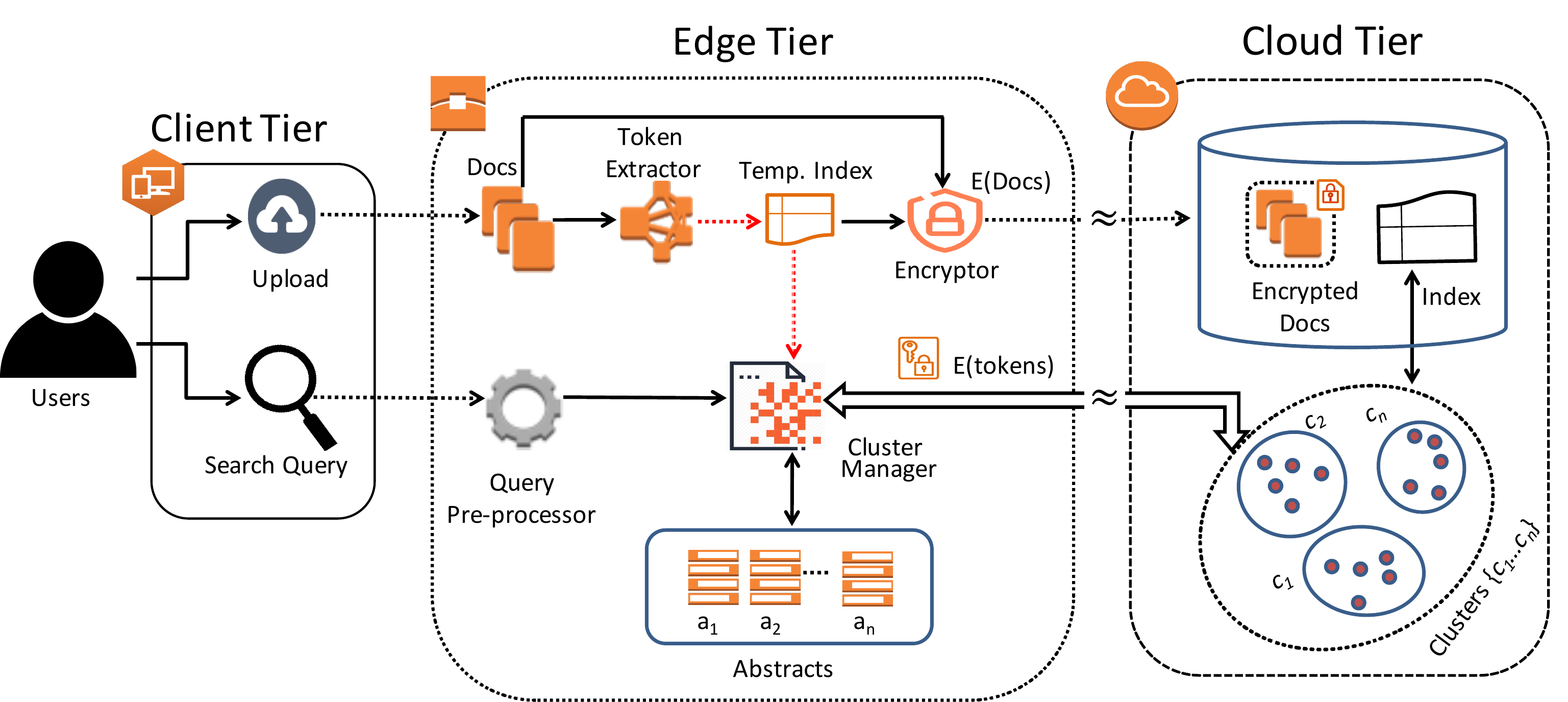} 
	\caption{Overview of the context where ClusPr is deployed in a three-tier architecture (of client, edge,
and cloud) to facilitate a secure cloud-based search service. The edge tier is assumed to be on the user premises and trusted. It is used to ease the computational overheads imposed by privacy and clustering related processes.} 
	\label{fig:archi-cluspr}
\end{figure}

~\\
Figure~\ref{fig:archi-cluspr} presents an architectural overview of the context where ClusPr is developed. The architecture represents applying ClusPr for S3BD, a cloud-based secure semantic search system that requires clustering over encrypted data~\cite{S3BD}. The architecture represents a three-tier system based on a client device, edge system, and the central cloud system unlike original S3BD~\cite{S3BD} and ClustCrypt~\cite{clustcrypt} leveraged architecture of S3BD. The edge tier resides on the user's premises (hence, is considered trusted) to relieve the client tier from processing computationally intensive tasks. This is particularly important for non-static (\ie semi-, fully-dynamic) datasets where documents have to be processed as they are uploaded to the cloud tier over time. 

In the specific context of S3BD, upon uploading a document by the user, the document is passed through \emph{Token Extractor} on the edge tiers to retrieve the keywords (aka tokens) semantically representing the document. For dynamic datasets, a temporary index structure is used to store the extracted tokens representing the occurrences of each new token in different documents. Next, the document is encrypted by the user's key and is securely stored on the cloud repository. 
Next, a \emph{Temporary Index} structure is formed based on the extracted tokens of the documents in question before encrypting and uploading them to the cloud. The Temporary Index structure shows the tokens, their frequency, and their appearances across the uploaded batch. Tokens of the Temporary Index are encrypted by the \emph{Encryptor} using the user's key. By encrypting documents as well as the extracted tokens, Encryptor preserves the data privacy on the cloud. Note that, although we can technically use homomorphic encryption to maintain the statistical properties (frequency and co-appearances), for efficiency reasons, in the current implementation, we keep the properties unencrypted. We assume that such properties do not reveal meaningful information about the data. In fact, in~\cite{wang2019research}, \textit{K}-means clustering was used over homomorphically encrypted big data and showed that the time overhead of clustering can be prohibitively expensive. In the next step, the Temporary Index is fed to the \emph{Cluster Manager} to make the suitable clustering decision on the cloud. Cluster Manager may decide to keep the existing clusters and only update them by the entries of the Temporary Index. Alternatively, upon observing a major update in the Temporary Index, the Cluster Manager decides to exhaustively re-cluster all of the tokens. Though a few of the aforementioned prior works can cluster encrypted data, they fall-short in clustering dynamic datasets, whereas, ClusPr can cluster both static and dynamic data while ensuring privacy. We explain the updating and re-clustering procedures ClusPr in Section~\secref{sec:clus-dyna}. Cluster Manager is also in charge of generating and maintaining \emph{Abstracts}. Each abstract $a_i$ is a sampled summary of a corresponding cluster $C_i$ on the cloud tier~\cite{sahan2019edge}. Abstracts are to prune the search operation and navigate the search only to clusters that are topically-related to the query. Further details about Abstracts are described in Section~\secref{cluspr:cluster-abs}.  

For static datasets, the architecture is streamlined such that the extracted tokens are encrypted and directly fed into the Index structure on the cloud tier. Once the data uploading procedure is completed, the cloud tier initiates the clustering procedure. As there is no re-clustering procedures defined for static clusters, the Cluster Manager is only in charge of generating and maintaining the abstracts~\cite{sahan2019edge}. It is noteworthy that, in the architecture of Figure~\ref{fig:archi-cluspr}, the dashed arrows located in the edge tier are to highlight the differences for dynamic datasets. Further details of the proposed static (S-Cluspr) and dynamic (SD-, FD-Cluspr) data clustering schemes are presented in Section~\ref{sec:clus-stat-cluspr} and~\ref{sec:clus-dyna} respectively. 

Similar to ClustCrypt, ClusPr uses RSA encryption technique for the encryption purpose and forms the Abstract set from the clusters. As an use case of the clustering policy in the context of a search system, upon issuing a search query by the user, the abstracts with the highest similarity to the search query are identified. Then, only the clusters associated with the abstracts are searched.

%% file: C3-4-static.tex
\section{ClustCrypt: Privacy-preserving Clustering Scheme for 
Static Unstructured Data }
\label{sec:clus-stat-clustcrypt}

In this part, first (in Section~\secref{subsec:kestimate}), we elaborate on how to estimate the appropriate number of clusters that should be formed to represent a static big dataset. Second, in Section~\secref{3.3_3.3}, we provide an algorithm to form the center of each cluster. Then, in Section~\secref{cluster-wise}, we explain methods to distribute the indexed terms across clusters. Finally, in Section~\secref{cluspr:cluster-abs}, we describe the way pruning is achieved, \ie the method that navigates a search query to relevant cluster(s).  

\subsection{\textit{Estimating the Number of Clusters for Static Datasets }}~\\
\label{subsec:kestimate}

Depending on the characteristics of a dataset and distribution of tokens in its documents, the appropriate number of clusters ($K$) can vary significantly. However, optimally determining $K$ directly impacts the accuracy of topic-based clustering and, subsequently, the efficiency of the system (\eg search application) that uses the clusters. Encrypted tokens and their metadata, including documents they appear in and their frequency, are the only available parameters to estimate $K$. The tokens and their metadata are generated by
a keyword extractor that retrieves \textit{n} single or multi-phrase tokens from each document. We assume that all documents are treated equally and the value of \textit{n} is the same across all documents in a given static dataset.

Estimating $K$ for the static dataset is performed based on the following two steps: (1) building Token-Document Frequency Matrix; and (2) constructing Normalized Matrix.

{\textbf{Step-1: Building Token-Document Frequency Matrix.}}
To be able to follow the scheme, we consider an example using \textit{five} tokens and \textit{six} documents in Table~\ref{ini mat}. We initialize a token-document matrix \mtx A from the index structure. In the matrix, each row represents a token and each column represents a document. Although our approach does not deal with plain-text tokens, just for further readability, in the Table~\ref{ini mat}, we redundantly show the plain-text tokens (in ``Word" column) along with their encrypted forms (in ``Hash" column).
Each entry $a_{i,j}$ of matrix \mtx A represents the frequency of $i^{th}$ token in $j^{th}$ document (denoted as $f(i,j)$).  
\begin{table}[h]
	\centering
	\caption{Token-Document Frequency Matrix \mtx A, built based on the index structure}
	\label{ini mat}
	\begin{tabular}{|l|l|c|c|c|c|c|c|}
		\hline
		\textbf{Word}       & \textbf{Hash}                                      & \multicolumn{1}{l|}{\textbf{d\textsubscript{1}}} & \multicolumn{1}{l|}{\textbf{d\textsubscript{2}}} & \multicolumn{1}{l|}{\textbf{d\textsubscript{3}}} & \multicolumn{1}{l|}{\textbf{d\textsubscript{4}}} & \multicolumn{1}{l|}{\textbf{d\textsubscript{5}}} & \multicolumn{1}{l|}{\textbf{d\textsubscript{6}}} \\ \hline
		
		\textcolor {gray}{Book}  & \textbf{Uh5W}                                   & 30                               & 0                                & 23                               & 4                                & 40                               & 0                                \\ \hline
		\textcolor {gray}{Solve} & \textbf{/Vdn}                                    & 5                                & 0                                & 0                                & 60                               & 34                               & 0                                \\ \hline
		\textcolor {gray}{Traffic}    & \textbf{oR1r}                                  & 0                                & 23                               & 0                                & 30                               & 0                                & 0                                \\ \hline
		\textcolor {gray}{Net}    & \textbf{vJHZ} & 52                               & 49                               & 0                                & 23                               & 0                                & 26                               \\ \hline
		\textcolor {gray}{Enter}    & \textbf{tH7c}                                  & 0                                & 45                               & 68                               & 0                                & 3                                & 5                                \\ \hline
		
	\end{tabular}
\end{table}

For a big dataset, the matrix size can be prohibitively large and sparse. To avoid this, we trim the matrix to include only the tokens that are influential in building clusters. We define \textit{document co-occurrences} as the number of documents containing a particular token. Then, to build the token-document frequency matrix \mtx A, we only take into account tokens whose document co-occurrences are either greater than or equal to the mean value of the document co-occurrences across the whole dataset. 

\noindent \textbf{Step-2: Constructing Normalized Matrix.}
To make the relationship among tokens and documents quantifiable and comparable, we need to normalize the token-document frequency matrix. Considering that $a_{i,j}$ represents the strength of association between token $t_i$ and document $d_j$, the maximum value in column $j$ of the token-document frequency matrix represents the token with the highest association with document $d_j$. 
Hence, for normalization, we divide the value of each entry of \mtx A to the highest value in the corresponding column of the matrix and the result is stored in a new matrix, called matrix \mtx N. The value for each entry $n_{i,j}$ is formally calculated based on Equation \ref{fe2}.
\begin{equation}
\label{fe2}
n_{i,j}=\frac{a_{i,j}}{\displaystyle\max_{\forall i} a_{i,j}} 
\end{equation}

For the example provided in Table~\ref{ini mat}, the normalized matrix \mtx N is presented in Table \ref{Normalized_mat}. 
\begin{table}[H]
	\centering
	\caption{Normalized Token-Document matrix \mtx N}
	\label{Normalized_mat}
	\begin{tabular}{|l|l|c|c|c|c|c|c|}
		\hline
		\textbf{Word}       & \textbf{Hash}                                      & \multicolumn{1}{l|}{\textbf{d\textsubscript{1}}} & \multicolumn{1}{l|}{\textbf{d\textsubscript{2}}} & \multicolumn{1}{l|}{\textbf{d\textsubscript{3}}} & \multicolumn{1}{l|}{\textbf{d\textsubscript{4}}} & \multicolumn{1}{l|}{\textbf{d\textsubscript{5}}} & \multicolumn{1}{l|}{\textbf{d\textsubscript{6}}} \\ \hline
		\textcolor {gray} {\scriptsize{Book}}  & \textbf{\scriptsize{Uh5W}}                                   & 0.58                              & 0                                & 0.34                              & 0.07                              & 1                                & 0                                \\ \hline
		\textcolor {gray}{\scriptsize{Solve}} & \textbf{\scriptsize{/Vdn}}                                    & 0.1                               & 0                                & 0                                & 1                                & 0.85                              & 0                                \\ \hline
		\textcolor {gray}{\scriptsize{Traffic}}    & \textbf{\scriptsize{oR1r}}                                  & 0                                & 0.47                              & 0                                & 0.5                               & 0                                & 0                                \\ \hline
		\textcolor {gray}{\scriptsize{Net}}    & \textbf{\scriptsize{vJHZ}} & 1                                & 1                                & 0                                & 0.38                              & 0                                & 1                                \\ \hline
		\textcolor {gray}{\scriptsize{Enter}}    & \textbf{\scriptsize{tH7c}}                                  & 0                                & 0.92                              & 1                                & 0                                & 0.08                              & 0.2                              \\ \hline
	\end{tabular}
\end{table}
\noindent \textbf{Step-3: Building Probabilistic Matrices \mtx R and \mtx S}
The goal, in this step, is to calculate the topic similarity among encrypted tokens. 
For that purpose, we need to calculate the probability that topic of a token shares similarity with other tokens. We hypothesize that tokens that co-occur across documents are likely to share the same topic. Besides, the magnitude of similarity between two tokens could be influenced by the tokens' distribution across the dataset. For instance, specific terms 
appear only in a few documents and are not widely distributed throughout the dataset. 
Such sparsely distributed tokens have low co-occurrences with other tokens which increases the diversity of topics in a dataset and potentially raises the required number of clusters ($K$). 
We leverage the normalized matrix (\mtx N) to perform a two-phase probability calculation that
 yields a matrix (denoted as \mtx Q) representing token-to-token topic similarity. 
\begin{table}[H]
\centering
\caption{Matrix \mtx R is built based on normalized matrix \mtx N to represent the importance of each token across all documents}
\label{R mat}
\begin{tabular}{|l|l|c|c|c|c|c|c|}
\hline
\textcolor {gray}{\footnotesize{Word}}       & \textbf{\footnotesize{Hash} }                                     & \multicolumn{1}{l|}{  \textbf{d\textsubscript{1}}} & \multicolumn{1}{l|}{\textbf{ d\textsubscript{2}}} & \multicolumn{1}{l|}{\textbf{  d\textsubscript{3}}} & \multicolumn{1}{l|}{\textbf{ d\textsubscript{4}}} & \multicolumn{1}{l|}{\textbf{  d\textsubscript{5}}} & \multicolumn{1}{l|}{  \textbf{ d\textsubscript{6}}} \\ \hline
\textcolor {gray}{\footnotesize{Book}}  & \textbf{\footnotesize{Uh5W}}                                   & \footnotesize{0.29}                              & \footnotesize{0}                                & \footnotesize{0.17}                              & \footnotesize{0.04}                              & \footnotesize{0.50}                              & \footnotesize{0}                                \\ \hline
\textcolor {gray}{\footnotesize{Solve}} & \textbf{\footnotesize{/Vdn}}                                    & \footnotesize{ .05}                        & \footnotesize{ 0}                          & \footnotesize{ 0}                          & \footnotesize{ 0.51}                        & \footnotesize{0.43}                        & \footnotesize{ 0}                          \\ \hline
\textcolor {gray}{\footnotesize{Traffic}}    & \textbf{\footnotesize{oRir} }                                 & \footnotesize{ 0}                          & \footnotesize{ 0.48}                        &\footnotesize { 0}                          & \footnotesize{ 0.52}                        & \footnotesize{ 0}                          & \footnotesize{ 0}                          \\ \hline
\textcolor {gray}{\footnotesize{Net}}    & \textbf{\footnotesize{vJHZ}} & \footnotesize{0.29}                        & \footnotesize{0.29}                        & \footnotesize{0}                          & \footnotesize{ 0.11}                          & \footnotesize{ 0}                        & \footnotesize{0.29}                        \\ \hline
\textcolor {gray}{\footnotesize{Enter} }   & \textbf{\footnotesize{tH7c}}                                  &\footnotesize { 0}                          &\footnotesize { 0.42}                        & \footnotesize{ 0.45}                        &\footnotesize { 0}                          & \footnotesize{ 0.03}                        & \footnotesize{ 0.09}                        \\ \hline
\end{tabular}
\end{table}
In the first phase, we calculate the \emph{importance} of each token to each document. The importance of token $t_i$, in document $d_j$, denoted as $\tau_{i,j}$, is defined based on Equation~\ref{eq:cont}.
 \begin{equation}\label{eq:cont}
\tau_{i,j} = \frac{n_{i,j}}{\displaystyle\sum_{\forall k} n_{i,k}}   
\end{equation}

Considering Equation~\ref{eq:cont} and matrix \mtx N, we generate matrix \mtx R whose entries represent the importance of each token across all documents. In fact, each entry $r_{i,j}$ of \mtx R represents the probability of choosing a document $d_j$, having token $t_i$. That is, $r_{i,j}=\mathbb P(t_i,d_j)$. 
In our example, Table \ref{R mat} shows the matrix \mtx R obtained from the matrix \mtx N (shown in Table~\ref{Normalized_mat}).

In the second phase, we calculate the importance of each document to each token. The importance of document $d_j$ for term $t_i$, denoted by $\delta_{j,i}$ and is defined 
based on Equation~\ref{frt}. 
\begin{equation}
\label{frt}
\delta_{j,i} = \frac{n_{j,i}}{\displaystyle\sum_{\forall q} n_{q,i}}
\end{equation}

\begin{table}[H]
	\centering
	\caption{Matrix \mtx S is built from \mtx{N} to represent the importance of   each  document  with  respect  to each token}
	\label{s mat}
	\begin{tabular}{|l|c|c|c|c|c|}
		\hline
		\textbf{Docs} & \multicolumn{1}{l|}{\textbf{\begin{tabular}[c]{@{}l@{}}\textcolor {gray}{Book}\\
					Uh5W\end{tabular}}} & \multicolumn{1}{l|}{\textbf{\begin{tabular}[c]{@{}l@{}}\textcolor {gray}{Solve}\\
					/Vdn\end{tabular}}} & \multicolumn{1}{l|}{\textbf{\begin{tabular}[c]{@{}l@{}} \textcolor {gray}{Traffic}\\
					oRir\end{tabular}}} & \multicolumn{1}{l|}{\textbf{\begin{tabular}[c]{@{}l@{}} \textcolor {gray}{Net}\\
					vJHZ\end{tabular}}} & \multicolumn{1}{l|}{\textbf{\begin{tabular}[c]{@{}l@{}} \textcolor {gray}{Enter}\\
					tH7c\end{tabular}}} \\ \hline
		\textbf{\ \ d\textsubscript{1}} & 0.34                                                                                       & 0.06                                                                                       & 0                                                                                      & 0.60                                                                                                                       & 0                                                                                        \\ \hline
		\textbf{\ \ d\textsubscript{2}} & 0                                                                                         & 0                                                                                         & 0.19                                                                                    & 0.49                                                                                                                       & 0.38                                                                                      \\ \hline
		\textbf{\ \ d\textsubscript{3}} & 0.17                                                                                       & 0                                                                                         & 0                                                                                      & 0                                                                                                                         & 0.45                                                                                      \\ \hline
		\textbf{\ \ d\textsubscript{4}} & .04                                                                                       & 0.51                                                                                       & 0.25                                                                                    & 0.19                                                                                                                       & 0                                                                                        \\ \hline
		\textbf{\ \ d\textsubscript{5}} & 0.52                                                                                       & 0.44                                                                                       & 0                                                                                      & 0                                                                                                                         & 0.04                                                                                      \\ \hline
		\textbf{\ \ d\textsubscript{6}} & 0                                                                                         & 0                                                                                         & 0                                                                                      & 0.84                                                                                                                       & 0.16                                                                                      \\ \hline
	\end{tabular}
\end{table}
Considering each $\delta_{j,i}$  
and \mtx N, we generate \mtx S whose entries represent the importance of each document with respect to each token. In fact, each entry $s_{i,j}$ represents the probability of choosing $t_i$ from $d_j$ (\ie we have $s_{i,j}=\mathbb P(d_j,t_i)$). 
In our example, Table~\ref{s mat} shows \mtx S obtained from \mtx N. 

\noindent\textbf{Step 4- Constructing Matrix \mtx Q to Determine the Number of Clusters}

Recall that \mtx R is a token-to-document matrix and \mtx S is a document-to-token matrix. To identify the similarity among the encrypted tokens, we multiply \mtx R and \mtx S. As the number of columns and rows of \mtx R and \mtx S are equal, it is possible to multiply matrix \mtx R with \mtx S. The resultant matrix, denoted as \mtx Q, is a token-to-token matrix and serves as the base to determine the number of required clusters. Each entry $q_{i,j}$ denotes the topic similarity between token $i$ and $j$. More specifically, $q_{i,j}$ indicates the magnitude to which token $i$ shares similar topic with token $j$ for $i\neq j$ and is calculated as $q_{i,j}= \displaystyle\sum_{\forall i,j}r_{i,j}\cdotp s_{j,i}$. 
Table \ref{c mat} shows matrix \mtx Q for the example we discuss in this section.

\begin{table}[H]
	\centering
	
	\caption{Cluster decision matrix \mtx Q is built based on the multiplication of \mtx R and \mtx S matrices}
	\label{c mat}
	\begin{tabular}{|l|c|c|c|c|c|}
		
		\hline
		\textbf{\textcolor{gray}{Word}-Hash}                                                                                       & \multicolumn{1}{l|}{\textbf{\begin{tabular}[c]{@{}l@{}}\textcolor {gray}{Book}\\ Uh5W\end{tabular}}} & \multicolumn{1}{l|}{\textbf{\begin{tabular}[c]{@{}l@{}} \textcolor {gray}{Solve}\\ /Vdn\end{tabular}}} & \multicolumn{1}{l|}{\textbf{\begin{tabular}[c]{@{}l@{}} \textcolor {gray}{Traffic}\\ oRir\end{tabular}}} & \multicolumn{1}{l|}{\textbf{\begin{tabular}[c]{@{}l@{}} \textcolor {gray}{Net}\\ vJHZ\end{tabular}}} & \multicolumn{1}{l|}{\textbf{\begin{tabular}[c]{@{}l@{}} \textcolor {gray}{Enter}\\  tH7c\end{tabular}}} \\ \hline
		\textbf{\begin{tabular}[c]{@{}l@{}}\scriptsize{\textcolor {gray}{Book}- Uh5W}\end{tabular}}                                       & 0.39                                                                                  & 0.25                                                                                  & 0.01                                                                                     & 0.18                                                                                                                       & 0.09                                                                                     \\ \hline
		\textbf{\begin{tabular}[c]{@{}l@{}}\scriptsize{\textcolor {gray}{Solve}- /Vdn}\end{tabular}}                                      & 0.26                                                                                  & 0.45                                                                                  & 0.12                                                                                     & 0.12                                                                                                                       & 0.02                                                                                     \\ \hline
		\textbf{\begin{tabular}[c]{@{}l@{}}\scriptsize{\textcolor {gray}{Traffic}-\ oRir}\end{tabular}}                                  & 0.02                                                                                  & 0.26                                                                                  & 0.21                                                                                     & 0.33                                                                                                                       & 0.18                                                                                     \\ \hline
		\textbf{\begin{tabular}[c]{@{}l@{}}\scriptsize{\textcolor {gray}{Net}- vJHZ}\end{tabular}} & 0.10                                                                                  & 0.07                                                                                  & 0.08                                                                                     & 0.58                                                                                                                       & 0.15                                                                                     \\ \hline
		\textbf{\begin{tabular}[c]{@{}l@{}}\scriptsize{\textcolor {gray}{Enter}-  tH7c}\end{tabular}}                                  & 0.09                                                                                  & 0.01                                                                                  & 0.08                                                                                     & 0.28                                                                                                                       & 0.37                                                                                     \\ \hline
	\end{tabular}
\end{table}


Diagonal entries of \mtx Q signify the topic similarity of each token with itself and dissimilarity (\ie separation) from other topics. More specifically, the value of $q_{i,i}$ indicates the magnitude that term $t_i$ does not share its topic with other terms. Therefore, we define diagonal entries ($q_{i,i}$) as \textit{separation factor}, because for each token, it represents the token's tendency to stay separate from other topics. As such, summation of the separation factors can approximate the number of clusters ($K$) needed to partition topics of a dataset. Let $m$ denote the total number of tokens in \mtx Q. Then, Equation~\ref{eq:trace} is used to approximate $K$ for a given dataset. We use the ceiling function to make $K$ an integer value.

\begin{equation}
k = \lceil\sum_{i=1}^{m} q_{i,i}\rceil
\label{eq:trace}
\end{equation}

Correctness of $K$ is verified using a hypothesis that states $K$ for a set should be higher if individual elements of the set are dissimilar, otherwise $K$ should be low~\cite{Can1990,cutting2017scatter}. Equation~\ref{eq:trace} is the core of approximating $K$. According to this equation, the maximum $K$ value can reach to $M$, when the documents are highly distinct and each individual token of the documents represents a unique topic, otherwise it is lower than $M$. 
Hence, our approach conforms with the clustering hypothesis.

%% file: C3-4-static-1.tex
\subsection{\textit{Determining Clusters' Centers }}~\\ \label{3.3_3.3}

In \textit{k}-means clustering, generally, the clusters' centers are arbitrarily chosen~\cite{aggarwal2019performance, LiuCroft}. Then, based on a distance measure function (\eg Euclidean distance~\cite{aggarwal2019performance} or semantic graph \cite{LiuCroft}), dataset elements are distributed into clusters. \textit{K}-means operates based on iteratively shifting clusters' centers until convergence. However, we realized that the extremely large number of tokens make the iterative center shifting step (and therefore \textit{k}-means clustering) prohibitively time consuming for big data~\cite{aggarwal2001surprising}. Accordingly, in this part, we are to propose a big-data-friendly method to cluster encrypted tokens. 
 
The key to our clustering method is to dismiss the iterative center shifting step. This change entails initial clusters' centers not to be chosen arbitrarily, instead, they have to be chosen proactively so that they cover various topics of the dataset. For that purpose, a na\"{i}ve method can be choosing the top \textit{k} tokens that have the highest number of associated documents. Although this approach chooses important (highly associated) tokens, it ends up selecting centers that have high document and topical overlap. To choose appropriate center tokens, we propose to choose tokens that not only have highly document association, but also cover diverse topics exist in the dataset. 

We define \emph{centrality} of a token $i$, denoted $\Phi_i$, as a measure to represent a topic and relatedness to other tokens of the same topic. Assume that tokens are sorted descendingly based on the degree of document association. Let $U$ represent the union of documents associated to the currently chosen centers. Also, for token $i$, let $A_i$ represent the set of documents associated to $i$. Then, \emph{uniqueness}~\cite{S3BD} of token $i$, denoted $\omega_i$, is defined as the ratio of the number of documents associated to $i$ but not present in $U$ (\ie $|A_i-U|$) to the number of documents associated to $i$ and are present in $U$ (\ie $|A_i\cap U|$). Uniqueness indicates the potential of a token to represent a topic that has not been identified by other tokens already chosen as centers. Particularly, tokens with uniqueness value greater than $1$ have high association to documents that are not covered by the currently chosen centers, hence, can be chosen as new centers.

Recall that each entry $c_{i,j}$ of matrix $C$ represents the topic similarity between tokens $i$ and $j$. Besides, diagonal entry $c_{i,i}$ measures separation of token $i$ from others. Therefore, the total similarity token $i$ shares with others can be obtained by $\Sigma_{\forall j | j \neq i}c_{i,j}$. Note that for token $i$, we have $\Sigma_{\forall j}c_{i,j}=1$, hence, the total similarity for token $i$ is equal to $1-c_{i,i}$. Centrality of a token is measured by the uniqueness of the token, the magnitude of similarity the token shares with others, and the magnitude of it being isolated. That is, for token $i$, centrality is defined as $\Phi_i=\omega_{i}\times c_{i,i}\times (1-c_{i,i})$.

\begin{algorithm}
\SetAlgoLined\DontPrintSemicolon
\SetKwInOut{Input}{Input}
\SetKwInOut{Output}{Output}
\SetKwFunction{algo}{algo}
\SetKwFunction{proc}{Procedure}{}{}
\SetKwFunction{main}{\textbf{ChooseCenter}}
\SetKwFunction{quant}{\textbf{CalculateUniqueness}}
\Input{$k$, $C$ matrix, and $central$ $index$ (with tokens sorted descendingly based on the degree of document association)}
\Output{Set $centers$ that includes at most $k$ center tokens}

\SetKwBlock{Function}{Function \texttt{ Choose Center($k, C, Index$):}}{end}

\Function{
	$centers \gets \emptyset$ \;
	$U \gets \emptyset$ \; 
	$\Theta \gets \{(\emptyset,\emptyset)\}$ //\small{Pairs of tokens and centrality values} \;
	\ForEach{token $i \in index$} {
		
		$\omega_i \gets \quant(i, U)$\;
     \If {$\omega_i >1$} {	
       	        	        	$U \gets U \cup U_i$ \; 
       	        	        	$\Phi_i \gets (\omega_i \times c_{i,i} \times (1-c_{i,i})) $\;
					Add pair ($i,\Phi_i$) to max-heap $\Theta$ based on $\Phi_i$ \; 
       	        	        }
	}

	$centers \gets$ \small{Extract $k$ max pairs from $\Theta$ heap  } \;
	\Return{centers} \;

}
\caption{Pseudo-code to determine clusters' centers}
\label{alg:centr}

\end{algorithm}
\vspace{-1pt}
Algorithm~\ref{alg:centr} shows the high-level pseudo-code to select maximum of $k$ centers from the set of indexed tokens of a dataset. In addition to $k$, the algorithm receives the central index and the $C$ matrix as inputs. The algorithm returns a set of at most $k$ center tokens, denoted $centers$, as output. In the beginning, the output set is initialized to null. $U$ represents the set of documents covered with the chosen centers. A heap structure, denoted $\Theta$, is used to store a pair for each token and its centrality value. 
For each token $i$, the uniqueness and centrality values are calculated (Steps 5 to 12) and the corresponding pair is inserted to the heap. Note that tokens with uniqueness lower than one do not have the potential to serve as a cluster center. In the next step, we select at most $k$ center tokens that have the highest centrality values.

\subsection{\textit{Clustering Tokens }}~\\
\label{cluster-wise}

Once $k$ tokens are chosen as cluster centers, the tokens 
 are distributed among the clusters. The distribution is performed based on the relatedness (aka distance) between the center tokens and remaining tokens. Established techniques exist to calculate such relatedness, however, most of them (\eg semantic graph \cite{LiuCroft} and Euclidean distance~\cite{aggarwal2019performance}) are not suitable for tokens sparsely distributed across the dataset~\cite{aggarwal2019performance}.
Besides, these are not designed to apply on encrypted data~\cite{LiuCroft}. 

In S3BD \cite{S3BD}, a method based on document co-occurrence is proposed to measure relatedness and cluster encrypted tokens. In this method, if two tokens are present in the same set of documents, they are considered related~\cite{S3BD}. We utilize that to measure the relatedness of tokens with cluster centers and distribute tokens to the most related cluster. To determine the relatedness between a particular token and a center, we need to calculate the \emph{contribution} and \emph{co-occurrences} metrics for the token. Let $t$ be a token in document $d$ of dataset $D$ with frequency denoted as $f(t,d)$. Then, contribution of $d$ to $t$, denoted as $\kappa(d,t)$, is defined based on Equation~\ref{contribution}. 

 \vspace{-5pt}
 \begin{equation} \label{contribution}
   \vspace{-5pt}
 	\kappa(d, t) = \frac{f(t, d)}{\sum\limits_{j \in D}{f(t,j)}}
 \end{equation}

Co-occurrence of token $t$ with center token $\gamma_x$ in document $d$ (denoted $\rho(t,d,\gamma_x)$ ) is defined as a ratio of the sum of frequencies of $t$ and center $\gamma_x$ in $d$ to the total frequencies of $t$ and $\gamma_x$ throughout the dataset. The formal presentation of co-occurrence is provided in Equation~\ref{eq:neweq1}. 

 \begin{equation} \label{eq:neweq1}
   \vspace{-5pt}
 \rho(t,d,\gamma_x) = \frac{f(t,d)+f(\gamma_x,d) }
 { \sum\limits_{j \in D}(f(t,j)+f(\gamma_x,j))} 
 \end{equation}

Based on the contribution and co-occurrence metrics, relatedness between token $t$ and $\gamma_x$ (denoted $r(\gamma_x, t)$), is defined as multiplication of these two metrics (\ie $r(\gamma_x, t) = \sum_{j\in D} \kappa (j,t) \cdotp \log{(\rho(t,\gamma_x,j))}$).

%% file: C3-4-static-2.tex
\section{S-ClusPr: Privacy-preserving Clustering Scheme For Static Unstructured Datasets}
\label{sec:clus-stat-cluspr}

In this section, we provide a detailed description of S-ClusPr scheme to cluster privacy-preserving static big datasets.
Note that S-ClusPr uses similar method to estimate suitable number of clusters ($k$) that is used in ClustCrypt in Section~\secref{subsec:kestimate}. However, we proposed more robust heuristics for the center selection and  token distribution method in ClusPr to obtain more topically segmented clusters. In Section \secref{cluspr-3.3} and~\secref{cluspr-cluster-wise}, we explain the center selection and token distribution method respectively.

\subsection{\textit{Center Selection }}~\\
\label{cluspr-3.3}

In \textit{K}-means clustering, generally, the clusters' centers are arbitrarily chosen~\cite{aggarwal2019performance, LiuCroft}. Then, based on a distance measure function (\eg Euclidean distance~\cite{aggarwal2019performance} or semantic graph ~cite{LiuCroft}), dataset elements are distributed into the clusters. \textit{K}-means operates based on iteratively shifting clusters' centers until it converges. However, we realized that the extremely large number of tokens make the iterative center shifting step (and therefore \textit{K}-means clustering) prohibitively time-consuming for big data~\cite{aggarwal2001surprising}. Accordingly, in this part, we are to propose a big-data-friendly method to cluster encrypted tokens. 

The key to our clustering method is to dismiss the iterative center shifting step. This change entails initial clusters' centers not to be chosen arbitrarily, instead, they have to be chosen proactively so that they cover various topics of the dataset. For that purpose, a na\"{i}ve method can choose the top \textit{K} tokens that have the highest number of associated documents. Although this approach chooses important (highly associated) tokens, it ends up selecting centers that have a high topical overlap. We propose to choose tokens that not only have high document association but also cover diverse topics exist in the dataset. 

We define \emph{centrality} of a token $i$, denoted $\Phi_i$, as a measure to represent a topic and relatedness to other tokens of the same topic. Assume that tokens are sorted in a descending manner, based on the degree of document association. Let $U$ represent the union of documents associated to the currently chosen centers. Also, for token $i$, let $A_i$ represent the set of documents associated to $i$. Then, \emph{uniqueness}~\cite{S3BD} of token $i$, denoted $\omega_i$, is defined as the ratio of the number of documents associated to $i$ but not present in $U$ (\ie $|A_i-U|$) to the number of documents associated to $i$ and are present in $U$ (\ie $|A_i\cap U|$). Uniqueness indicates the potential of a token to represent a topic that has not been identified by other tokens already chosen as centers. Particularly, tokens with uniqueness value greater than $1$ have high association to documents that are not covered by the currently chosen centers, hence, can be chosen as new centers.

Recall that each entry $q_{i,j}$ of matrix \mtx Q represents the topic similarity between tokens $i$ and $j$. Besides, diagonal entry $q_{i,i}$ measures separation of token $i$ from others. Therefore, the total similarity token $i$ shares with others can be obtained by $\Sigma_{\forall j | j \neq i}q_{i,j}$. Note that for token $i$, we have $\Sigma_{\forall j}q_{i,j}=1$, hence, the total similarity for token $i$ is equal to $1-q_{i,i}$. Centrality of a token is measured by the uniqueness of the token, the magnitude of similarity the token shares with others, and the magnitude of it being isolated. That is, for token $i$, centrality is defined as: $\Phi_i=\omega_{i}\times q_{i,i}\times (1-q_{i,i})$.

\begin{algorithm}
	\SetAlgoLined\DontPrintSemicolon
	\SetKwInOut{Input}{Input}
	\SetKwInOut{Output}{Output}
	\SetKwFunction{algo}{algo}
	\SetKwFunction{proc}{Procedure}{}{}
	\SetKwFunction{main}{\textbf{ChooseCenter}}
	\SetKwFunction{quant}{\textbf{CalculateUniqueness}}
	\SetKwFunction{docassoc}{\textbf{CalculateDocumentAssoc}}
	\Input{$K$, \mtx C matrix, and $Index$ (with tokens sorted descendingly based on the degree of document association)}
	\Output{ $centers$ set that includes at most $K$ center tokens}
	
	\SetKwBlock{Function}{Function \texttt{ Choose Center($k, \mtx Q, Index$):}}{end}
	
	\Function{
		$centers \gets \emptyset$ \;
		$U \gets \emptyset$ \; 
		$\Theta \gets \{(\emptyset,\emptyset)\}$ //\small{Pairs of tokens and centrality values} \;
		\ForEach{token $i \in Index$} {
			
			$\omega_i \gets \quant(i, U)$\;
			\If {$\omega_i >1$} {	
			    $A_i \gets \docassoc(i,Index)$\;
				$U \gets U \cup A_i$ \; 
				$\Phi_i \gets (\omega_i \times q_{i,i} \times (1-q_{i,i})) $\;
				Add pair ($i,\Phi_i$) to max-heap $\Theta$ based on $\Phi_i$ \; 
			}
			
		}
		
		$centers \gets$ \small{Extract $K$ max pairs from $\Theta$ heap  } \;
		\Return{centers} \;	
		
	}
	\caption{Pseudo-code to determine clusters' centers}
	\label{alg:ccentr}
	
\end{algorithm}
\vspace{-1pt}
Algorithm~\ref{alg:ccentr} shows the high-level pseudo-code to select maximum of $K$ centers from the set of indexed tokens of a dataset. In addition to $K$, the algorithm receives the central index and the \mtx Q as inputs. The algorithm returns a set of at most $K$ center tokens, denoted $centers$, as output. In the beginning, the output set is initialized to null. $U$ represents the set of documents covered with the chosen centers. A heap structure, denoted $\Theta$, is used to store a pair for each token and its centrality value. 
For each token $i$, the uniqueness and centrality values are calculated (Steps $5 - 13$) and the corresponding pair is inserted to the heap. Note that tokens with uniqueness lower than one do not have the potential to serve as a cluster center. In the next step, we select at most $K$ center tokens that have the highest centrality values.  

\subsection{\textit{Distributing Encrypted Tokens Across Clusters }}~\\
\label{cluspr-cluster-wise}

Once $K$ tokens are nominated as cluster centers, the remaining tokens of the index 
are distributed across the clusters with respect to their \emph{relatedness} (aka distance) with the center tokens. 

\begin{figure} 
 	\centering
 	\includegraphics[width=.75\linewidth]{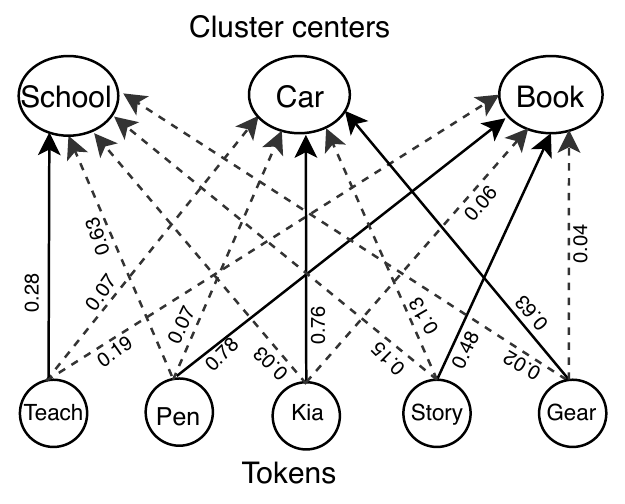}
 	\caption{\small{A bipartite graph representing the relatedness among centers and remaining tokens. The weight of each edge represents the relatedness of a token and a center. Solid lines show centers that offer the maximum relatedness for a token.}}
 	\label{fig: bipartite}
 \end{figure}

Because there is no intersection between the non-center tokens and members of the $centers$ set, we can model the token distribution across the clusters as a weighted bipartite graph where the weight of each edge represents the relatedness between a token and a center. 
Figure~\ref{fig: bipartite} depicts an example of a bipartite graph to show the relationship of each token and centers. Solid lines show the edge with the highest weight for each token that represent the cluster that a token should be distributed to. Established techniques (\eg semantic graph \cite{LiuCroft}, Euclidean distance~\cite{aggarwal2019performance}) are to calculate the relatedness, however, these methods are not appropriate for encrypted tokens that are sparsely distributed~\cite{aggarwal2019performance}~\cite{LiuCroft}. 

As encrypted tokens lose their semantics, we ought to define the relatedness between tokens based on their statistical characteristics and then leverage it to distribute each token to the cluster that offers the maximum relatedness. 

Intuitively, the relatedness measure between tokens $t_i$ and $t_j$, denoted $r(t_i,t_j)$, is defined based on the magnitude of their \emph{co-occurrences}, \ie the number of documents where the two tokens appear together \cite{S3BD,clustcrypt}. Let $F_i$ and $F_j$ respectively denote the sets of documents that $t_i$ and $t_j$ are appeared in. Then, the intuitive co-occurrence of the two tokens is $F_{co}=F_i \cap F_j$. However, a deeper analysis reveals that quantifying the relatedness only based on the cardinality of co-occurrence (\ie $|F_{co}|$) can be misleading for the two following reasons:

\emph{First}, intuitive co-occurrence ignores the magnitude of disparity across $F_i$ and $F_j$ that negatively impacts the relatedness between $t_i$ and $t_j$. The disparity is determined based on the symmetric difference (\ie we have $F_{dis} = F_i \oplus F_j$). Accordingly, to consider the impact of both co-occurrence and disparity, we define a new measure, called \emph{relative co-occurrence}, and leverage it to determine the relatedness between $t_i$ and $t_j$. 

\emph{Second}, intuitive co-occurrence ignores the importance of $t_i$ and $t_j$ in each document $d\in F_{co}$. Accordingly, to measure the co-occurrence value in each document $d$, denoted $\upsilon(t_i,t_j,d)$, we consider the importance of each one of the tokens relative to their importance across all documents of $F_{co}$. We use frequency of a token in a document to measure its importance in that document. Formally, in document $d$, we calculate the value of co-occurrence based on Equation~\ref{ceq:posCor}.

\begin{equation} \label{ceq:posCor}
\vspace{-5pt}
\upsilon(t_i,t_j,d) = \frac{f(t_i,d)}
{ \sum\limits_{\forall m \in F_{co}}f(t_i,m)} \cdotp  \frac{f(t_j,d)}
{ \sum\limits_{\forall m \in F_{co}}f(t_j,m)}
\end{equation}

Similarly, we utilize Equation~\ref{ceq:negCor} to measure the impact of disparity between two tokens in each document $d\in F_{dis}$, denoted $\varphi(t_i,t_j,d)$.

\begin{equation} \label{ceq:negCor}
\vspace{-5pt}
\varphi(t_i,t_j,d) = \frac{f(t_i,d)}
{ \sum\limits_{\forall m \in F_{dis}}f(t_i,m)} +  \frac{f(t_j,d)}
{ \sum\limits_{\forall m \in F_{dis}}f(t_j,m)}
\end{equation}

In document $d$, once we know the co-occurrence and disparity between $t_i$ and $t_j$, we can calculate the relative co-occurrence as $\rho(t_i,t_j,d)=\upsilon(t_i,t_j,d)-\varphi(t_i,t_j,d)$. Then, the relative co-occurrence across all documents of the two tokens (\ie $F_i \cup F_j$) is leveraged to calculate the relatedness between them.   

Assuming $c$ as the token that represents center of a given cluster (\ie $t_i=c\in centers$), we define relatedness between $c$ and token $t$, according to Equation~\ref{ceq:distance}. Token $t$ is distributed to the cluster whose center offers the maximum relatedness. Note that, in this equation, to emphasize the importance of token $t$ in document $d$, we also consider its frequency ratio.

\begin{equation} \label{ceq:distance}
\vspace{-5pt}
r(c, t) = \sum_{d\in (F_t \cup F_c)} \rho(t,c,d)\cdotp \frac{f(t,d)}{\sum\limits_{\forall m \in F_t}{f(t,m)}}  
\end{equation}

\subsection{\textit{Pruning Clusters to Expedite the Search Operation }}~\\
\label{cluspr:cluster-abs}
The purpose of building topic-based clusters is to achieve scalable search over big data via limiting (pruning) the search scope based on the query topic, instead of exhaustively traversing the whole index structure. For pruning, we need to identify the clusters that are semantically relevant to the search query and discard the irrelevant ones. However, pruning is a challenging task when we operate on the encrypted data in the cloud. 

To overcome the challenge, we require the topic of each cluster in plain-text, such that we can identify the clusters whose topics are semantically related to the search query and only consider those clusters for searching. For that purpose, in our previous work \cite{sahan2019edge}, we established a method to represent the topic of each cluster $C_x$ (denoted $\alpha_x$) by considering the top-n most-frequent tokens of $C_x$. The tokens of $\alpha_x$ are decrypted and maintained on the edge tier of ClusPr~in a structure called \emph{Abstract}.
Abstracts are leveraged to measure the topic similarity between a query and their corresponding clusters. In the next step, the search is conducted on the clusters that are most relevant to the query. For further details about creating abstracts and pruning operation, interested readers can refer to our earlier study~\cite{sahan2019edge,S3BD}.    

%% file: C3-5-dyna.tex
\section{Privacy-preserving Clustering Scheme For Dynamic Unstructured 
datasets }
\label{sec:clus-dyna}

\subsection{\textit{Overview }}~\\ 
In the previous section, we explained clustering of static (\eg archive) encrypted big datasets. However, many big datasets are dynamic (\eg healthcare data, criminal records)~\cite{mary2012density} and their contents change over time. In this section, we deal with clustering and subsequently searching over such datasets. We consider two types of dynamic datasets: First is the \emph{semi-dynamic} datasets whose contents are updated in batch over time (\eg Museum of Modern Art (MoMA) dataset \cite{moma21}); Second is \emph{fully-dynamic} datasets whose contents are constantly updated (\eg Twitter streams \cite{twitter}).

The latest changes on the dataset have to be reflected in the clusters. Otherwise, altered documents are not retrieved by the search system, even if they include relevant contents. In fact, the updates on the dataset affect the tokens' co-occurrences and, subsequently, the clustering arrangement. As such, the challenge is to know \emph{how the addition or deleting documents change the topics and number of clusters}. 

Given the size of big datasets, reconstructing clusters (called \emph{re-clustering}) upon arrival of every single document or a small batch of documents is time-prohibitive. Moreover, the small updates generally cause negligible changes in the co-occurrences of tokens that are unlikely to modify the arrangement of clusters. Only significant updates can cause decisive changes on the magnitude of co-occurrence and relatedness that entail re-clustering. Accordingly, the two followup questions are: \emph{when to perform re-clustering?} and \emph{how to re-cluster the tokens?} 
To address these questions, based on the type of dynamic datasets, we propose two clustering schemes in ClusPr: Semi-dynamic data clustering scheme (SD-ClusPr) and Fully-dynamic data clustering scheme (FD-ClusPr).

\subsection{\textit{Semi-Dynamic Data Clustering Scheme (SD-ClusPr) }}~\\
\label{subsec:sdcs}
In semi-dynamic datasets, topic-based clustering can be initially achieved on the first batch of documents in the dataset according to the method described in the previous section. Then, the re-clustering decisions are made depending on the changes caused by the new batch of documents. That is, we need to determine whether the change caused by the  extracted tokens of the new batch is significant or not.

To determine the significance of changes caused by the tokens of the new batch, we utilize ${\chi}^2$ (chi-square) distribution test \cite{chil19} that can identify significant changes observed in a variable of a given population. The ${\chi}^2$ test is known as \emph{testing goodness of fit} and it is represented by Equation~\ref{chi}, where $O_i$ is the observed and $E_i$ is the expected value of a particular variable in $K$ trials.

\begin{equation}
\label{chi}
{\chi}^2 = \sum \limits_{ i=1} ^{ k} [ (O_i - E_i)^2 / E_i ]
\end{equation}

We consider the number of the extracted tokens in the new batch and the number of tokens in the existing clusters. Our null hypothesis ($H_0$) is to perform re-clustering and ${\chi}^2$ test is employed to check the validity of $H_0$. If the difference between the number of new tokens and existing tokens is small, a low value of ${\chi}^2$ is obtained. For one degree of freedom with $95\%$ confidence interval, the value of ${\chi}^2=3.841$ fails to reject $H_0$.
Alternatively, if the number of tokens in the new batch is significantly smaller than the number of existing tokens, ${\chi}^2$ value becomes higher that denotes significant deviation from ${H_0}$. Then, the decision is to reject ${H_0}$ and keep the existing clusters. 

Once the re-clustering decision is made, we use the method explained in Section~\secref{sec:clus-stat-cluspr} to cluster tokens of the updated dataset. In the event that re-clustering is not achieved, the new tokens are accumulated with the of tokens of the next batches. As a result, the total number of new tokens becomes significant that leads to a lower ${\chi}^2$ value and subsequently acceptance of $H_0$.


\subsubsection{\textbf{Updating Clusters. }}
Let $U_1$ a new batch of documents that introduces a set of new tokens $T=\{t_1, t_2,..., t_n\}$ that does not exist in the existing clusters. Assume that based on the re-clustering decision method, mentioned in the previous part, we determine to keep the existing clusters $\{C_1, C_2,..., C_n\}$ to accommodate $T$. 

To distribute $t_i\in T$ to a cluster, we can measure the relatedness as explained in Section~\secref{cluster-wise}. Alternatively, we can leverage the set of abstracts $\{A_1, A_2,..., A_n\}$. As they are in the plain-text format, a more accurate relatedness measurement can be conducted using the semantic similarity, as opposed to inferring the relatedness based on token co-occurrences in documents. In this case, we use Word2Vec~\cite{mikolov2013efficient} model to calculate the relatedness of $t_i$ and abstract $A_j$. Then, $t_i$ is assigned to a cluster that offers the highest relatedness. To avoid poor assignments, we define $\theta$ as the relatedness threshold that should be reached to assign $t_i$ to $C_j$. In the event that $t_i$ cannot join any cluster, a new cluster, called $C_{new} \in C$, is formed and $t_i$ is considered as its center. The above procedure is repeated for all $t_i \in T$.
 
\begin{algorithm}
	\SetAlgoLined\DontPrintSemicolon
	\SetKwInOut{Input}{Input}
	\SetKwInOut{Output}{Output}
	\SetKwFunction{algo}{algo}
	\SetKwFunction{proc}{Procedure}{}{}
	\SetKwFunction{main}{\textbf{ChooseCenter}}
	\SetKwFunction{quant}{\textbf{CalculateUniqueness}}
	\Input{set of abstracts $A$,  $tempIndex$ 
	, $\theta$}
	
	\Output{$H$, map of new tokens to clusters}
	
	\SetKwBlock{Function}{Function \texttt{ SD-ClusPr($A, tempIndex, \theta$):}}{end}
	\Function{
			
			$T \gets tempIndex \setminus Central Index$ \;
			$H \gets \emptyset$ \;
			$A \gets \cup_{i=1}^{n}A_i$ \;
			$\Phi\gets \emptyset$\; //\small{Max-heap to find the abstract with highest similarity} \;
			\ForEach{token $t \in T$} {
				\ForEach{$a_{ij}\in A$}{
				$s \gets$ sim $(a_{ij}, t)$ \;
				    \If { $s > \theta$} {
    					Add $(s,i)$ to $\Phi$ \;
					}
				}
				\If {$\Phi \neq \emptyset$} {
				 //\small{Allocate $t$ to existing cluster\;
				 $(t,i) \gets$ Extract max pair from $\Phi$ \;
					Add $(t,i)$ to $H$\;}
				
				}
				
				\Else {
				//\small{Forming a new abstract and cluster and add it to $H$}\;
				    $A_{n+1} \gets \{t\}$ \;
			        $A \gets \cup_{i=1}^{n+1}A_i$ \;
					Add $(t,n+1)$ to $H$\;
				}
			}
		Encrypt $H$ and push it to the cloud tier \;
		
	}

	\caption{Pseudo-code to update clusters in SD-ClusPr.}
	\label{alg:semi}  
      
\end{algorithm}

\subsubsection{\textbf{Determining the value of $\theta$ Threshold. }}
We estimate the value of $\theta$ threshold by leveraging the abstracts $\{A_1, A_2,...A_n\}$. Recall that the elements of abstract $A_i$ are the ones that best represent the topic of its corresponding cluster $C_i$. We define coherency of $A_i$ as the average similarity distance across pairs of its elements. Let $\{a_{i1},..., a_{ip}\}$ be the set of elements of $A_i$. Then, coherency of $A_i$, denoted $K_i$, is defined based on Equation~\secref{eq:koh} where $sim(x,y)$ shows the similarity distance between $(x,y)\in A_i\times A_i$.

\begin{equation}\label{eq:koh}
    K_i=\frac{\sum\limits_{\forall (x,y)\in A_i\times A_i | x\neq y} Sim(x,y)}{\binom p2}
\end{equation}

Then, we define $\theta$ as the global minimum across all abstracts (\ie $\theta=\min_{\forall i} K_i$). This implies that a new token can join a cluster only if its distance does not worsen the coherency of current clusters. 
Otherwise, the new token forms its own cluster.  

Algorithm~\secref{alg:semi} shows the pseudo-code of how to update clusters in SD-ClusPr, in case we choose not to perform re-clustering. In addition to the set of abstracts ($A$) and $\theta$, the algorithm receives the set of tokens for a new document batch, which is stored in form of a temporary index. 
The algorithm returns the $H$ structure that includes the mapping of new tokens to their respective clusters. In Steps $7-9$, for each new token, we calculate the similarity distance with respect to all abstract elements $a_{ij}$ and check whether the similarity distance exceeds $\theta$ or not. If it exceeds $\theta$,
we make a pair of similarity distance and corresponding abstract number, denoted as $(a_{ij},t)$ and build max-heap $\Phi$ based on the distance (in Step $10-12$). If $\Phi$ contains any value, we extract from it the pair that has the largest value (\ie the abstract that offers the most topic similarity for $t$). Then, in Step 17, the pair of $(t,i)$ is added to $H$. 
On the contrary, if $\Phi$ is null, it implies that no cluster offers a considerable similarity to $t$, and so, in Steps $19-24$, we build a new abstract and cluster using $t$.  
Finally, we encrypt the tokens of $H$ and push it to the cloud tier. On the cloud end, \emph{cluster manager} updates its clusters based on $H$.    


\subsection{\textit{Fully-Dynamic Data Clustering Scheme (FD-ClusPr) }}~\\
\label{subsec:fdcs}
Unlike SD-ClusPr, for fully-dynamic datasets, clusters have to be formed or updated upon arrival of the documents. That is, continuous or burst arrival of new documents should trigger FD-ClusPr. Accordingly, in FD-ClusPr, we consider two cases in forming clusters: (A) \emph{initial case} that occurs when first document arrives and there is no existing cluster and (B) \emph{update case}, where the existing clusters have to be updated based on the new changes in the dataset. 

In the initial case, the edge tier extracts the set of new tokens from the uploaded document(s).
We designate the token with the highest frequency to represent the topic and choose it as the cluster center too. 
Then, the second most frequent token is clustered based on its similarity distance with the designated cluster center, according to the method discussed in Section~\secref{subsec:sdcs}. Also, to determine joining the existing cluster or forming a new one, we initialize the threshold to $\theta = 0.1$. This procedure continues until all tokens are clustered. In the update case, we apply the same method as SD-ClusPr. That is, upon uploading a document, the system decides to either perform re-clustering or updating existing clusters. 

%% file: C3-6-secu.tex
\section{Security Analysis of the Proposed Clustering Works}
\label{sec:clus-secu}
In this section, we only cover the security analysis of ClusPr. In this regard, explaining security analysis of the three-tiered architecture also covers the analysis of two-tiered ClustCrypt. 

The proposed clustering schemes are applicable in the context of searchable encryption and document retrieval systems. According to the three-tier architecture, described in Figure~\ref{fig:archi-cluspr}, client- and edge tiers are in the user premises, hence, the activities conducted and the user's key on these tiers are considered safe and trusted. The \emph{Abstract} structures are kept on the edge tier in plain-text to enable us to measure the similarity with the search phrase and performing pruning. 

On the other hand, activities performed on the cloud-tier are considered as dishonest and prone to different types of attacks. We are concerned about both internal (\ie affiliated parties) and external (\ie unaffiliated outside intruders) attackers who desire to learn the encrypted clustered tokens and documents. To explain the threats of the attackers, we provide the following preliminaries:
\textit{View}: This term denotes the portion that is visible to the cloud during any given interaction among client, edge, and server. The central index and the set of clusters $C_{1}...C_{n}$,  the trapdoor of the given encrypted search query $Q^{'}$, and the collection of encrypted documents $D^{'}$. In some models, $Q^{'}$ also contains a particular weight for each term. The search results related to $Q^{'}$ are considered as $I_{c}$. The view of expanded $Q^{'}$ and $I_c$ are symbolized as $V(Q^{'})$ and $V(I_{c})$ respectively.

\textit{Trace}: This term denotes the information exposed about $I_c$. Our aim is to allow the attacker to infer the information of $I_c$ as little as possible.

The View and Trace enclose all the information that the attacker would gain. To encrypt the document set we use probabilistic encryption model that is considered to be one of the most secure encryption techniques~\cite{S3BD,cryptdb}. This does not utilize one-to-one mapping and so, $D^{'}$ is not prone to dictionary-based attacks~\cite{dictionaryAttacks}. Each token in a cluster is deterministically encrypted. Thus, each cluster in the View, only shows an encrypted mapping of the tokens and their co-occurrences in the plain-text format.

If any type of attacker can gain access to the cloud, he/she could only understand the importance of a particular encrypted token by observing the co-occurrences. It is technically possible to encrypt co-occurrences using homomorphic encryption \cite{homomorphic:modern} and perform computation on the co-occurrences while it is in the encrypted form. However, in Section~\secref{sec:back-privacy-clust}, we discuss that this technique practically falls short on performance~\cite{Naehrig:2011} and affects the real-time behavior of the search system. As such, in the current implementation, we use co-occurrence information in the plain-text format. Note that, even when the co-occurrences are not encrypted, the attacker cannot decrypt the token. 

An attacker could obtain a Trace regarding $V(Q^{'})$. From that view, the attacker could only understand the importance of each search term from $Q^{'}$ by analyzing the associated weights of the query terms. Similar to the previous consideration, the attacker is not able to reveal the search terms from $Q^{'}$. In spite of a minimally trusted computing base, an attacker may still intend to access the system through
man-in-the-middle, either \emph{honest but compromised} or \emph{untrusted} cloud providers to attack the confidentiality of the user data.
By any means, if the attacker successfully performs a man-in-the-middle attack, he/she can access the document list $V(I_{c})$ resulting from searching $Q^{'}$ with Trace. At this point, the attacker may only obtain the documents' names with encrypted contents that are unreadable.

There are methods (\eg \cite{george2021structured}) that can be used to tackle
frequency attacks when the searches and cluster updates are predictable.
Theoretically, an attacker could build a dictionary considering all the clusters' tokens by performing frequency attack. Eventually, the attacker tries to build a clone document set $D'$ utilizing the dictionary. Although all of the tokens extracted from a particular document are sufficient to learn the topic of the document, it is not possible to unveil the whole document as we do not use all of the keywords of the document set to build the encrypted index. Besides, we encrypt the whole document at once instead of word level encryption before outsourcing it to the cloud. This procedure ensures that even if the document set is compromised on the cloud tier, it is impossible to perform a dictionary attack.

Even if the attacker knows the trace, he/she cannot understand what exactly the retrieved encrypted documents convey. Moreover, attacks can be occurred in the communication between the edge and cloud tiers. In this case, by monitoring the search process, an attacker could obtain the resultant document list for $Q'$. However, the attacker is not able to decrypt the documents, since they can be decrypted only when they are downloaded on the edge system.  

An attacker could also attempt to modify data (\eg encrypted tokens and documents) in the clusters. Such attacks can potentially tamper with the integrity of user data. However, this type of attack could be detected, because neither the edge will be able to decrypt the modified tokens to form or update \emph{Abstracts}, nor the user will be able to decrypt the retrieved documents in the original plain-text form. This is because of applying symmetric encryption (\eg AES encryption) on the user’s data with keys managed by the user. Hence, in the event that
the encrypted data are altered by an attacker, such data cannot be decrypted by the users' keys. 
Actually, protecting the user's key is crucial to restrain possible attacks. If the key is compromised, the system cannot detect the attacker and, therefore, both tokens and documents can be exposed. 

%% file: C3-7-perf.tex
\section{Performance Evaluation of Clustering}
\label{sec:clus-perf}

\subsection{\textit{Experimental Setup }}~\\
We developed working versions of ClustCrypt and ClusPr and made it available publicly in our Github\footnote{\url{https://git.io/fjDsq}},\footnote{\url{https://github.com/hpcclab/ClustCrypt}}.
We evaluate the performance of ClusPr using three distinct datasets that have different properties and volumes. We compare and analyze the clustering quality with other approaches that operate in encrypted or unencrypted domains. The experiments were conducted on a machine with two $10$-core $2.8$ GHz E5 Intel Xeon processors
and $64$ GB of memory. 

To evaluate the performance of ClusPr in handling big data, we used a subset of \texttt{Amazon Common Crawl Corpus (ACCC)} dataset~\cite{comcrawl}. The whole dataset size $\approx$ $150$ terabytes that contains different web-based contents, such as blogs and social media contents. We randomly selected $6,119$
documents that collectively form a $\approx$ $500$ GB document set. The second
dataset, named \texttt{Request For Comments (RFC)}~\cite{rfc}, is domain-
specific and includes documents about the internet and communication networks. RFC includes $2,000$ documents and its total size is
$\approx$ 247 MB. The third dataset is \texttt{BBC}~\cite{BBC} that is not domain-specific and includes news in certain categories such as technology, politics, sports, entertainments, and business. It contains $2,225$ documents and is $\approx$ $5$ MB. The reason for choosing this small
dataset is that, unlike ACCC and RFC, each document of BBC is short and we can verify clusters’ coherency manually. For each dataset, the documents are passed through Maui keyword extractor~\cite{Maui} to identify keywords semantically represent the document.

\subsection{\textit{Evaluation Metrics and Baselines from Prior Works }}~\\

For performance evaluation of the proposed works, we compare them against four other schemes, where two schemes cluster plain-text data and the other two schemes cluster encrypted data. Among the first two, one of the schemes \emph{W2V Kmeans}) is based on \textit{K}-means clustering~\cite{kmeans} where feature extraction is done based on Word2Vec~\cite{mikolov2013efficient} embedding.

Another scheme, \textit{WordNet}~\cite{millerwordnet}, is an enhanced version of \textit{K}-means that generates synonym set based on the input data and then, applies $K$-means clustering on the sets.
Token distribution in WordNet is performed based on edge counting method, proposed by Wu and Palmer~\cite{millerwordnet}.

Two encrypted clustering schemes that have been used in the comparison are namely, S3BD~\cite{S3BD}, and~\textit{HK}-means++~\cite{wang2019research}
. We have discussed S3BD and~\textit{HK}-means++ in Section~\ref{sec:back-privacy-clust}.
\emph{ClustCrypt} is the preliminary version of S-ClusPr. Their difference mainly lies in the way tokens are distributed across the clusters. In ClustCrypt, the relatedness is simply calculated based on contribution and co-occurrences metrics, whereas in S-ClusPr, the magnitude of both similarity and disparity are considered to measure the relatedness (see Section~\ref{cluspr-cluster-wise} for further details).

The goodness of clusters set can be quantified by a number of evaluation metrics. However,
evaluating the performance of a clustering scheme is not as simple as counting errors in classification algorithm. Specifically, instead of considering the absolute values of cluster labels, cluster evaluation metrics either measure the separation of clustered data similar to ground truth set of classes or internal cluster validation. Internal cluster validation denotes that members belong to the same class should be more similar than members of other classes and vice versa.  
In practice, class label information is not always available in most of the application scenarios and, therefore, internal validation metrics are the only option for validation in such situation~\cite{rodriguez2019clustering,kwon2017clustervision}.

As there is no ground truth for the considered datasets, we choose evaluation metrics that evaluate the clusters based on statistical analysis of the cluster members. We evaluate three widely-adopted clustering metrics, namely Silhouette coefficient (SC), Calinski-Harabasz index (CI), and Davies-Bouldin index (DI). 

\textit{Silhouette Coefficient (SC)} score interprets and validates intra-cluster consistency. In particular, the metric signifies how similar a cluster member is to its own cluster compared to the other clusters. The value of the SC score ranges from $-1$ to $+1$, where a high value indicates that a given member is well matched to its own cluster and poorly matched to the other ones. \textit{Calinski-Harabasz Index (CI)} denotes how well-defined (\ie well-separated) the clusters are. The CI value of clusters is calculated based on the ratio of the sum of between-clusters dispersion to the sum of inter-cluster dispersion. A higher CI value indicates a more topically separated (\ie less overlapping) clustering and vice versa. Similar to the CI metric, \textit{Davies Bouldin Index (DI)} is used to measure the goodness of separation across clusters and the reason we consider it in our evaluation is to verify the CI metric evaluation for the clusters. DI is calculated based on the ratio of within-cluster distances to the between-cluster distances. A lower DI value indicates a more topically-separated clustering and it is preferred. In addition to these metrics, we measure the \emph{clusters' coherency} to evaluate the quality of the topic-based clustering within each cluster. This is a similarity-based evaluation metric to calculate the average of all possible pair-wise token similarity for a given cluster. In fact, Coherency represents how the tokens in a cluster are related to a certain topic. Then, the average of coherency across all clusters is calculated to represent the overall quality of a certain clustering method.

We instrument the pre-trained Google News Word2vec model~\cite{mikolov2013efficient} to determine the similarity between any two given keywords. The model is a $300$-dimension vector representation of three million phrases.
The model requires a text dataset as input to build a vocabulary from the input dataset and learns vector representation of the words in the dataset. The model uses cosine similarity and provides the score ($-1\leq similarity\; score\leq 1$) for any two given tokens. We note that, the pre-trained Word2vec model operates only on plain-text tokens. Subsequently, we do not encrypt the tokens while uploading for evaluation purposes. However, the proposed schemes assume tokens to be encrypted and do not use the properties of plain-text tokens.

\subsection{\textit{Evaluation Results }}~\\
\label{subsec:evaluationres}

\subsubsection{\textbf{Evaluating Silhouette Coefficient (SC) Score. }}
\begin{figure} 
 	\centering
 	\includegraphics[width=\linewidth]{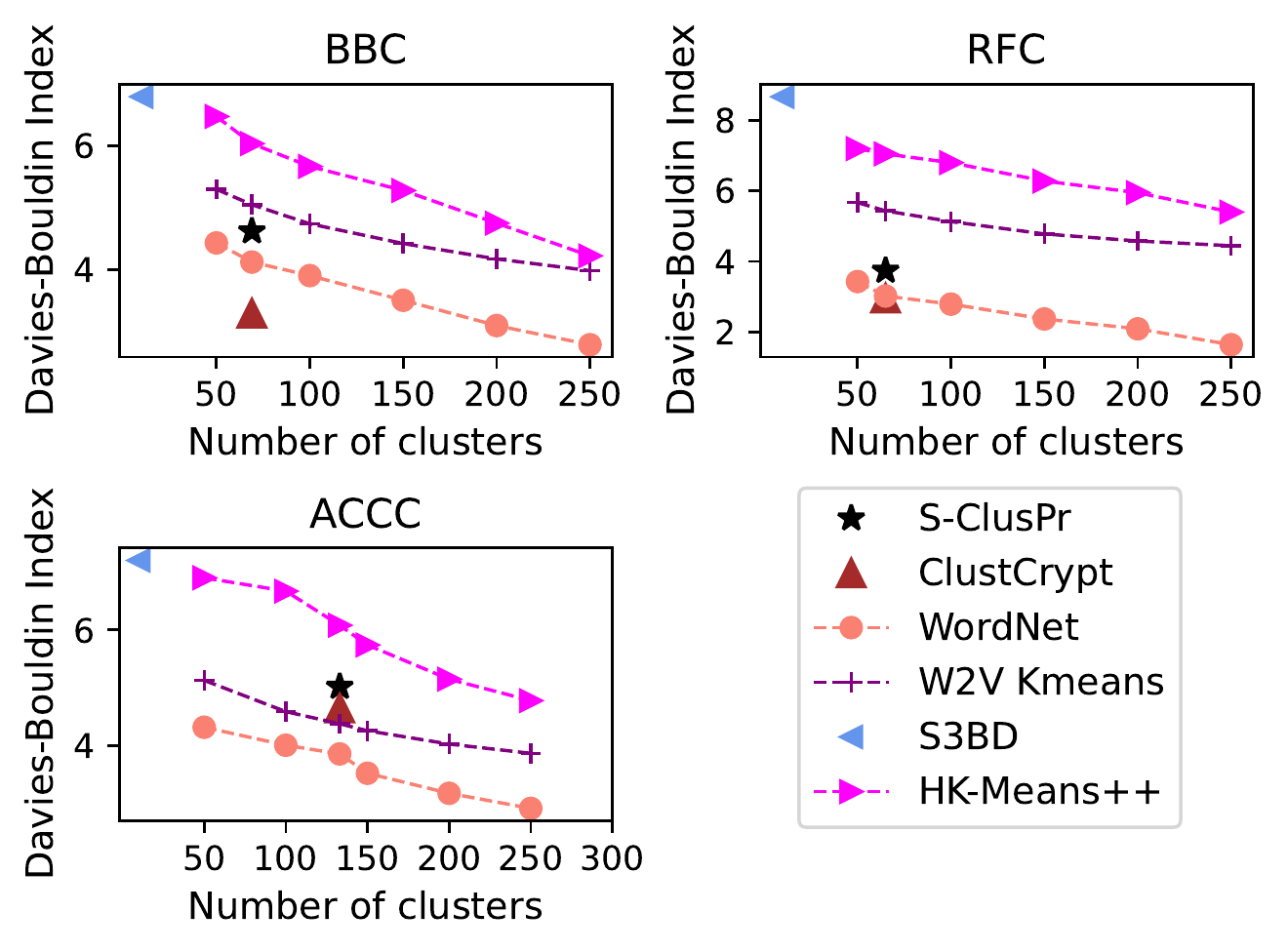}
 	\caption{\small{Silhouette Coefficient (SC) metric for each dataset. The results are obtained from S-ClusPr, \textit{HK-}means++, ClustCrypt (that are encrypted-based clustering schemes), W2V-Kmeans, and WordNet clustering schemes (that operate on plain-text tokens). }}
 	\label{fig: silh}
 \end{figure}

\noindent Figure~\ref{fig: silh}
shows the results of SC score evaluation on the three datasets and for varying number of clusters (in the horizontal axis). We note that, for this experiment, the value of $K$ in W2V Kmeans, WordNet, and \textit{HK-}means++ is randomly chosen and iteratively evolves. As such, we calculate the SC score for all the considered $K$ values and show them in multiple data points in the figure. However, other schemes (namely, S-ClusPr, ClustCrypt, S3BD) are not iterative and provide only one SC score for their determined $K$ values.

As the procedure of estimating the number of clusters is similar in ClustCrypt and S-ClusPr schemes, we can see that both of the schemes generate $69$, $65$, and $133$ clusters for the \texttt{BBC}, \texttt{RFC}, and \texttt{ACCC} datasets, respectively. As \texttt{ACCC} is the largest 
and broadest (\ie not domain-specific) dataset, it yields the highest $K$ value. \texttt{RFC} is not the smallest dataset, however, due to its domain-specific nature, it yields the lowest $K$ value.

Figure~\ref{fig: silh} represents SC metric outcomes for ClustCrypt, S-ClusPr and the four other compared schemes. According to the figure, considering all of the datasets, overall top performers are:
WordNet and S-ClusPr. Moreover, S-ClusPr outperforms others in the \texttt{RFC} dataset. On the contrary, ~\textit{HK-}means++ and S3BD underperform in most of the situation.       
The experiment indicates that the cluster sets generated by  \textit{HK}-means++ and S3BD contain less intra-cluster similarity. 
WordNet and S-ClusPr provide the highest intra-cluster similarity and hence, outperform others in all datasets.

 \begin{figure} 
 	\centering
 	\includegraphics[width=\linewidth]{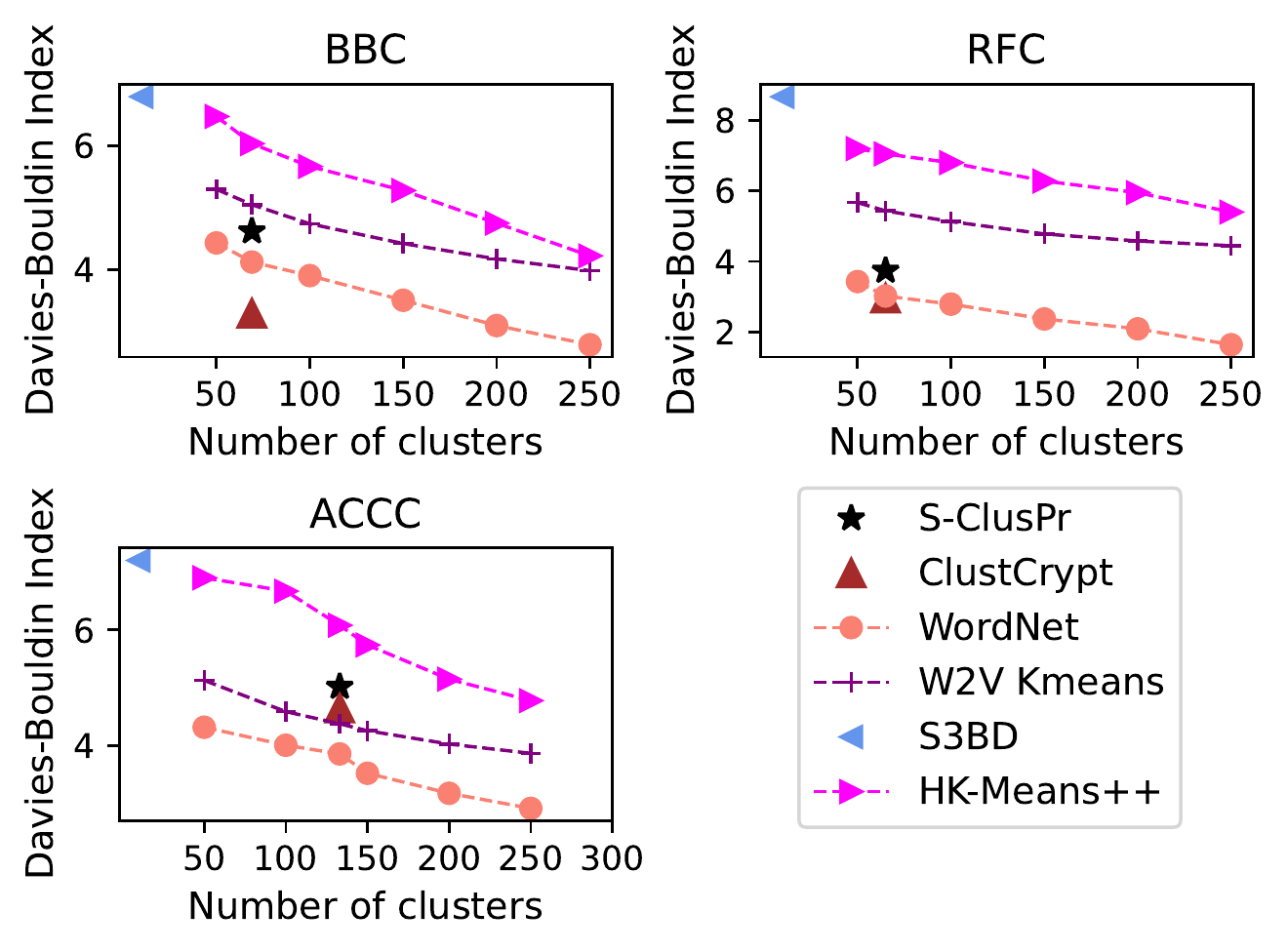}
 	\caption{\small{Davies-Bouldin Index (DI) for each dataset using different clustering schemes.}}
 	\label{fig: boud}
 \end{figure}

 \begin{figure} 
 	\centering
 	\includegraphics[width=\linewidth]{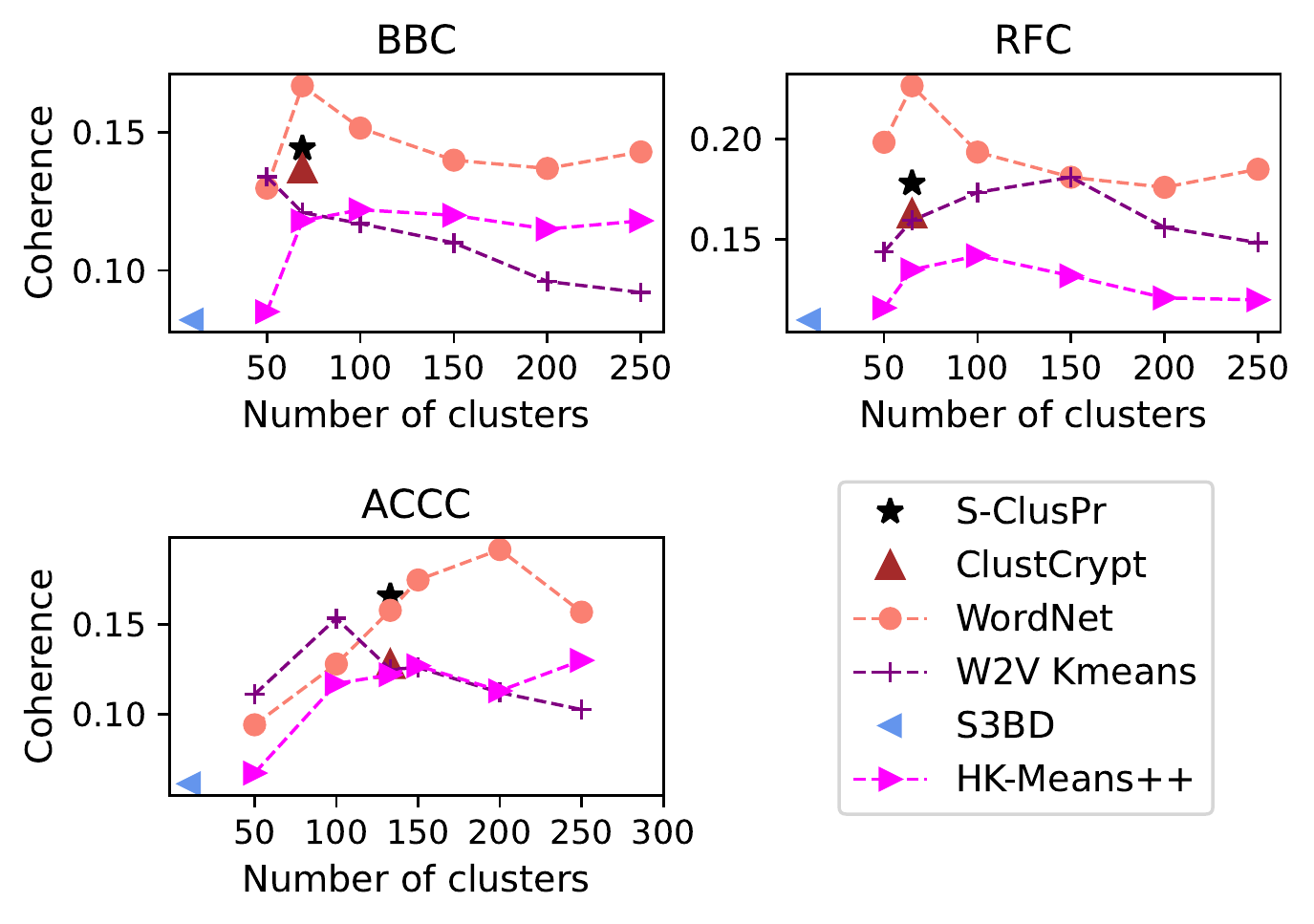}
 	\caption{\small{Cluster coherency for each dataset.}}
 	\label{fig: main}
 \end{figure}

\begin{table}[h]
\begin{subtable} {\linewidth}
\label{cal:bbc}
\subcaption{BBC}
\centering
\resizebox{\linewidth}{!}{
\begin{tabular}{c|c|c|c|l|c|c|}
\cline{2-7}
                                                                                & \multicolumn{6}{c|}{Approaches}                                                                                                                                                                                                                                                  \\ \hline
\multicolumn{1}{|c||}{\begin{tabular}[c]{@{}c@{}}No. of \\ Cluster\end{tabular}} & \begin{tabular}[c]{@{}c@{}}HK- \\ means++\end{tabular} & WordNet        & \begin{tabular}[c]{@{}c@{}}W2V \\ Kmeans\end{tabular} & S3BD                                 & ClustCrypt  & S-ClusPr             \\ \hline \hline
\multicolumn{1}{|c||}{10}                                                        &-                                                          &-                                                                               & -                                                      & \ \ \ \ 8.7                                          & \ -   & \ -                    \\ \hline
\multicolumn{1}{|c||}{50}                                                        & 25.43                                                   & \textbf{277.53}                                                                     & 11.16                                                 & \ \ \ \ -                                               & \ -   & \ -                   \\ \hline

\multicolumn{1}{|c||}{69}                                                        & 18.47                                                   & \textbf{253.60}                                                                     & 9.22                                                  & \ \ \ \ -                                             & 11.70  & 13.58                 \\ \hline
\multicolumn{1}{|c||}{100}                                                       & 11.13                                                    & \textbf{203.87}                                                                   & 7.37                                                  & \ \ \ \ -                                               & -  & \ -                    \\ \hline
\multicolumn{1}{|c||}{150}                                                       & 14.05                                                    & \textbf{164.43}                                                                    & 5.81                                                  & \ \ \ \ -                                               &-    & \ -                   \\ \hline
\multicolumn{1}{|c||}{200}                                                       & 10.17                                                    & 122.51                                                                     & 4.93                                                  & \ \ \ \ -                                               & -       & \ -               \\ \hline
\multicolumn{1}{|c||}{250}                                                       & 12.02                                                    & 97.15                                                                      & 4.38                                                  &  \ \ \ \ -                                              &  -      & \ -               \\ \hline
\end{tabular}
}
\end{subtable}

\bigskip

\begin{subtable} {\linewidth} 
\subcaption{RFC}
\centering
\resizebox{\linewidth}{!}{
\begin{tabular}{c|c|c|c|l|c|c|}
\cline{2-7}
                                                                                & \multicolumn{6}{c|}{Approaches}                                                                                                                                                                                                                                                             \\ \hline
\multicolumn{1}{|c||}{\begin{tabular}[c]{@{}c@{}}No. of \\ Cluster\end{tabular}} & \begin{tabular}[c]{@{}c@{}}HK- \\ means++\end{tabular} & WordNet                       & \begin{tabular}[c]{@{}c@{}}W2V \\ Kmeans\end{tabular} & S3BD                                & ClustCrypt      & S-ClusPr              \\ \hline \hline
\multicolumn{1}{|c||}{10}                                                        & -                                                         & -                                                                                       & -                                                      & 1247.20                                            & -    & -                  \\ \hline

\multicolumn{1}{|c||}{50}                                                        & 1730.26                                                  & 4320.63                                                                          & \textbf{60380.05}                                              & \ \ \ \ -                         & \ -                & \ -               \\ \hline

\multicolumn{1}{|c||}{65}                                                        & {1945.42}                             & { 3980.75}                                     & \multicolumn{1}{l|}{\textbf{51564.61}}                                 & \multicolumn{1}{l|}{\ \ \ \ -}         & 23760.64 & \textbf{29439.30}              \\ \hline
\multicolumn{1}{|c||}{100}                                                       & 1834.64                                                  & 3660.78                                                                           & \textbf{24374.17}                                              & \ \ \ \ -   & -                                                 & -                      \\ \hline
\multicolumn{1}{|c||}{150}                                                       & 1684.47                                                  & 3110.25                                                                          & \textbf{18684.33}                                              & \ \ \ \ -                           & -                          &  -                     \\ \hline
\multicolumn{1}{|c||}{200}                                                       & 846.71                                                  & 2572.89                                                                         & 16746.74                                              &  \ \ \ \ -                                                  & -  & -                     \\ \hline

\multicolumn{1}{|c||}{250}                                                       & 436.43                                                   & 1834.58                                                                          & 15139.11                                              & \ \ \ \ -                                                    & -   & -                   \\ \hline
\end{tabular}
}
\label{tab:cal rfc}
\end{subtable}

\caption{Calinski-Harabasz Index for the datasets.}
\label{tab:cal}
\end{table}


\subsubsection{\textbf{Evaluating Calinski-Harabasz Index (CI). }}
\noindent Table~\ref{tab:cal} represents CI metric outcomes for S-ClusPr and the four other schemes. According to the table, the RFC clusters provide large CI values compared to the BBC dataset, regardless of the employed clustering scheme. It is noteworthy that, we had the same observation for the ACCC dataset, however, we do not show its table due to the shortage of space. The superiority of RFC is because it is a domain-specific dataset with a few topics compared to the other two. 
Within Table~\ref{tab:cal rfc}, we can see that although W2V-Kmeans significantly outperforms the other schemes for most of the $K$ values, WordNet, ClustCrypt, and S-ClusPr also provide satisfactory CI values that imply well-partitioned clusters.

\subsubsection{\textbf{Evaluating Davies Bouldin Index (DI). }}
\noindent The DI values for the clusters, obtained by S-ClusPr and the compared schemes are expressed in Figure~\ref{fig: boud}. In most of the scenarios, we observe that increasing the number of clusters reduces the DI value. This is because, typically,
configuring clustering schemes to build more clusters on a given dataset leads to a higher coherency within each of the clusters. 

According to the figure, we observe that WordNet scheme outperforms others. The DI value for S-ClusPr is in the acceptable range, which indicates that the scheme can offer a competitive goodness of separation across clusters in compared to the most of other schemes. On the other hand, higher DI value yielded by \textit{HK-}means++ signifies poor cluster separation.

\subsubsection{\textbf{Evaluating Cluster Coherency.  }}
\noindent Figure~\ref{fig: main} shows the clusters' coherency on the three datasets using various clustering schemes. 
Using S-ClusPr, $69$, $65$, and $133$ clusters are created for the \texttt{BBC}, \texttt{RFC}, and \texttt{ACCC} datasets, respectively. As \texttt{ACCC} is the largest 
and broadest (\ie not domain-specific) dataset, it yields the highest $K$ value. \texttt{RFC} is not the smallest dataset, however, due to its domain-specific nature, it yields the lowest $K$ value. For the same reason, across the three datasets, S-ClusPr offers the highest coherency value ($\approx 0.16$) for the \texttt{RFC} dataset.

In compare to ClustCrypt, we notice that S-ClusPr offers a negligible coherency improvement ($\approx 6\%$) for the \texttt{BBC} and \texttt{RFC} datasets. However, for the \texttt{ACCC} dataset, S-ClusPr improves the coherency by approximately $31\%$.    

Analysis of the plain-text-based schemes reveal that, WordNet clusters offer the highest coherency value. This is expected, because it is difficult for an encrypted clustering scheme (\eg S-ClusPr) to outperform the unencrypted ones, since they do not have access to the semantics of the tokens~\cite{millerwordnet} to build the clusters. However, we observe that the coherency offered by S-ClusPr competes with the one offered by the $K$-means scheme. In particular, S-ClusPr provides a higher coherency value than $K$-means for the \texttt{RFC} and \texttt{BBC} datasets.

To evaluate the suitability of estimated number of clusters ($K$) by S-ClusPr, we configure both $K$-means and WordNet to use the estimated $K$ number of clusters for the studied datasets. According to the figure, for \texttt{RFC} and \texttt{BBC}, S-ClusPr suggested sets of $K$ clusters offer a higher coherency than $K$-means and a comparable one to WordNet. In the case of \texttt{ACCC}, S-ClusPr even outperforms WordNet in terms of coherency.


%% file: C3-7-1.tex
\subsubsection{Analyzing the Impact of S-ClusPr on Searchable Encryption Systems. }
\noindent One objective of this research is to enhance the performance of S3BD secure search system. As such, we instrumented S-ClusPr in S3BD and compared the coherency of resulting clusters with its original clustering scheme that predetermines a value for $k=10$. Moreover, its center selection only considers the co-occurrences. In this experiment, we intend to evaluate the improvement that S-ClusPr achieves within S3BD on the three studied datasets. In this experiment, the estimated values of $K$ for \texttt{BBC}, \texttt{RFC}, and \texttt{ACCC} are $69$, $65$, and $133$, respectively.
\begin{figure}
	\centering
	\includegraphics[width=.7\linewidth]{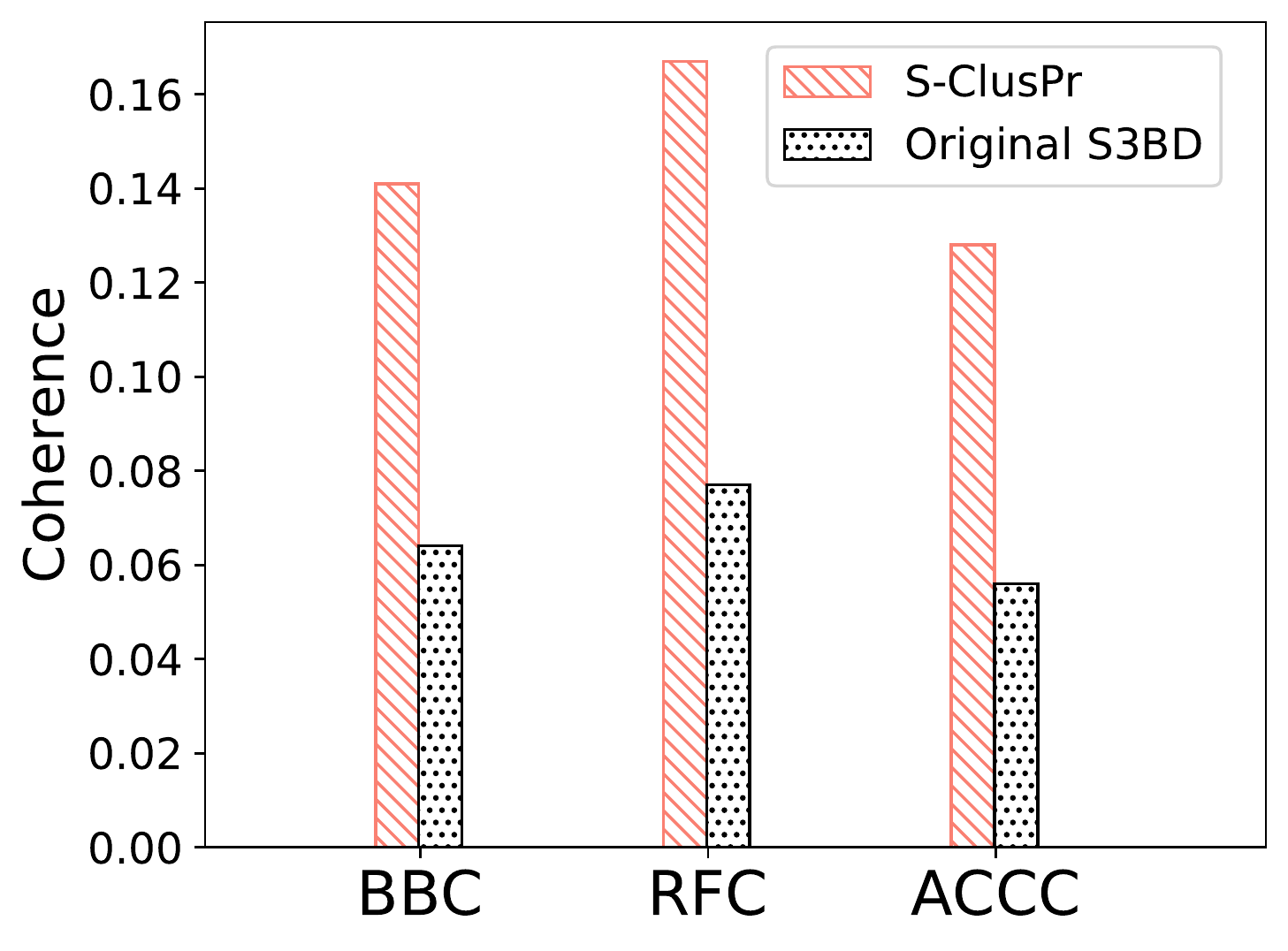}
	\caption{\small{Comparing the impact of clustering using S-ClusPr against original clustering of S3BD for the studied datasets.}}
	\label{fig: Pro_en}
\end{figure}

\noindent\textbf{ Impact on the Clustering Coherency of S3BD. }
Figure \ref{fig: Pro_en} shows that for all the studied datasets, clusters generated by S-ClusPr have remarkably higher coherency than the original clustering scheme of S3BD. This shows determining number of clusters based on dataset characteristics and choosing center tokens based on the centrality concept is effective. Our hypothesis is that, such efficiency improves the accuracy and offers more relevant semantic search results. This is because tokens of the clusters are more congruent to the clusters' topics, hence, more effective pruning is accomplished. For further evaluation of this hypothesis, next experiments concentrate on the impact of S-ClusPr on the search quality.

\begin{figure}
	\centering
	\includegraphics[width=\linewidth]{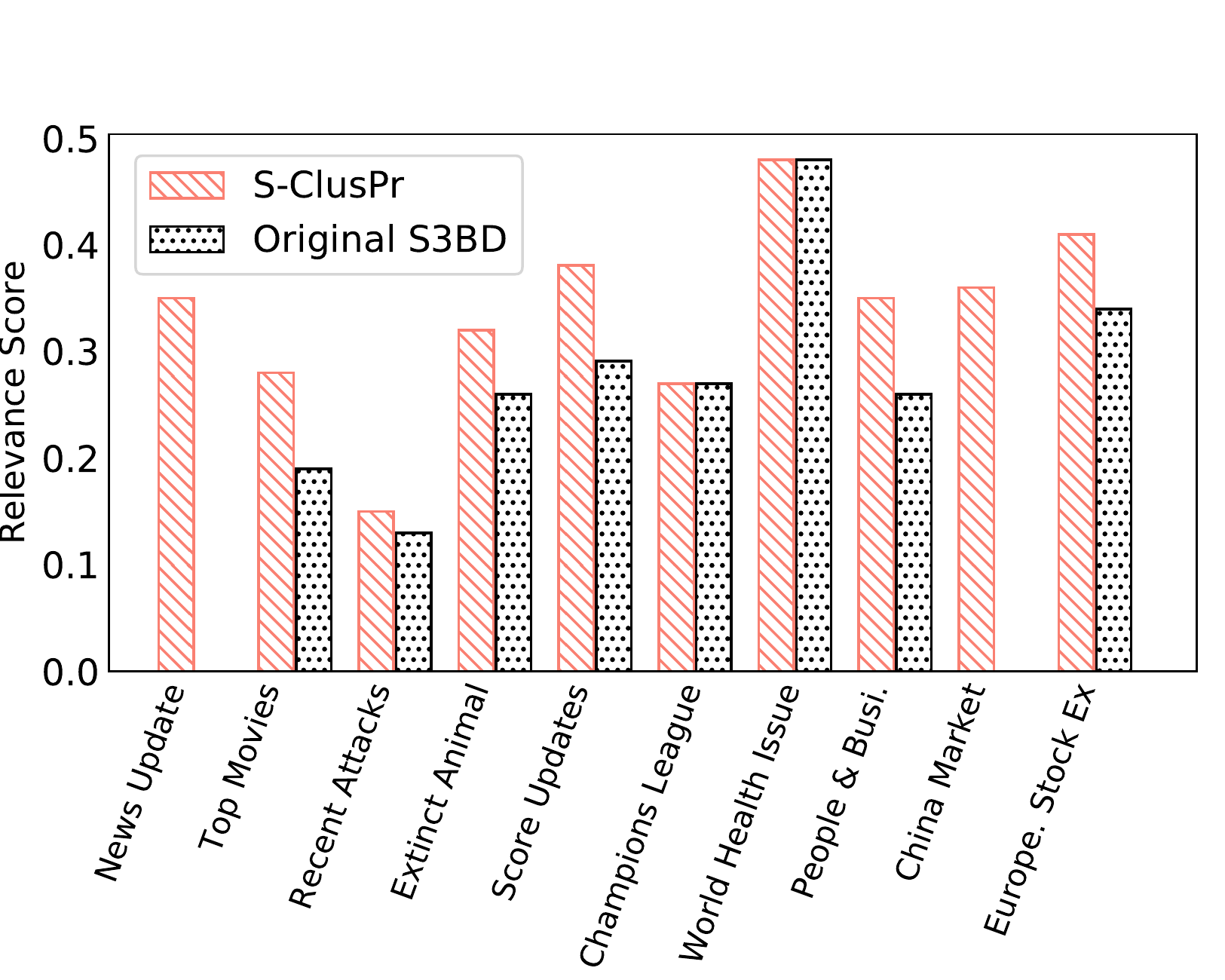}
	\vspace{-5pt}
	\caption{\small{Comparing the relevancy of search results using S-ClusPr vs original S3BD clustering in BBC dataset. The value of relevancy is calculated based on TSAP@10 scoring metric.}}
	\label{fig: search_ac}
	
\end{figure}

\noindent\textbf{ Impact on the Search Accuracy of S3BD. }
The purpose of improving the clusters' coherency in this study is to ultimately enhance the search accuracy by retrieving more relevant documents. To evaluate the impact of such improvement, in this part, we compare and analyze how the search accuracy of S3BD system is affected by utilizing S-ClusPr's clusters against the circumstance where its original clustering method is utilized.
For the evaluation, we generated a set of $10$ benchmark search queries that are listed in Table~\ref{bench}. 

\begin{table}
	
	\centering
	\resizebox{\linewidth}{!}{
	
	\begin{tabular} {|p{3.4cm}|p{3.1cm}|p{3.1cm}|}
		
		\hline
		
		{\texttt{\textbf{ACCC Dataset}}  }                  & {\texttt{\textbf{BBC Dataset}}}                & {\texttt{\textbf{RFC Dataset}}}                   \\ \hline
		
		\texttt{{Orlando Magic}}                  & \texttt{{News Update}}        & \texttt{{Internet}}              \\ \hline
		
		\texttt{{Samsung Galaxy}}         & \texttt{{Top Movies}}             & \texttt{{TCP}}                   \\ \hline
		
		\texttt{{Baseball routine}}        & \texttt{{Recent Attacks}}        & \texttt{{Fiber Doctor}} \\ \hline
		
		\texttt{{Recommendation}}                & \texttt{{Endangered Animals}} & \texttt{{Wifi}}                  \\ \hline
		
		\texttt{{North America}}     & \texttt{{Score Updates}}             & \texttt{{IoT}}    \\ \hline
		
		\texttt{{Tennis Tournament}} & \texttt{{Champions League}}       & \texttt{{Radio Frequency}}       \\ \hline
		
		\texttt{{Holy Martyr}}       & \texttt{{World Health Issue}}     & \texttt{{UDP}}                   \\ \hline
		
		\texttt{{Library}}            & \texttt{{People and Business }}         & \texttt{{Edge Computing}}        \\ \hline
		
		\texttt{{Stardock}}               & \texttt{{China Market}}          & \texttt{{Encryption Schemes}} \\ \hline
		
		\texttt{{Orthodox Church}}    & \texttt{{European Stock Exchange}} & \texttt{{Broadcasting}}          \\ \hline
		
	\end{tabular}
	}
	\caption{\small{Benchmark queries for each one of the studied datasets.}}
	
	\label{bench}
\end{table}


To measure the relevancy of search results for each query, we use \textit{TREC-Style Average Precision} scoring method~\cite{mariappan}. This method works based on the recall-precision concept and the score is calculated by $\sum_{i=0}^N r_i/N$, where $r_i$ denotes the score for $i^{th}$ retrieved document and $N$ is the cutoff number (number of elements in the search results) that we consider as $10$. Therefore, we call it \emph{TSAP@10}. 



We measure TSAP@10 score only for the \texttt{RFC} dataset and its benchmark queries. The reason is that it is domain-specific and feasible to determine the relevancy of the retrieved documents. 
To compare the relevancy provided by S-ClusPr against the original S3BD clustering, we apply the benchmark queries to the S3BD search system. In Figure~\ref{fig: search_ac}, the relevancy score of the results for each query when the two clustering schemes are applied are measured and presented.
According to the Figure, for most of the queries, S-ClusPr clustering offers a higher relevancy score. For the two queries that have identical \emph{TSAP@10} score, their retrieved document lists are equivalent. Also, S-ClusPr clusters provide score for \textit{News Update} and \textit{China Market} benchmark queries, whereas original S3BD clusters do not retrieve any relevant documents for these queries.

\begin{figure}
	\centering
	\includegraphics[width=.7\linewidth]{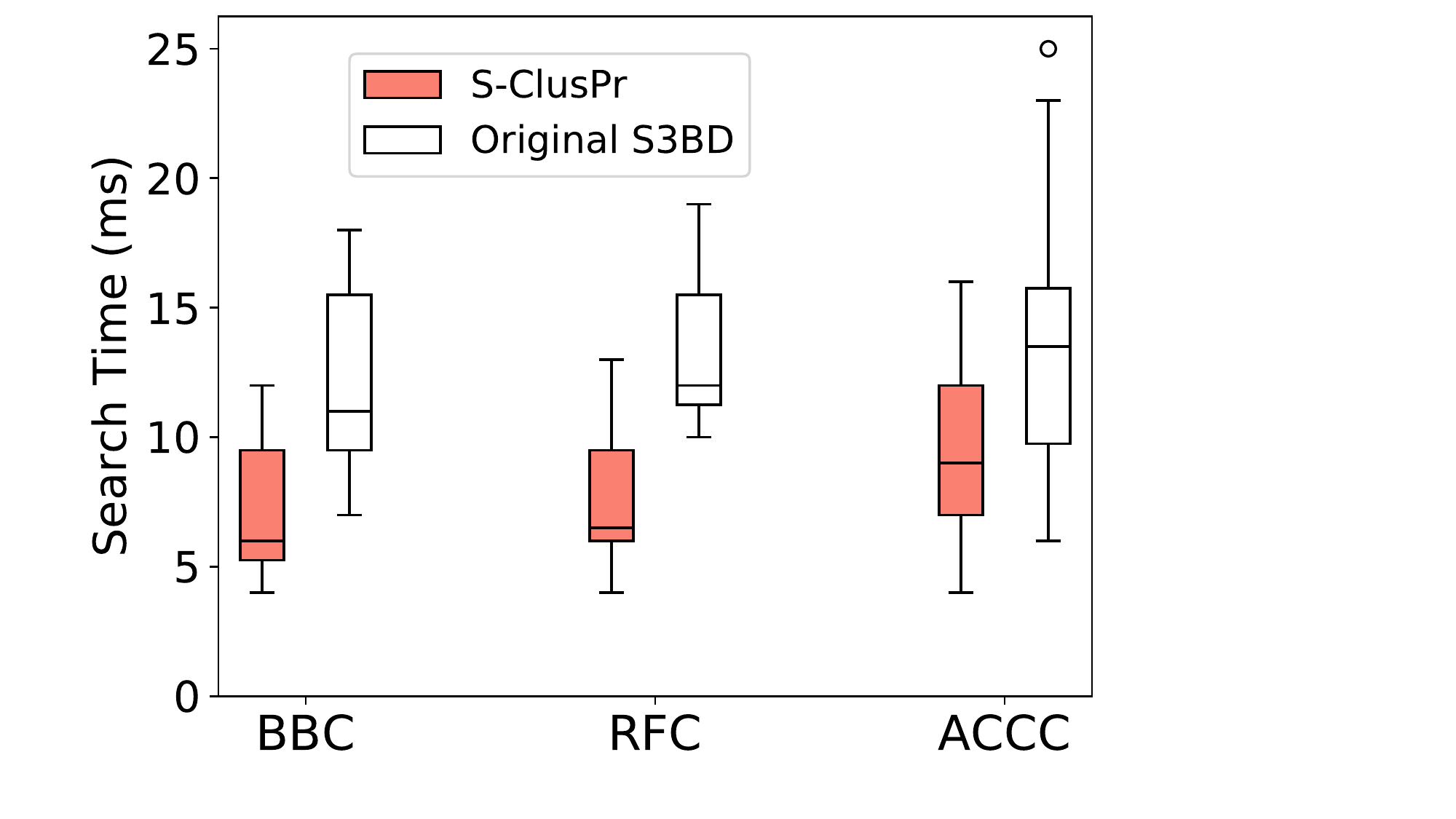}
	\caption{\small{Search time of S3BD when S-ClusPr is used for clustering versus when the original S3BD clustering is used.}}
	\label{fig: time_en}
\end{figure}

\noindent\textbf{ Impact on the Search Time of S3BD. }
Figure~\ref{fig: time_en} presents the total search time of the benchmark queries for each dataset. The search time is measured as the turnaround time of searching each query---from the time a query is issued until the result set is received. To eliminate the impact of any randomness in the computing system, we searched each set of benchmarks 10 times and reported the results in form of box plots.
The figure indicates that  when S-ClusPr clustering is utilized, the search time is significantly shorter than the circumstance where the original S3BD clustering is used. Longer search time impacts the scalability and real-time quality of the search operation on big data. 
Analyzing Figures~\ref{fig: Pro_en} to~\ref{fig: time_en} reveals that integrating S-ClusPr in the search system, not only makes it more accurate, but makes it faster and more scalable too.   

\begin{figure*} [!htbp]
\begin{subfigure}{.33\textwidth}
\centering
\includegraphics[width=\linewidth]{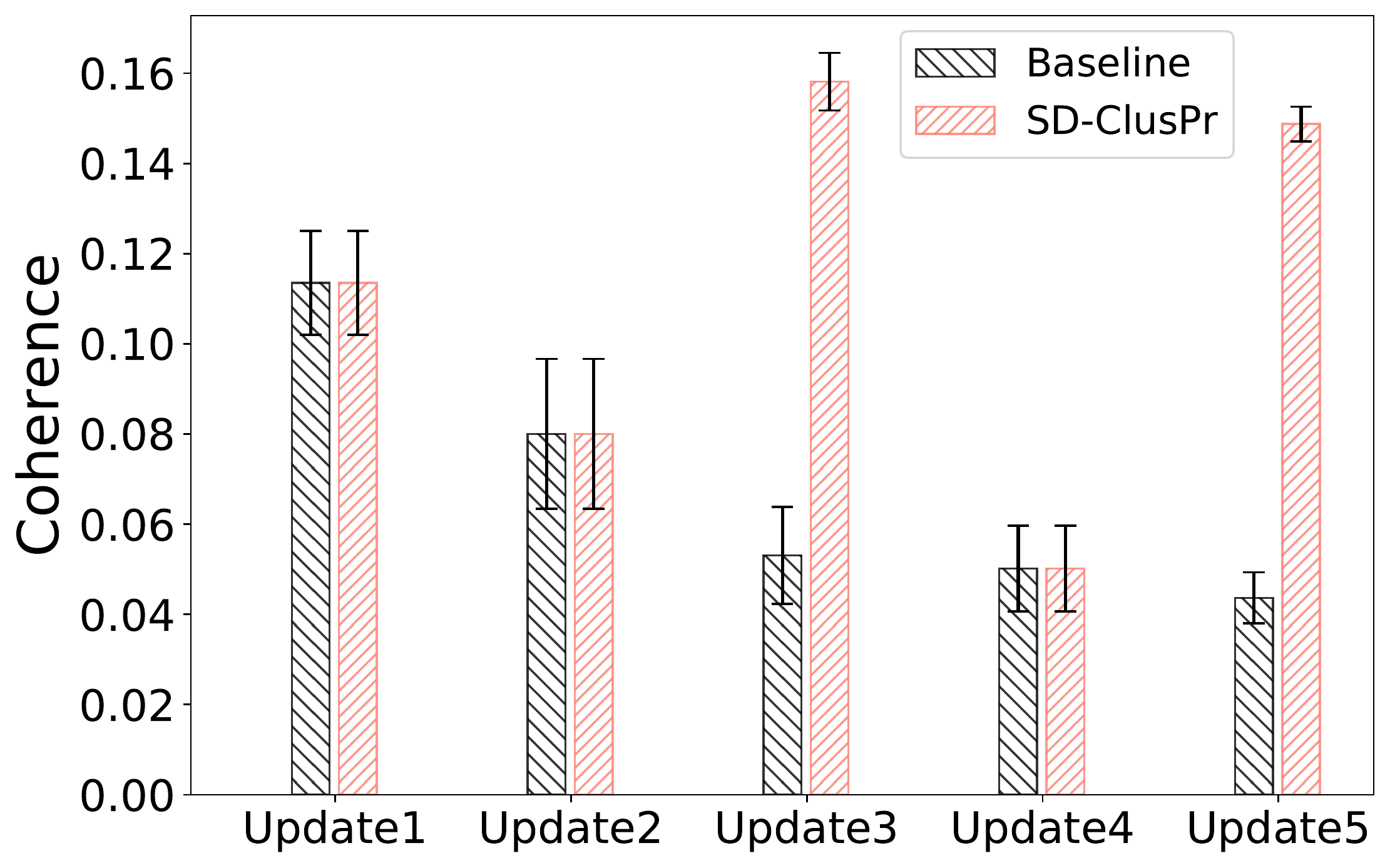}
\caption{BBC Dataset}
\label{fig: semidyna_bbc}
\end{subfigure}\hfill
\begin{subfigure}{.33\textwidth}
\centering
\includegraphics[width=\linewidth]{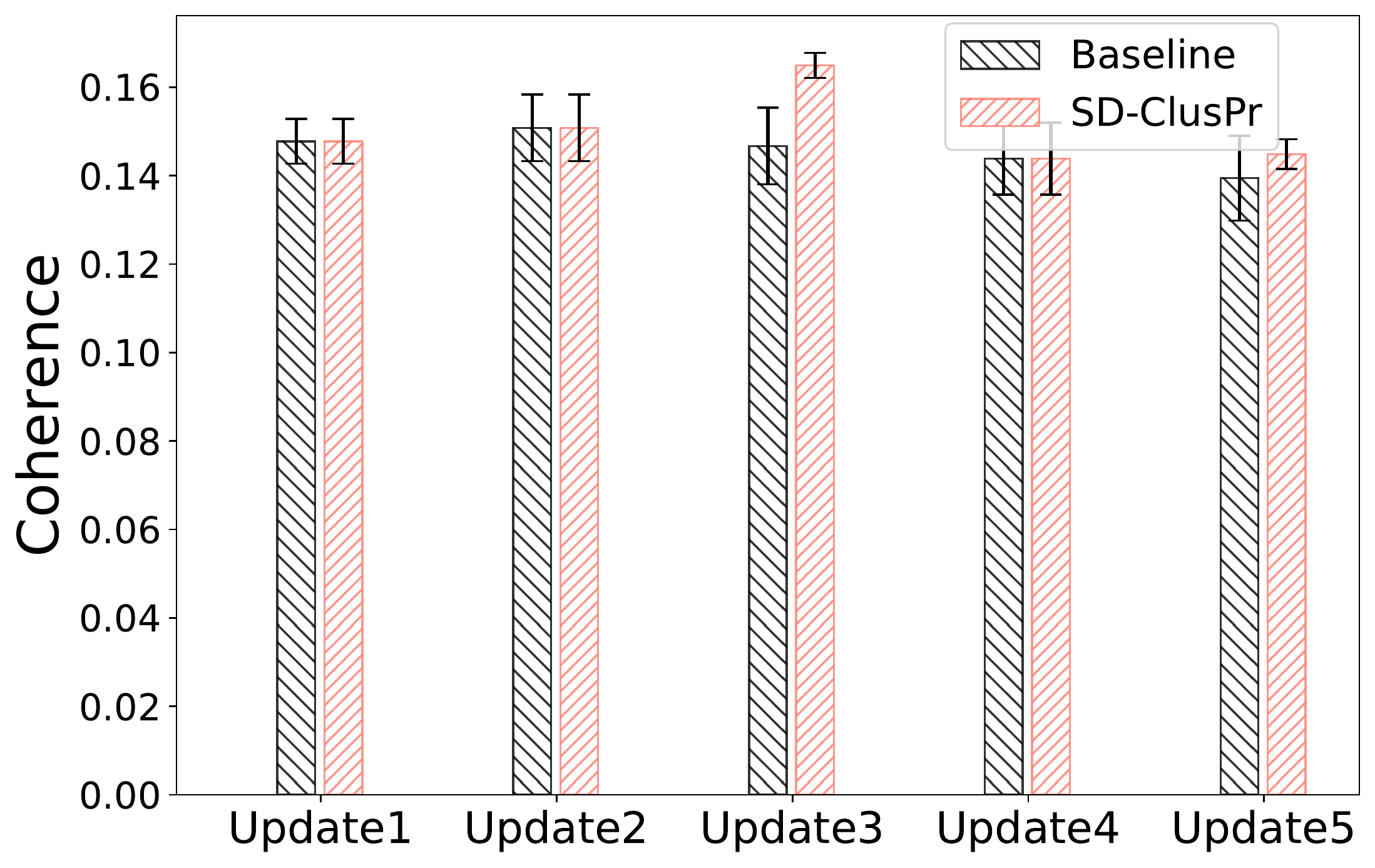}
\caption{RFC Dataset}
\label{fig: semidyna_rfc}
\end{subfigure}\hfill
\begin{subfigure}{.33\textwidth}
\centering
\includegraphics[width=\linewidth]{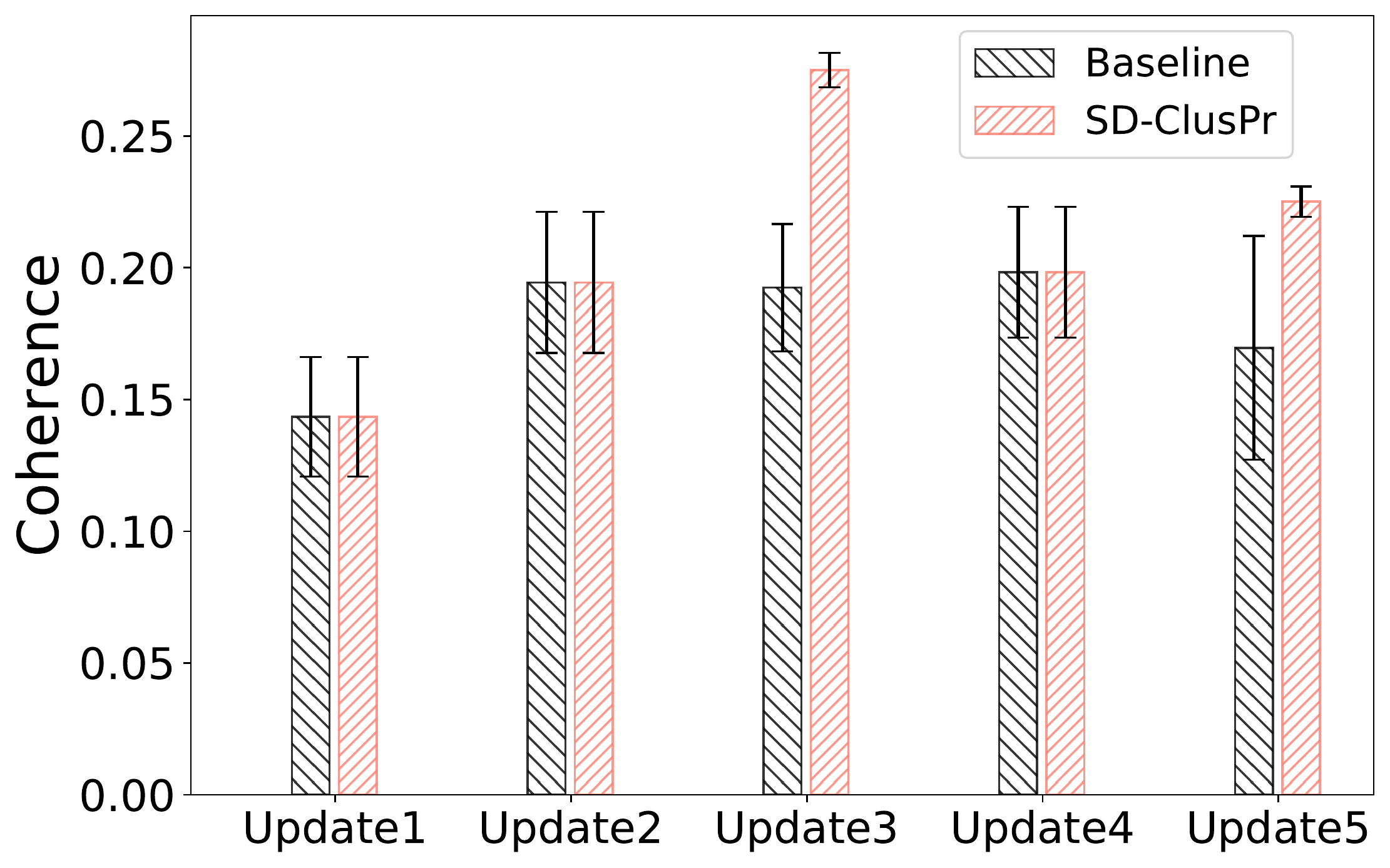}
\caption{ACCC Dataset}
\label{fig: semidyna_accc}
\end{subfigure}
\caption{Clusters' coherency for different updates of the three studied datasets when SD-ClusPr is applied with and without re-clustering option.}
\end{figure*}

\subsubsection{\textbf{Evaluation of Clustering Coherency for Dynamic Schemes. }}
In this part, we analyze the effectiveness of dynamic clustering schemes (SD-ClusPr and FD-ClusPr). We mention in Section~\secref{sec:clus-dyna} that FD-ClusPr is a specific case of SD-ClusPr. Hence, we only consider the SD-ClusPr scheme for evaluation. To this end, we leverage the three studied datasets and build subsets that each one serves as a batch update. Specifically, we consider an existing set of clusters based on $500$ documents for each dataset. Then, we sample five times to create a list of five updates that each one includes a set of documents. List $U$ includes the pairs of update names and the size of each update as follows: $U=<(U_1,25), (U_2,50), (U_3,100), (U_4,20), (U_5,200)>$. To assure that the results are not biased to any particular sample, we performed the sampling procedure 10 independent times and report the mean and 95\% confidence interval of the analysis in the results. The reason we designated $U_3$ and $U_5$ to be larger is to examine SD-ClusPr decision in re-clustering. To evaluate the scheme in terms of the cluster coherency, we build a \emph{baseline} version from SD-ClusPr that does not consider re-clustering. The baseline only performs clustering based on existing clusters (as explained in Algorithm~\ref{alg:semi}) to accommodate the new updates.

Figures~\ref{fig: semidyna_bbc},~\ref{fig: semidyna_rfc}, and~\ref{fig: semidyna_accc}, respectively, present cluster coherency of five different batch updates of \texttt{BBC},~\texttt{RFC}, and~\texttt{ACCC} respectively applying SD-ClusPr scheme. In Figure~\ref{fig: semidyna_bbc}, we observe that the coherency of clusters are decreased in baseline for $U_3$ whereas the coherency obtained for SD-ClusPr beats the previous by around 105\%. We observe the similar pattern of coherency variation for $U_5$. For baseline, the lowest coherency is obtained in $U_5$. On the contrary, in SD-ClusPr, we observe around $115\%$ improvement in coherency for $U_5$.   

According to Figure~\ref{fig: main}, clusters formed for the \texttt{RFC} dataset shows the highest coherency. Similarly, in Figure~\ref{fig: semidyna_rfc}, we observe the highest coherency for all updates in compare with other datasets. With respect to baseline, we observe that SD-ClusPr causes minor improvements in coherency of both $U_3$ and $U_5$. Since the documents are more domain-specific, clusters do not lose coherency significantly from one update to the other. As such, we do not observe significant improvements by SD-ClusPr. 
Similar to \texttt{BBC} and \texttt{RFC}, in Figure~\ref{fig: semidyna_accc}, we observe improvement in the coherency for \texttt{ACCC} dataset. In particular, the improvement in coherency for $U_3$ and $U_5$ is approximately $45\%$ and $35\%$, respectively. 

From these experiments, we conclude that ClusPr scheme can improve the coherency of clustering even for dynamic datasets. Specifically, we observed that for sufficiently large batches, such as $U_3$ and $U_5$, SD-ClusPr decides to re-cluster that remarkably improves the clustering coherency. 

%% file: C3-8-sum.tex
\section{Summary}
\label{sec:clus-sum}

In this chapter, we propose two secure clustering solutions, namely ClustCrypt and ClusPr in the form of trusted applications for three forms of unstructured datasets, namely static, semi-dynamic, and dynamic. The proposed clustering functions based on statistical characteristics of the datasets to: \textbf{(A)} determine the suitable number of clusters; \textbf{(B)} populate the clusters with topically relevant tokens; and \textbf{(C)} adapt the cluster set based on the dynamism of the underlying dataset. Experimental results, obtained from evaluating ClusPr against other schemes in the literature, on three different test datasets demonstrate between $30\%$ to $60\%$ improvement on the cluster coherency. Moreover, we notice that employing ClusPr within a privacy-preserving enterprise search system can reduce the search time by up to $78\%$, while improving the search accuracy by up to $35\%$.

In the next chapter, we explore how to enable secure enterprise search over unstructured data without jeopardizing its confidentiality. 


%% file: C4-Saed.tex
\chapter{Edge-Based Intelligence for Privacy-Preserving Enterprise Search on the Cloud}
\label{chap:saed}

\section{Overview }
Cloud-based enterprise search services (\eg AWS Kendra) have been entrancing big data owners by offering convenient and real-time search solutions to them. However, to offer an intelligent search over the privacy-preserving data, these services have to access the user's search history that further jeopardizes his/her privacy. To overcome the privacy problem, the main idea of this research is to separate the intelligence aspect of the search from its pattern matching aspect. According to this idea, the search intelligence is provided by an on-premises edge tier and the shared cloud tier only serves as an exhaustive pattern matching search utility. We propose \emph{Smartness at Edge} (SAED mechanism) that offers intelligence in the form of semantic and personalized search at the edge tier while maintaining privacy of the search on the cloud tier. At the edge tier, SAED uses a knowledge-based lexical database to expand the query and cover its semantics. SAED personalizes the search via an RNN model that can learn the user's interest. A word embedding model is used to retrieve documents based on their semantic relevance to the search query.

\section{Problem Statement }
Ideally, data owners desire a privacy-preserving cloud service that offers semantic and personalized searchability in a real-time manner, without overwhelming their resource-constrained (thin) client devices (\eg smartphones). A large body of research has been undertaken on privacy-preserving enterprise search services in the cloud \cite{li,sun,amini14,S3C,S3BD} whose goals are to protect user's sensitive data from internal and external attackers. However, most of these works fall short in retrieving search results that are semantically relevant to the context and  user's interest (\ie personalized search)~\cite{S3BD,S3C}.  
In addition, these works often rely on the client device and impose significant overhead on it to perform a secure query processing or to encrypt/decrypt user documents.

To satisfy all of the aforementioned desires of a particular user, our main idea in this research is to separate the intelligence aspect of the enterprise search from its pattern matching aspect.








%% file: C4-3-archi.tex
\section{SAED: Smart Edge-Leveraged Enterprise Search System}
\label{sec:saed-main}

\subsection{\textit{Architectural Overview }}~\\
In this part, we provide a bird-eye view of the SAED system, that enables intelligent and secure enterprise search on the cloud. The system is structured around three tiers, shown in Figure~\ref{fig:C1intro1.1}, and explained as follows:

\begin{itemize}
    \item \emph{Client tier} (\eg smartphone, tablet) contains a lightweight application that provides a user interface for uploading documents and to search over
them in the cloud. Datasets are either uploaded by the user or by the organization that owns the data. 

\item \emph{Edge tier} extracts representative keywords of the documents being uploaded to the cloud tier and builds an index on the cloud tier. Upon receiving a search query from the client tier, the SAED system on the edge tier offers intelligence by considering the query semantics and the user's interest. The edge tier is located in the client's premises, hence, deemed as an honest and secure system. To offer a secure enterprise search service, the edge tier encrypts both the uploaded data and the search query. In addition, it decrypts the result set before delivering it back to the client tier. 

\item \emph{Cloud tier} contains numerous high-end servers that are utilized for storing (encrypted) data and performing the large-scale computation required to exhaustively search against the index \cite{S3C,S3BD}. The index can be clustered based on the underlying topics of its keywords (please refer to our prior works~\cite{S3BD,clustcrypt} for further details).                   
\end{itemize}

\begin{figure*} [!htb]

	\centering
	\includegraphics[width=.95\linewidth]{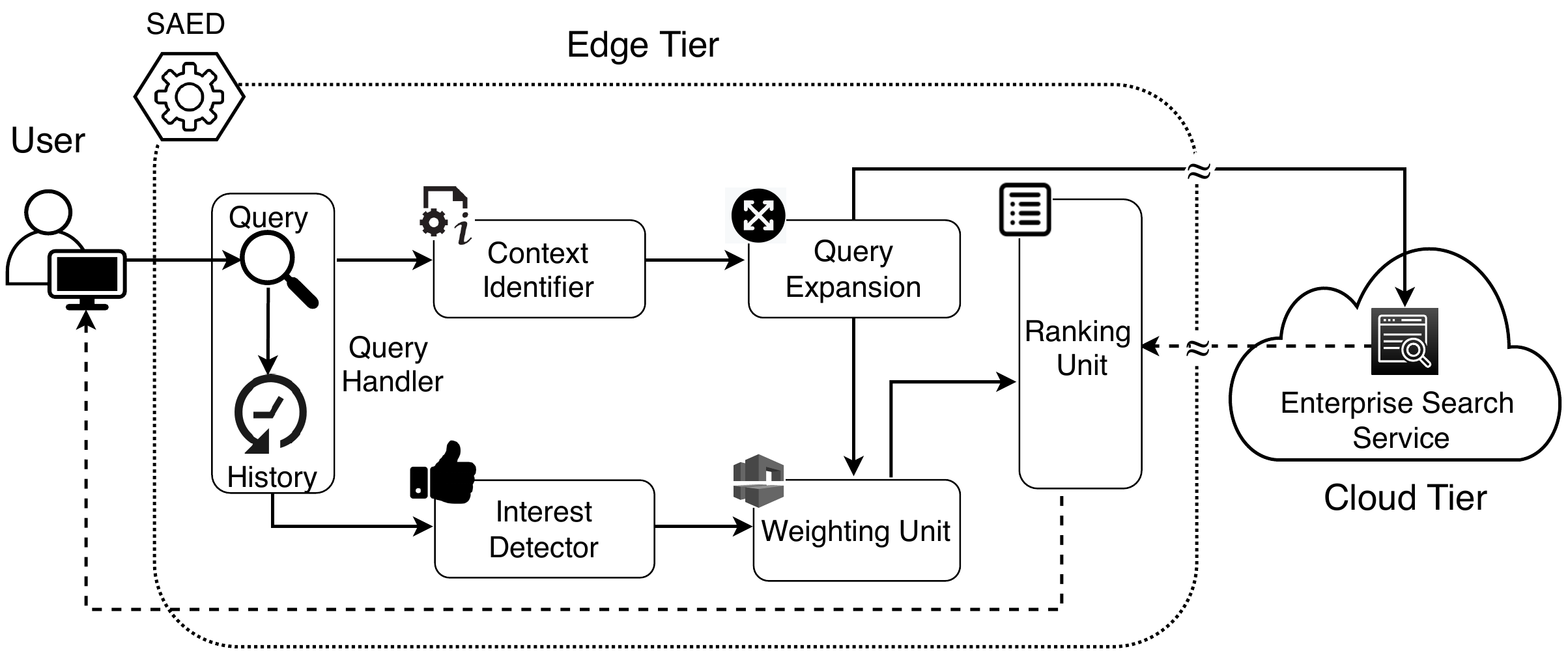}
	\caption{Architectural overview of the SAED system within edge tier and as part of the three-tier enterprise search service. SAED provides semantic search via identifying the query context and combining that with the user's interests. Then, Query Expansion and Weighting unit of SAED, respectively, incorporate the semantic and assure the relevancy of the results. Solid and dashed lines indicate the interactions from user to the cloud tier and from the cloud tier to the user respectively. }
	\label{fig:C4_archi}
\end{figure*}

In Figure~\ref{fig:C4_archi}, we depict the components of SAED and show the interactions between them. At first, a user-provided search query is received by the \emph{Query Handler} that keeps track of the user's search history and initializes the \emph{Context Identifier} unit whose job is to extract the context and disambiguate the query phrase. Then, according to the extracted context, the query is proactively expanded by the \emph{Query Expansion} unit and a \emph{query set} is constructed. To achieve the personalized search, the \emph{Interest Detector} unit of SAED leverages the user's search history to recognize his/her interest and weight each element of the query set (\ie expanded queries) based on its relatedness to the user interest. Once the pattern matching phase is accomplished on the cloud tier, the resulted documents are returned to SAED on the edge tier. Next, the \emph{Ranking Unit} utilizes the assigned weights to order the retrieved documents based on their relevance to the user's interest and generates a retrieved document list, denoted as $D_\theta$, that is sent to the user's device. 
In the next parts, we elaborate on each unit of the SAED system.

\subsection{\textit{Query Context Identification }}~\\
Identifying the context of a given search phrase is vital to navigate the search to the semantics intended by the user. Considering the example of \texttt{cloud computing} as the search query, without a proper context identification the returned document set can potentially include documents about \texttt{sky} and \texttt{climate}, whereas, an efficient context identifier can recognize the right semantic and navigate the search to the topics around \texttt{distributed, edge, fog,} and \texttt{cloud computing}. In fact, identifying the context helps the Query Expansion unit to form a query set diversified around relevant keywords that semantically represent the search query and subsequently improve the relevancy of the results.

Prior context identification works (\eg \cite{silva2020improving,kuzi2016query, diaz2016query}) have the following shortcomings: \emph{first,} they often assume each keyword has the same importance in the query and recognize the query context via averaging the embeddings of its keywords. However, not all keywords in a query necessarily help in identifying the context. For example, the keyword \texttt{various} in \texttt{various cloud providers} does not bring any significance to the context and can be eliminated. \emph{Second,} the embedding methods used by the existing works always provide the same representation for a given keyword, irrespective of the underlying context. This is particularly problematic for ambiguous keywords whose meaning vary based on the query context. For instance, the embedding of \texttt{cloud} in the aforementioned example should be different when it is used along with the \texttt{computing} as opposed to when it is used along with the \texttt{weather} in a given query. \emph{Third,} existing methods only consider the embeddings of the common keywords, while discarding most of the name-entities (\eg names and locations) that do not exist in the vocabulary of Word2Vec \cite{miller1995wordnet,fellbaum2017wordnet}. For instance, consider \texttt{best selling books of J.K. Rowling} as the query; \texttt{Book} and \texttt{Sell} are identified as the query context and \texttt{J.K. Rowling} is discarded. However, our analysis suggests that the context of a short query phrase often has contextual association with the discarded name-entities. 

To overcome the shortcomings and identify the actual context of a given query, we propose to take a holistic approach and extract the \emph{semantic across query keywords, proportionate to the importance of each keyword}. The main output of the Context Identification unit is a set of keywords, denoted as \emph{C}, that collectively represent the context of the query.

Specifically, to eliminate unimportant keywords that do not contribute to the semantic of query $Q$, the Context Identification unit utilizes \emph{Yake}~\cite{yake}, which is a unsupervised keyword extractor that discards unimportant keywords of the query. The remaining keywords (\ie the trimmed query, denoted as the $Q^\prime$ set) are considered for context identification. To learn the true semantic of $Q^\prime$, the unit leverages the Lesk algorithm \cite{fellbaum2017wordnet} of WordNet to disambiguate each keyword $q\in Q^\prime$. Lesk algorithm works based on the fact that keywords in a given sentence (query) tend to imply a certain topic. For keyword $q$, Lesk can determine its true semantics via comparing the dictionary definitions of $q$ against other keywords in $Q^\prime$ (\ie $Q^\prime - \{q\}$). Let $c_q$ be the set of keywords representing the context of $q$. Then, the context of $Q$ is determined as $C=\cup_{\forall q\in Q^\prime} c_q$. Lastly, the Context Identifier recognizes name-entities from $Q$ using WordNet and considers them as part of the context, but in a separate set, denoted as $N$. The reason for considering a separate set is that we apply a different treatment on $N$ and $C$ in the other units of SAED.

\begin{algorithm}
	\SetAlgoLined\DontPrintSemicolon
	\SetKwInOut{Input}{Input}
	\SetKwInOut{Output}{Output}
	\SetKwFunction{algo}{algo}
	\SetKwFunction{proc}{Procedure}{}{}
	\Input{query $Q$}
	\Output{$C$: set of keywords representing context of $Q$, \\$N$: set of name-entity in $Q$ }

	\SetKwBlock{Function}{Function \texttt{contextIdentification($Q$):}}{end}
	
	\Function{
	$Q^\prime \gets$ extract keywords from $Q$ using Yake alg.\;
	
	\ForEach {$q \in Q$}{ 
	  
	  \If {$q \in$ {Name-entity}}
	  {
	    $N\gets N \cup \{q\}$ \;
	  }
	  \Else
	  {
	   
	    \If{$q \in Q^\prime $} { 
	    $E_q\gets$ define $q$ based on $Q^\prime -q$ using Lesk alg.\;
	    $c \gets$ extract set of keywords of $E_q$ using Yake alg. \;
        $C \gets C \cup c$\;
	   } 
	  }
	  
	  }
	   
	  return  {$C, N$} \;
	}
	\caption{Pseudo-code to detect the context of a given query in the Context Identification unit of SAED.}
	\label{alg:con_iden}
	\end{algorithm}
	
Algorithm~\ref{alg:con_iden} provides a pseudo-code for identifying the context of incoming query $Q$. The outputs of the pseudo-code are two sets, namely $C$ and $N$, that collectively represent the context of $Q$. In Step 2 of the pseudo-code, Yake algorithm is used to filter $Q$ by extracting its important keywords and generate the $Q^\prime$ set.
Name-entities of $Q$ are identified by checking against WordNet and form the set $N$ (Steps 4--6).
Next, in Steps 8--12, for each keyword $q\in Q^\prime$, the Lesk algorithm is employed to disambiguate $q$ and find its true definition with respect to the rest of keywords in $Q^\prime$. Important keywords of the definitions form the context set ($C$) for $Q$.

\subsection{\textit{Query Expansion Unit }}~\\
\label{query_expansion}
The \emph{Query Expansion} unit is in charge of proactively expanding the query keywords based on their relevant synonyms that are in line with their identified context. Neglecting the query context and blindly considering all the synonyms, as achieved in \cite{silva2020improving, kuzi2016query, diaz2016query, S3BD},  leads to finding irrelevant documents. Accordingly, the unit leverages the context of $Q$ (\ie $C$ and $N$) to only find the set of synonyms, denoted as $P$, that are semantically close to the query context. 

Word2Vec \cite{mikolov2013efficient} is a shallow neural network model that can be trained to generate vector representation of keywords, such that the cosine similarity of two given keywords indicates the semantic similarity between them. Accordingly, to proactively expand each keyword $q\in Q$, the Query Expansion unit instruments Word2Vec, pre-trained with Google News dataset \cite{khatua2019tale}, to form the set of nominated synonyms, denoted as $s_q$. 
Let $s_q^i$ be a synonym of $q$ (\ie $s_q^i\in s_q$). Then, the similarity of $s_q^i$ and the query context, denoted as $sim(s_q^i,C)$, is defined based on the sum of similarities with each element of $C$, as shown in Equation~\ref{eq:queryexp}.

\begin{equation}
\label{eq:queryexp}
    sim(s_q^i , C)=\sum_{\forall C_{j}\in C}sim(s_q^i,C_j)
\end{equation}

Then, $s_q^i$ is chosen as an element of $P$, only if it is semantically close enough to the query context. To determine the sufficient closeness, we consider $sim(s_q^i,C)$ to be greater than the mean of the pair-wise similarity across all members of $s_q$ (\ie $sim(s_q^i,C) > \mu_{\forall q \forall j}(sim(s_q^j,C))$). We note that because the elements of $C$ and $N$ represent the context of $Q$, they as well are added to $P$.

Algorithm~\ref{alg:queryExpansion} provides a high level pseudo-code for generating the expanded query set $P$. In Steps 2--7 of the pseudo-code, the synonym set for each $q$ is generated. Next, the similarity between each word $s_q^i$ and $C$ is calculated. The similarity values are used to calculate the mean similarity of all nominated queries in Step 8. In Steps 9--15, expanded query set $P$ is formed by including nominated synonyms whose semantic closeness is greater than $\mu$. Lastly, in Step 16, set $P$ is expanded by including context set and name-entities.

\begin{algorithm}
	\SetAlgoLined\DontPrintSemicolon
	\SetKwInOut{Input}{Input}
	\SetKwInOut{Output}{Output}
	\SetKwFunction{algo}{algo}
	\SetKwFunction{proc}{Procedure}{}{}
	\SetKwFunction{main}{\textbf{ChooseCenter}}
	\Input{$Q$, $C, N$}
	\Output{$P$: the expanded query set  }
	
	\SetKwBlock{Function}{Function \texttt{ QueryExpansion($Q$, $C$, $N$)}}{end}
	
	\Function{
    	\ForEach{$q\in Q$} {
            $s_q \gets$ use WordNet to obtain synonym set of $q$ \;  
            \ForEach {$s_q^i \in s_q$} {
    	        $sim(s_q^i,C) \gets \displaystyle\sum_{\forall C_{j}\in C}sim(s_q^i,C_j)$ \;
    	}
    	}
         $\mu\gets$ calculate mean $sim(s_q^j,C)$ across all $q\in Q, \forall s_q^j\in s_q$ \;
        \ForEach{$q\in Q$}{
            \ForEach {$s_q^i \in s_q$} {
             \If{$sim(s_q^i,C) > \mu$}  
                {Add  $s_q^i$ to set $P$}
            }  
        }
        $P \gets P \cup C \cup N$ \;
	  return  $P$ \;
	}
	\caption{Pseudo-code to expand query based on the context in the Query Expansion unit of SAED}
	\label{alg:queryExpansion}
	
\end{algorithm}

\subsection{\textit{User Interest Detection }}~\\  
Detecting the user's search interest is essential to deliver personalized search. In SAED, interest detection is achieved by analyzing two factors: (A) the user's search history; and (B) the user's reaction to the retrieved results of prior search queries. This can be detected based on the results chosen by the user or the time spent for browsing them.  

Let $\Delta^\prime$ represent the whole resulted documents that are sent to the user and $\tau$ represent the documents where the user is interested in. We have $\tau \subseteq \Delta_\prime$. Accordingly, the user's interest can be derived from the topics of $\tau$.  
The Interest Detector unit uses an existing document classification model \cite{kastrati2019impact}, operating based on Na\"ive Biased (NB) method, to determine the topics of $\tau$, denoted as $t_\tau$. We also perform majority voting on $t_\tau$ to find the user's main interest. The process is repeated to store \textit{n}-prior search interests data of the user. The data is characterized as sequential as it is harvested from each successful search.  
By analyzing the user's prior search interests, the edge tier trains a recurrent neural network-based prediction model~\cite{pang2020innovative} that can predict the user's search interest. In case of SAED, as the data does not contain long dependency and to keep the model simple and to maintain real-timeliness, instead of a stacked (\ie deeper) model, we feed the harvested user-specific historical search data to train a many-to-one vanilla RNN model~\cite{rnn2019}.

\subsection{\textit{Weighting Unit }}~\\
Once SAED learns the user interest, the next step to accomplish a context-aware and personalized enterprise search is to determine the closeness of contextually-expanded queries (\ie elements of $P$) to the user's interest. In fact, not all expanded queries have the same significance in the interpretation of the query. Accordingly, the objective of the \emph{Weighting unit} is defined as quantifying the closeness of each expanded query to the user's interest. 
Later, upon completion of the search operation on the cloud tier, the weights are used by the \emph{Ranking unit} of SAED to prune and sort the result set. 

Prior weighting schemes (\eg~\cite{S3C,S3BD,diaz2016query,wang2019query,kuzi2016query}) often use the word frequency-based approach (\eg TF-IDF~\cite{S3BD}) and discard the user interests. Alternatively, the weighting procedure of SAED quantifies the importance of each expanded query $p\in P$ based on two factors: (A) The \emph{type} of $p$, which means if it directly belongs to the context ($C$ and $N$ sets) or is derived from them; and (B) The \emph{semantic similarity} of $p$ to the user interest. 

In particular, those elements of $P$ that directly represent the query context or name-entities (\ie $\forall p | p\in P\cap (C \cup N)$) explicitly indicate the user's search intention, hence, weighting them should be carried out irrespective of the user interest. A deeper analysis indicates that name-entities that potentially exist in a query represent the search intention, thus, biasing the search results to them can lead to a higher user satisfaction. As such, the highest weight is assigned to $\forall p | p \in (P \cap N$). The highest weight is determined by the domain expert, however, in the experiments we consider it as $\eta_{max}=1$. 
We define the \textit{contribution} of $q\in Q$ as the ratio of the number of keywords added to $C$ because of $q$ (denoted $C_q$) to the cardinality of $C$. Let $\eta_p$ denote the weight of $p\in P$. 
Then, for those elements of $P$ that are in the query context (\ie $\forall p \in (P \cap C)$), $\eta_p$ is calculated based on the contribution of the query keyword $q$ corresponding to $p$. Equation~\ref{eq:weight2} formally represents how $\eta_p$ is calculated.
\begin{equation}\label{eq:weight2}
    \eta_p=\frac{\eta_{max}\cdotp |C_q|}{|C|}
\end{equation}

The weight assignment for those $p$ that are derived from elements of $C$, as explained in Section~\ref{query_expansion}, (\ie $\forall p| p\in P-(C \cup N)$) is carried out via considering semantic similarity of $p$ with the user interest $\theta$. That is, $\eta_p=sim(p,\theta)$.


 Algorithm~\ref{alg:weight} provides the high level pseudo-code for distributing weight to the expanded query set $P$. The algorithm considers $P, C, N$, and highest weight value $\eta$ as the inputs. After assigning weights to $\forall p$ iteratively, it returns the weights mapped with corresponding $p$ as a hash map denoted as $\varpi$.
 In Step 2 of the pseudo code, $\theta$ gets the user's search interest that is identified by leveraging a pre-trained document classifier and a vanilla RNN model. 
 In the following Step, 
 hash map $\varpi$ is initialized to contain the weights that mapped with corresponding $p$. 

Overall, in Steps 4--15, weight of each $p$ denoted as $\varpi_p$ is calculated according to its type. Specifically, in Steps 5--7, $\varpi_p$, where $p \in (P \cap N)$ is set by directly assigned $\eta$.  
In Steps 8--11, $p$, where $p \in (P \cap C)$ is weighted based on its contribution towards context $C$. At first, $q$ is determined that generates $p$ and weight $\varpi_p$ of $p$ is calculated by the ratio between $\eta$ and total number of keywords added in $C$ for the corresponding $q$.
In the following Steps (12--14), $\varpi_p$, where $p \in P- (C \cup N)$ is calculated by its semantic similarity with $\theta$.
Lastly, the algorithm is finished by returning hash map $\varpi$ filled with weights corresponding to their $q$ (Step 16).

\begin{algorithm}
	\SetAlgoLined\DontPrintSemicolon
	\SetKwInOut{Input}{Input}
	\SetKwInOut{Output}{Output}
	\SetKwFunction{algo}{algo}
	\SetKwFunction{main}{\textbf{ChooseCenter}}
	\Input{$P$, $C$, $N$, $\eta$}
	\Output{$\varpi$ }

	\SetKwBlock{Function}{Function \texttt{ weighting($P$, $C$, $N$, $\eta$)}}{end}
	
	\Function{
	 $\theta \gets$ predict a user's search interest \; 
	 $\varpi \gets $ initialize hash map to store weights mapped with their corresponding keywords \;
	 
	 
	 
	 
	 \ForEach{$p \in P$} 
	 {  
         \If { $p \in N$}{
            $\varpi_p \gets \eta$ \;}

        \ElseIf {$p \in C$}{
            $\varpi_p \gets \frac{\eta_{max}\cdotp |C_q|}{|C|} $ \;
            }
	    \Else 
        { 
             $\varpi_p \gets$ sim($p,\theta$)  /*Compute  similarity and store it in hash map */\;
        }

    }
	
	return $\varpi$ \;
	}
	
	\caption{Pseudo-code to weight expanded query}
	\label{alg:weight}
	\end{algorithm}

\subsection{\textit{Ranking Unit }}~\\
Once the expanded query set $P$ is formed, the cloud tier performs string matching for each $p\in P$ across the index structure. We note that, if the user chooses to perform a secure search, the elements of $P$ are encrypted before delivered to the cloud tier. In addition, in our prior works \cite{clustcrypt}, we proposed methods for the cloud tier to cluster the index structure and perform the pattern matching only on the clusters that are relevant to the query. 

The cloud tier returns the resulted document set, denoted as $\Delta$, to the edge tier where the Ranking unit of SAED ranks them based on the relevance and the user's interest and generates a document list, called $\Delta^\prime$ to show to the user. 
For a document $\delta_i \in \Delta$, the ranking score, denoted as $\gamma_i$, is calculated by aggregating the importance values of each $p \in P$ within $\delta_i$ and with respect to its weight ($\eta_p$). The importance of $p$ in $\delta_i$ is conventionally measured based on the \emph{TF-IDF} score \cite{viegas2019cluwords}. Accordingly,   
$\gamma_i$ is formally calculated based on Equation~\ref{eq:ranking engine}. 
\begin{equation}
\label{eq:ranking engine}
\gamma_i= \displaystyle\sum_{\forall p \in P}\Bigg(\eta_p \cdot TF\textrm{-}IDF(p, \delta_i )\Bigg)
\end{equation}

The TF-IDF score of $p$ in $\delta_i$ is defined based on the frequency of $p$ in $\delta_i$ versus the inverse document frequency of $p$ across all documents in $\Delta$. Details of calculating the tf-idf score can be found in \cite{viegas2019cluwords}.
Once the Ranking unit calculates the ranking score for all $\delta_i \in \Delta$, then the documents are sorted in the descending order based on their ranks and thus, the document list $\Delta^\prime$ are formed with each $\delta_i$ and displayed to the user.

%% file: C4-4-plug.tex
\section{SAED As a Pluggable Module  Enterprise Search Solutions}
\label{sec:saed-plug}

The advantage of SAED is to be independent from the enterprise search service deployed on the cloud tier. That is, using SAED neither interferes with nor implies any change on the cloud-based enterprise search service. SAED can be plugged into any enterprise search solution. It provides the search smartness on the on-premises edge tier and leaves the cloud tier only for large-scale  pattern matching. The whole SAED solution reforms the enterprise search to be semantic, personalized, and confidential services.

In this work, we set SAED to work both with AWS Kendra and S3BD. In the case of using AWS Kendra, the Query Expansion unit sends the expanded query set $P$ to Kendra to search each keyword $p$ against the dataset on the Amazon cloud. The resulted documents are received by SAED and ranked before being delivered to the client tier. In the implementation, we only show top 10 documents from the resulted list to the user. Similarly, we plugged SAED to S3BD to perform confidential semantic search on the cloud. Because S3BD maintains an encrypted index structure that has to be traversed against each search query, the elements of $P$ had to be encrypted before handing them over to the cloud tier. We also verified SAED when it is used along with AWS Kendra where the dataset was encrypted. We noticed that SAED can achieve smart search even when Kendra is set to work with encrypted dataset. The performance measurement and analysis of using SAED along with AWS Kendra and S3BD are elaborated in the next Section. 

%% file: C4-5-perf.tex
\section{Performance Evaluation of SAED}
\label{sec:saed-perf}

\subsection{\textit{Experimental Set up }}~\\
We have developed a fully working version of SAED and made it available publicly in our Github\footnote{\url{https://github.com/hpcclab/SAED-Security-At-Edge}} page. To conduct a comprehensive performance evaluation of SAED on the enterprise search solutions, we developed it to work with both S3BD~\cite{S3BD} and AWS Kendra~\cite{kendra}. S3BD already has the query expansion and weighting mechanisms, but we deactivated them and set it to use the expanded queries generated by SAED. In the experiments, the combination of SAED and S3BD is shown as SAED+S3BD. Likewise, the combination of SAED and AWS Kendra is shown as SAED+Kendra.

We evaluated SAED using two different datasets, namely \texttt{Request For Comments (RFC)} and \texttt{BBC} that have distinct properties and volume. The reason we chose the \texttt{RFC} dataset is that it is domain-specific and includes $4,951$ documents about the Internet and wireless communication network. Alternatively, the \texttt{BBC} dataset is more diverse. It includes $2,224$ news documents in five distinct categories, including politics, entertainment, business, sports, and technology. 

To conduct a comprehensive evaluation, we used both systematic metrics and human-based feedback as elaborated in Section~\ref{subsec:eval met}. We deployed and experimented SAED on a Virtual Machine (VM) within our local edge computing system. The VM had two 10-core 2.8 GHz E5 Xeon processors with 64 GB memory and Ubuntu 18.4 operating system.

\subsection{\textit{Benchmark Queries }}~\\
The datasets that we use to carry out the experiments are not featured with any benchmark. Therefore, we required to develop benchmark queries for the datasets before evaluating the performance of SAED. We developed $10$ benchmark queries, shown in Table~\ref{tab:benchmark}, for each one of the two datasets. The benchmark queries are proactively designed to explore the breadth and depth of the datasets in question. In addition, some of the queries intentionally contain ambiguous keywords to enable us examining the context detection capability of SAED.
For the sake of brevity, we provide one acronym for each benchmark query (see Table~\ref{tab:benchmark}).
For each benchmark query, we collected at most the top-20 retrieved documents. Then, the quality of the retrieved documents were measured via both automated script and human-based users.
\begin{table}[]

\caption{\small{Benchmark search queries developed for the RFC and BBC datasets.}}
\resizebox{\linewidth}{!}{%
\begin{tabular}{|l|l|}
\hline
\textbf{BBC Dataset}                  & \textbf{RFC Dataset}               \\ \hline
European Commission (\texttt{EC})          & Network Information (\texttt{NI})        \\ \hline
Parliament Archives (\texttt{PA})          & Host Network Configuration (\texttt{HNC}) \\ \hline
Top Camera Phones 2020 (\texttt{TCP})       & Data Transfer (\texttt{DT})              \\ \hline
Credit Card Fraud   (\texttt{CCF})          & Service Extension(\texttt{SE})          \\ \hline
Animal Welfare Bill (\texttt{AWB})          & Transport Layer (\texttt{TL})           \\ \hline
Piracy and Copyright Issues (\texttt{PCI})   & Message Authentication  (\texttt{MA})   \\ \hline
Car and Property Market (\texttt{CPM})      & Network Access (\texttt{NA})             \\ \hline
Rugby Football League  (\texttt{RFL})       & Internet Engineering  (\texttt{IE})      \\ \hline
Opera in Vienna  (\texttt{OV})             & Fibre Channel (\texttt{FC})             \\ \hline
Windows Operating System (\texttt{WOS})     & Streaming Media Service (\texttt{SMS})   \\ \hline
\end{tabular}
}
\label{tab:benchmark}
\end{table}

\subsection{\textit{Evaluation Metrics }}~\\
\label{subsec:eval met}
We have to measure the search relevancy metric to understand how related the resulted documents are with respect to the user's query and how they meet the his/her interests. For the measurement, we use TREC-Style Average Precision (TSAP) score, described by Mariappan \etal~\cite{mariappan}. TSAP provides a qualitative score in a relatively fast manner and without the knowledge of the entire dataset~\cite{S3BD}.
It works based on the precision-recall concept that is commonly used for judging text retrieval systems. The TSAP score is calculated
based on $\sum_{i=0}^N r_i/N$, where $r_i$ denotes score for $i^{th}$ retrieved document  and $N$ denotes the cutoff number (total number of retrieved documents). Since we consider $N=10$, we call the scoring metric as \emph{TSAP@10}.

To determine $r_i$ for retrieved document $\delta^\prime_i\in \Delta^\prime$, we conducted a human-based evaluation.
We engaged five volunteer students to judge the relevancy of each retrieved document. For every search query, the volunteers labeled each retrieved document as \texttt{highly relevant, partially relevant,} or \texttt{irrelevant}. After performing majority voting based on the provided responses for document $i$, the value of $r_i$ is determined as follows:
\begin{itemize}
\item $r_i = 1/i$ if a document is \texttt{highly relevant}
\item $r_i = 1/2i$ if a document is \texttt{partially relevant}
\item $r_i = 0$ if a document is  \texttt{irrelevant}
\end{itemize}
 
We report TSAP@10 score to show the relevancy of results for each benchmark query. In addition, mean TSAP score is reported to show the overall relevancy across each dataset. As we set the top 10 documents to be retrieved for each search, the highest possible for \emph{TSAP@10} score can be 0.292~\cite{mariappan}. 

In addition to the TSAP score, we measure \emph{Mean F-1} score too to compare the search quality offered by the SAED-plugged enterprise search solutions against the original enterprise search solutions (\ie without SAED in place). The F-1 score maintains a balance between the precision and recall metrics, which is useful for unstructured datasets with non-uniform topic distribution. 

\subsection{\textit{Evaluating Search Relevancy }}~\\ 
The purpose of this experiment is to evaluate the search relevancy of enterprise search systems that have SAED plugged into them and compare them against the original (unmodified) systems. To evaluate the personalized search, we set (assumed) \texttt{technology} as the user's interest for both datasets. We note that, in this part, the enterprise search solutions (S3BD and AWS Kendra) are set to work in the plain-text datasets.

\begin{figure*} [h]
\begin{subfigure} {.45\textwidth}
\includegraphics[width=.95\linewidth]{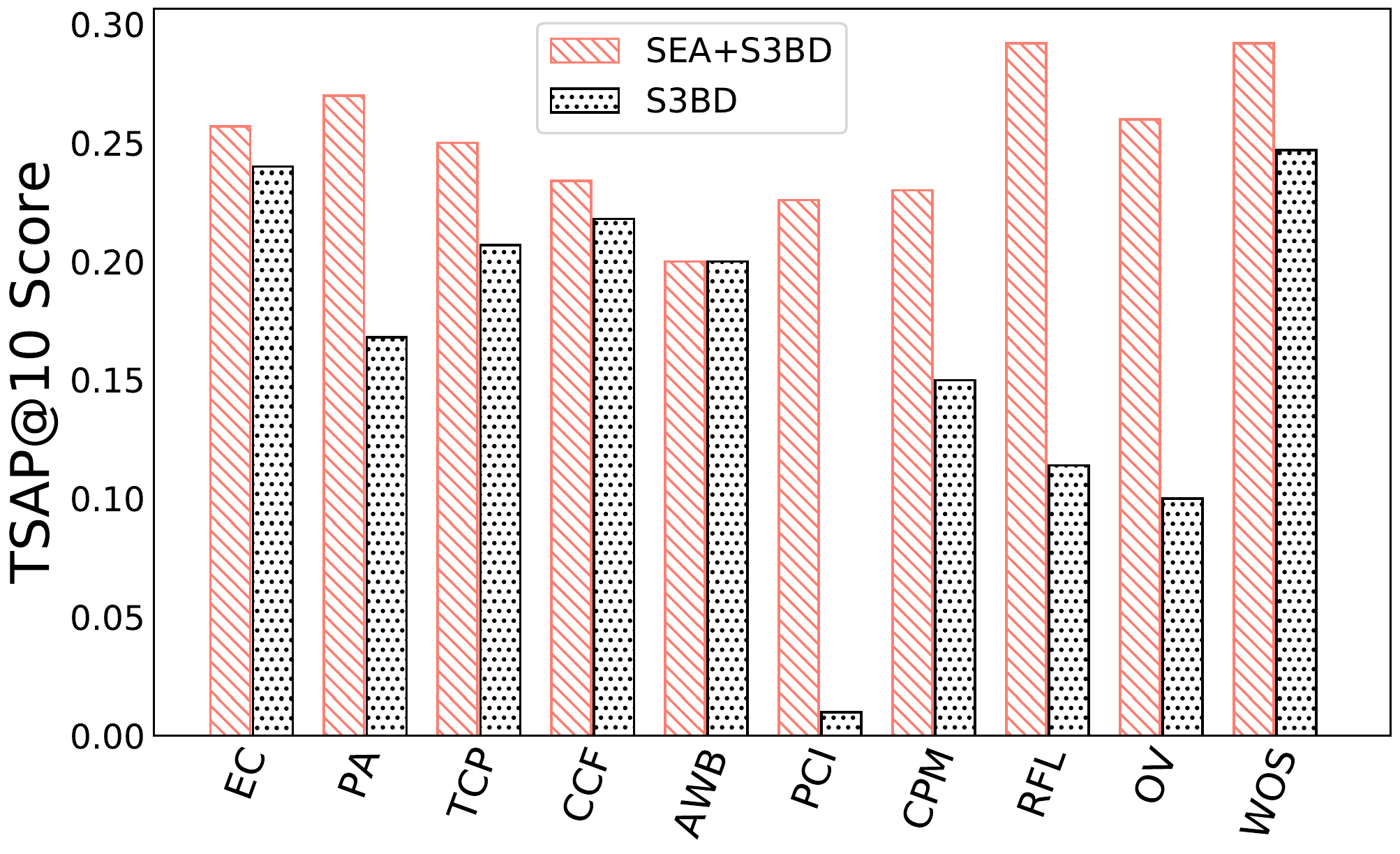}
\caption{BBC dataset}
\label{fig: s3bd bbc}
\end{subfigure} \hfill
\begin{subfigure} {.45\textwidth}
\includegraphics[width=.95\linewidth]{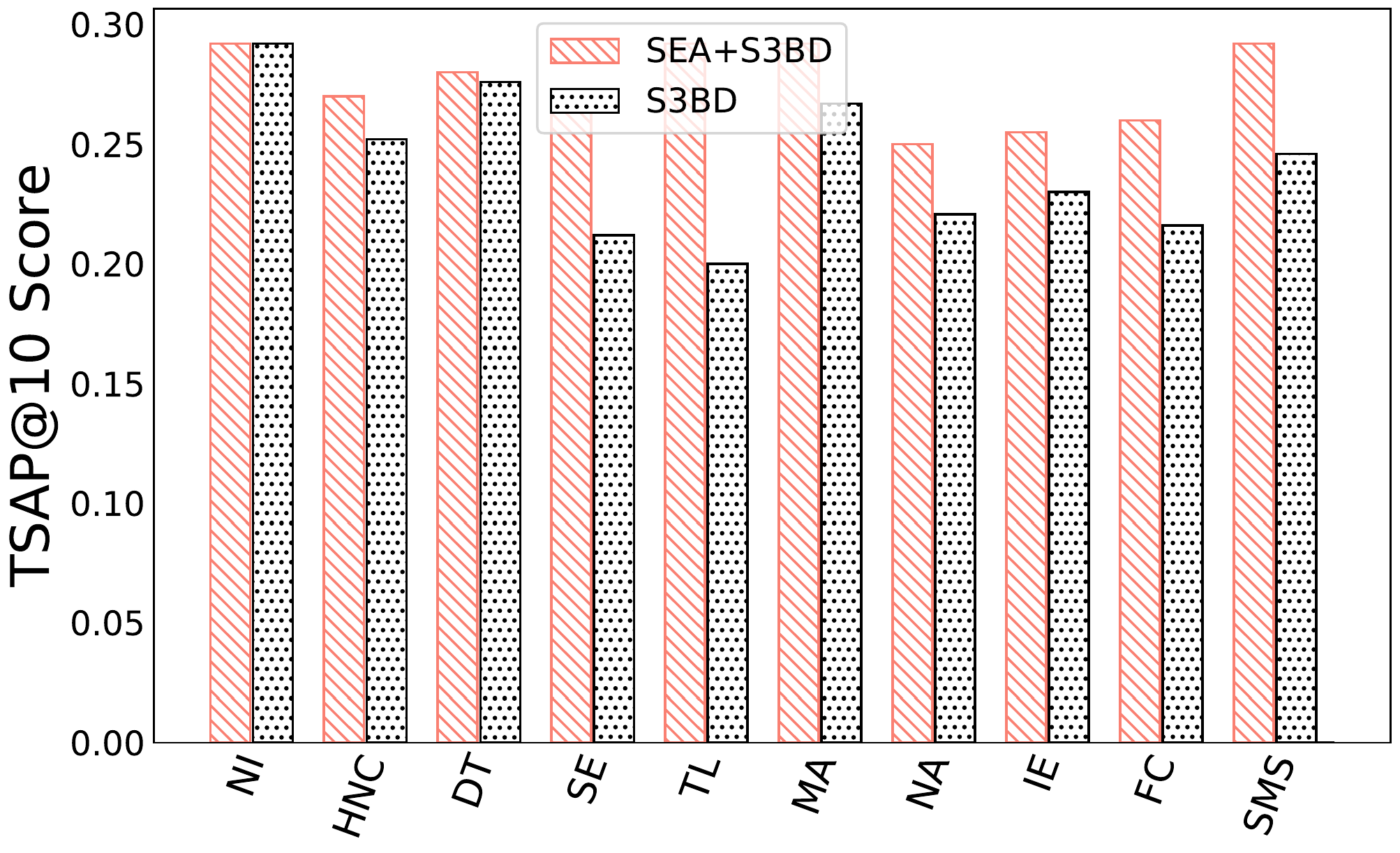}
\caption{RFC dataset}
\label{fig: s3bd rfc}
\end{subfigure} 
\caption{Comparing TSAP@10 scores of SAED+S3BD and S3BD systems. Horizontal axes show the benchmark queries.}
\end{figure*}

\noindent{\textbf{S3BD vs SAED+S3BD. } }
Figure~\ref{fig: s3bd bbc} shows the TSAP@10 score for the RFC and BBC datasets for the original S3BD and SAED+S3BD. The horizontal axes in both subfigures show the benchmark queries and the vertical axes show the search relevancy based on the TSAP@10 score. 

\begin{figure*} [h]
\begin{subfigure}  {.45\textwidth}
\includegraphics[width=.95\linewidth]{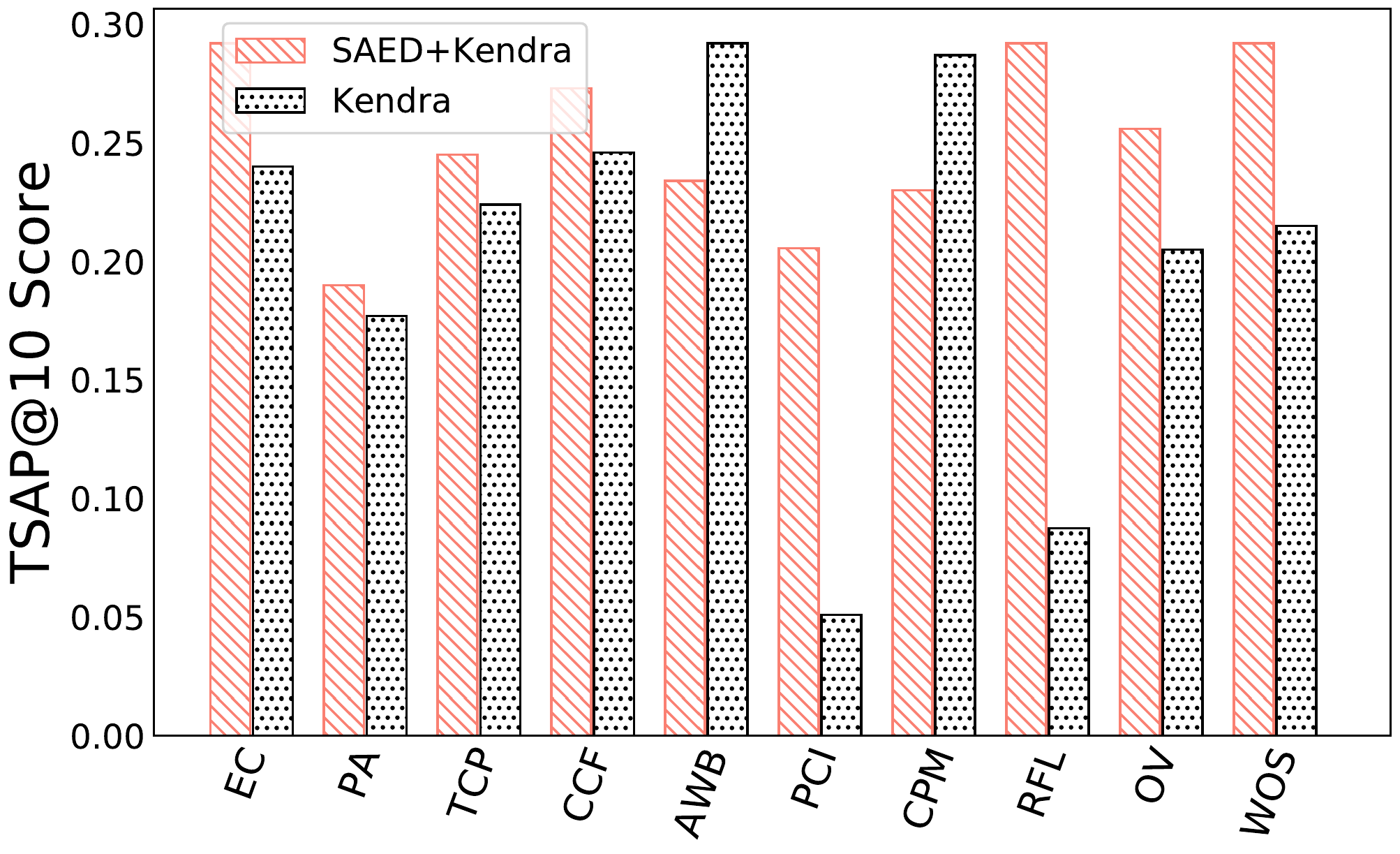}
\caption{BBC dataset}
\label{fig: kendra bbc}
\end{subfigure} \hfill
\begin{subfigure}  {.45\textwidth}
\includegraphics[width=.95\linewidth]{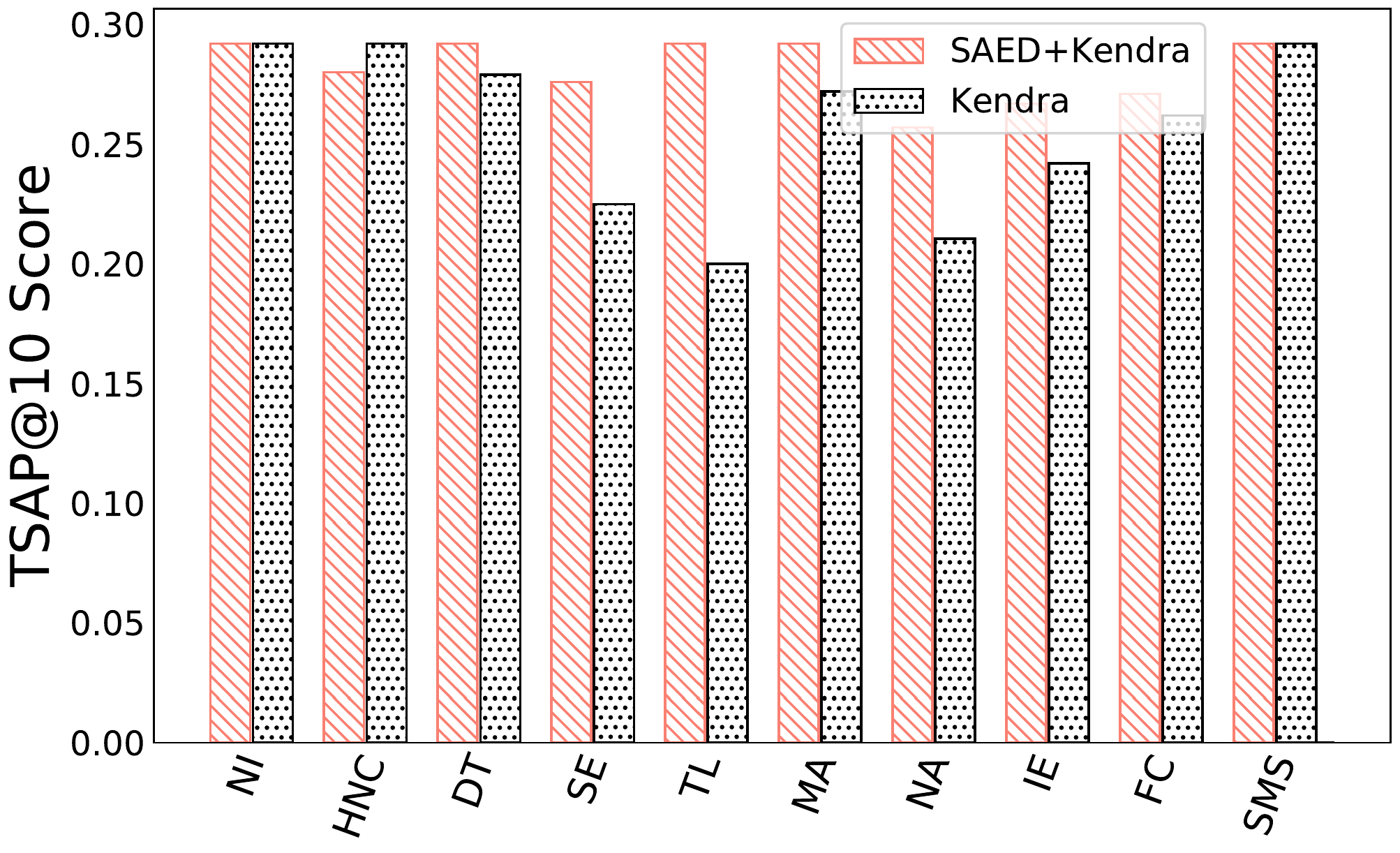}
\caption{RFC dataset} 
\label{fig: kendra rfc}
\end{subfigure} \hfill
\caption{Comparing TSAP@10 scores obtained from SAED+Kendra versus AWS Kendra in searching benchmark queries.}
\end{figure*}

In both Figure~\ref{fig: s3bd bbc} and~\ref{fig: s3bd rfc}, we observe that for all queries in both datasets, SAED+S3BD outperforms the S3BD system. In addition, we observe that S3BD produces less relevant results for the BBC dataset compared to the RFC dataset. This is because, unlike the RFC dataset, in several cases, the exact keywords of the benchmark queries do not exist in the BBC dataset. The worst case of these issues has occurred for the \texttt{PCI} query in S3BD, because its query expansion procedure could not capture the complete semantics. In contrast, SAED+S3BD is able to handle the cases where the exact keyword does not exist in the dataset, thus, we see that it yields to a remarkably higher relevancy.

Even if we consider \texttt{PCI} as an outlier and exclude that from the analysis, in Figure~\ref{fig: s3bd bbc}, we still notice that the TSAP@10 score of SAED+S3BD is on average $41.2\%$ higher than S3BD. Although the difference between S3BD and SAED+S3BD is less significant for the RFC dataset (in Figure~\ref{fig: s3bd rfc}), we still notice some $17\%$ improvement in TSAP@10 score. This is because RFC is a domain-specific dataset and the exact keywords of queries can be found in the dataset, hence, making use of smart methods to extract the semantic is not acute to earn relevant results. From these results, we can conclude that SAED can be specifically effective for generic datasets where numerous topics exist in the documents.

\noindent{\textbf{AWS Kendra vs SAED+Kendra.} }
\begin{figure*}
\begin{subfigure} {.45\textwidth}
\includegraphics[width=.95\linewidth]{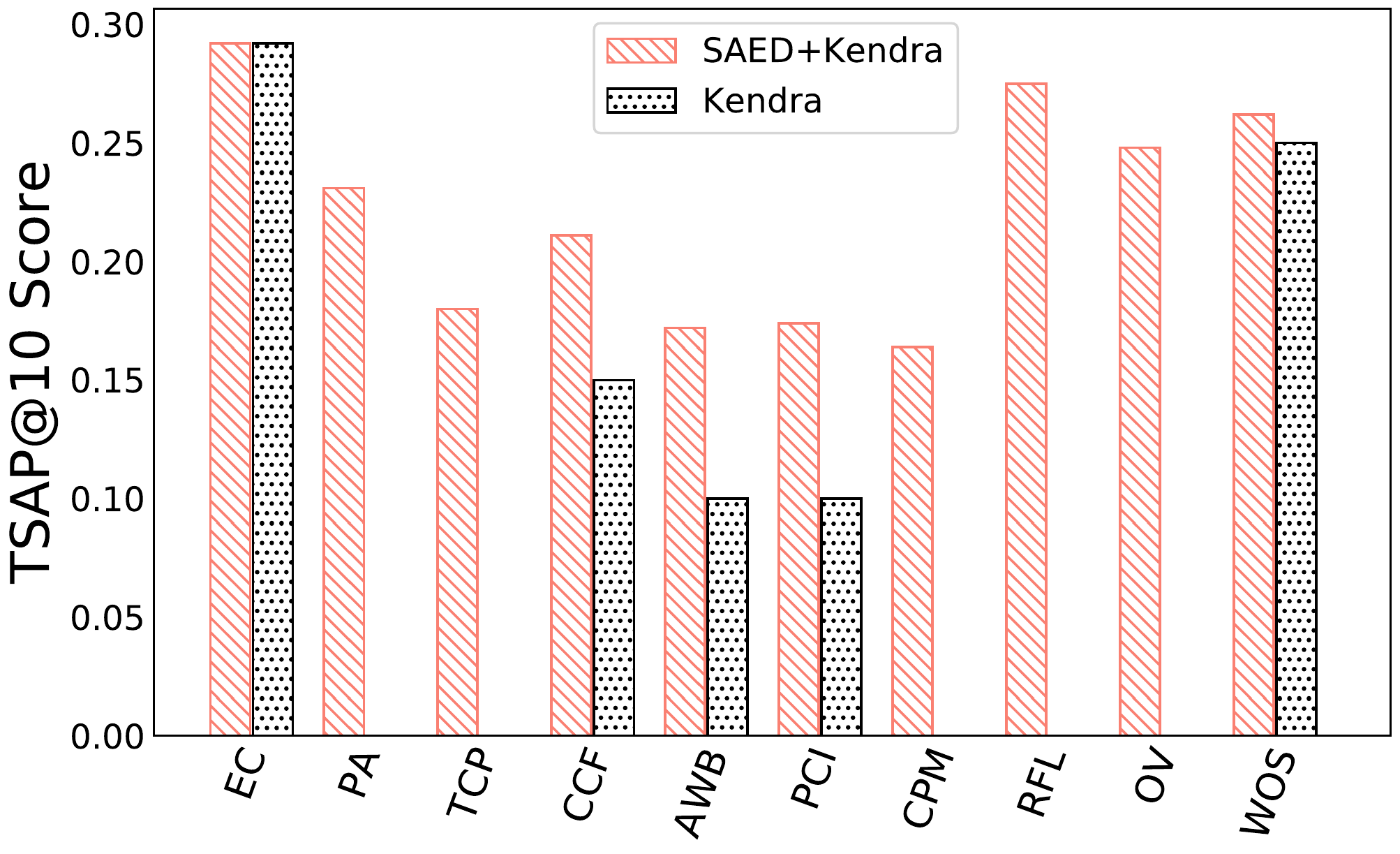}
\caption{Encrypted BBC dataset}
\label{fig: encry kendra bbc}
\end{subfigure} \hfill
\begin{subfigure} {.45\textwidth}
\includegraphics[width=.95\linewidth]{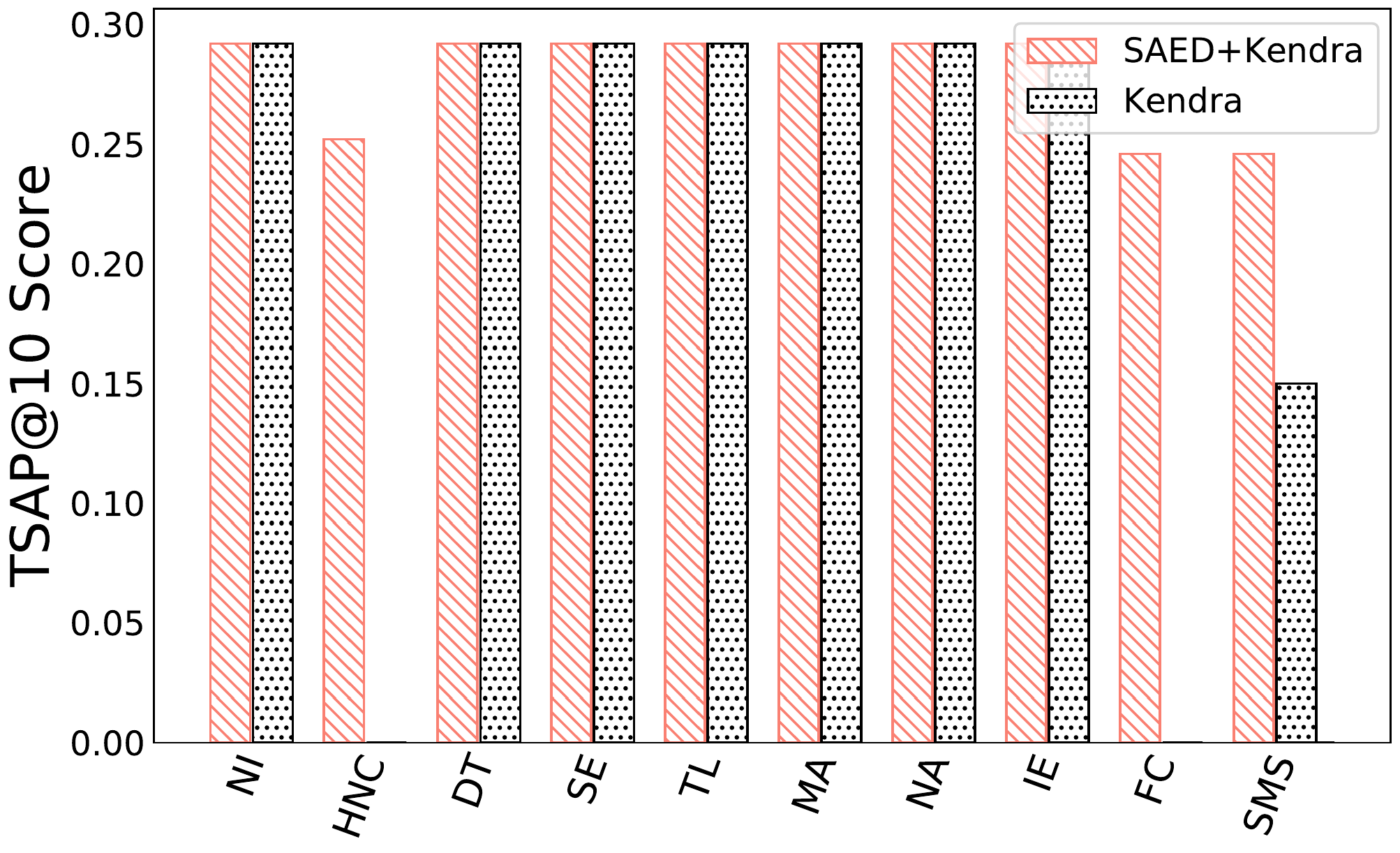}
\caption{Encrypted RFC dataset}
\label{fig: encry kendra rfc}
\end{subfigure} \hfill
\caption{Comparing TSAP@10 scores obtained from SAED+Kendra vs AWS Kendra systems in the encrypted domain.}
\end{figure*}
In Figures~\ref{fig: kendra bbc} and \ref{fig: kendra rfc}, we report TSAP@10 score obtained from AWS Kendra versus SAED+Kendra for BBC and RFC datasets, respectively.  
Specifically, in Figure~\ref{fig: kendra bbc} (BBC dataset), a significant improvement (on average $26.5\%$) is noticed in the TSAP@10 score of SAED+Kendra. However, unlike SAED+S3BD, SAED+Kendra does not beat Kendra for all the queries. The reason Kendra outperforms SAED+Kendra for \texttt{AWB} and \texttt{CPM} queries is that SAED injects extra keywords and sends the expanded query set to AWS Kendra. Then, Kendra returns documents that are related to the queries and to the expanded keywords. We realized that the Ranking unit of SAED occasionally prioritizes documents that include keywords of the expanded queries instead of those with the query keywords. 

Similar to the S3BD experiment, we observe that the relevancy resulted from Kendra and SAED+Kendra is less significant for RFC. However, we still obtain around $12\%$ improvement in TSAP@10 score according to Figure~\ref{fig: kendra rfc}.

\subsection{\textit{Relevancy of Privacy-Preserving Enterprise Search }}~\\

To examine the efficiency of SAED for privacy-preserving enterprise search systems, we conducted experiments using encrypted BBC and RFC datasets. The encrypted datasets were uploaded to the cloud tier and the expanded queries were also encrypted and searched on the cloud tier via Kendra.          

We use the TSAP@10 score, as shown in Figure~\ref{fig: encry kendra bbc} and~\ref{fig: encry kendra rfc}, for the BBC and RFC datasets, respectively. Figure~\ref{fig: encry kendra bbc} indicates that SAED+Kendra substantially outperforms Kendra for all the benchmark queries. We can see that for encrypted dataset Kendra cannot do anything except pattern matching and returning documents that exactly include the encrypted query. Therefore, searching for several queries (\eg \texttt{PA},\texttt{TCP}, \texttt {CPM}, etc.) does not retrieve any documents. We notice that, in both systems, the highest TSAP@10 score is in searching \texttt{EC}. The reason is the high number of documents in BBC that contain the exact phrase \texttt{European commission}.

The reported TSAP@10 scores for the RFC dataset in Figure~\ref{fig: encry kendra rfc} shows a clear improvement in compared with the BBC dataset. We observe that seven out of ten queries provide an equal TSAP@10 scores in both systems. The reason that makes Kendra competitive to SAED+Kendra is the exact availability of the benchmark queries in RFC. However, for \texttt{HNC} and \texttt{FC}, the exact query keywords are not present in the dataset, hence, Kendra fails to find any results.

\subsection{\textit{Discussion of the Relevancy Results }}~\\
In Table~\ref{tab:mean score}, we report \emph{mean F-1} and \emph{mean TSAP@10} scores for the SAED-plugged enterprise search systems along with their original versions upon utilizing the datasets both in the plain-text and encrypted forms. From the table, we notice that, regardless of the enterprise search system being employed, a higher search relevancy is consistently achieved for the RFC dataset as opposed to the BBC dataset. 

The search relevancy is consistently improved when SAED+Kendra is used and it provides on average of $23\%$ improvement in mean F-1 score and $21\%$ in the mean TSAP@10 score. Although original S3BD is the underperformer, using SAED+S3BD improves its mean F-1 and mean TSAP@10 scores by
on average of $40\%$ and $32\%$, respectively.

\begin{table}[h]

\centering
\resizebox{\linewidth}{!}{

\begin{tabular}{c|c|c|c|c|}
\cline{2-5}
\textbf{}                                  & \multicolumn{2}{c|}{\textbf{BBC}} & \multicolumn{2}{c|}{\textbf{RFC}} \\ \hline
\multicolumn{1}{|c|}{\textbf{Systems}} &
  \textbf{\begin{tabular}[c]{@{}c@{}}Mean\\ F-1\end{tabular}} &
  \textbf{\begin{tabular}[c]{@{}c@{}}Mean\\ TSAP@10\end{tabular}} &
  \textbf{\begin{tabular}[c]{@{}c@{}}Mean\\ F-1\end{tabular}} &
  \textbf{\begin{tabular}[c]{@{}c@{}}Mean\\ TSAP@10\end{tabular}} \\ \hline \hline
\multicolumn{1}{|c|}{S3BD}                 & 0.50            & 0.17            & 0.80            & 0.24            \\ \hline
\multicolumn{1}{|c|}{SAED+S3BD}            & 0.82            & 0.25            & 0.92            & \textbf{0.28}   \\ \hline
\multicolumn{1}{|c|}{Kendra}               & 0.67            & 0.20            & 0.88            & 0.26            \\ \hline
\multicolumn{1}{|c|}{SAED+Kendra}          & \textbf{0.90}   & \textbf{0.27}   & \textbf{0.93}   & \textbf{0.28}   \\ \hline  
\multicolumn{1}{|c|}{Kendra (Encry.)}      & 0.31            & 0.09            & 0.75            & 0.22            \\ \hline
\multicolumn{1}{|c|}{SAED+Kendra (Encry.)} & 0.73            & 0.22            & 0.90            & 0.27            \\ \hline
\end{tabular}
}
\caption{\small{Comparing the mean F-1 and the mean TSAP@10 scores obtained from SAED-plugged enterprise search systems versus their original forms. The highest resulted scores are shown in bold font.}}
\label{tab:mean score}
\end{table}

In the encrypted domain, we notice that SAED+Kendra offers a substantially higher (up to $130\%$) search relevancy for BBC dataset. As the exact keywords of the given search queries are not present in the encrypted form of BBC dataset, AWS Kendra fails to perform semantic search, rather does only a pattern matching, which makes it an underperformer for this dataset. 
On the other hand, search relevancy is improved for RFC dataset since mean F-1 and mean TSAP@10 scores are improved by
at least $20\%$. This is because, most of the queries are present exactly in the dataset and Kendra retrieves most of the relevant documents by relying only on pattern matching.

\subsection{\textit{Evaluating the Search Time }}
\begin{figure} [h]
\centering
\vspace{-6pt}
\includegraphics[width=.90\linewidth]{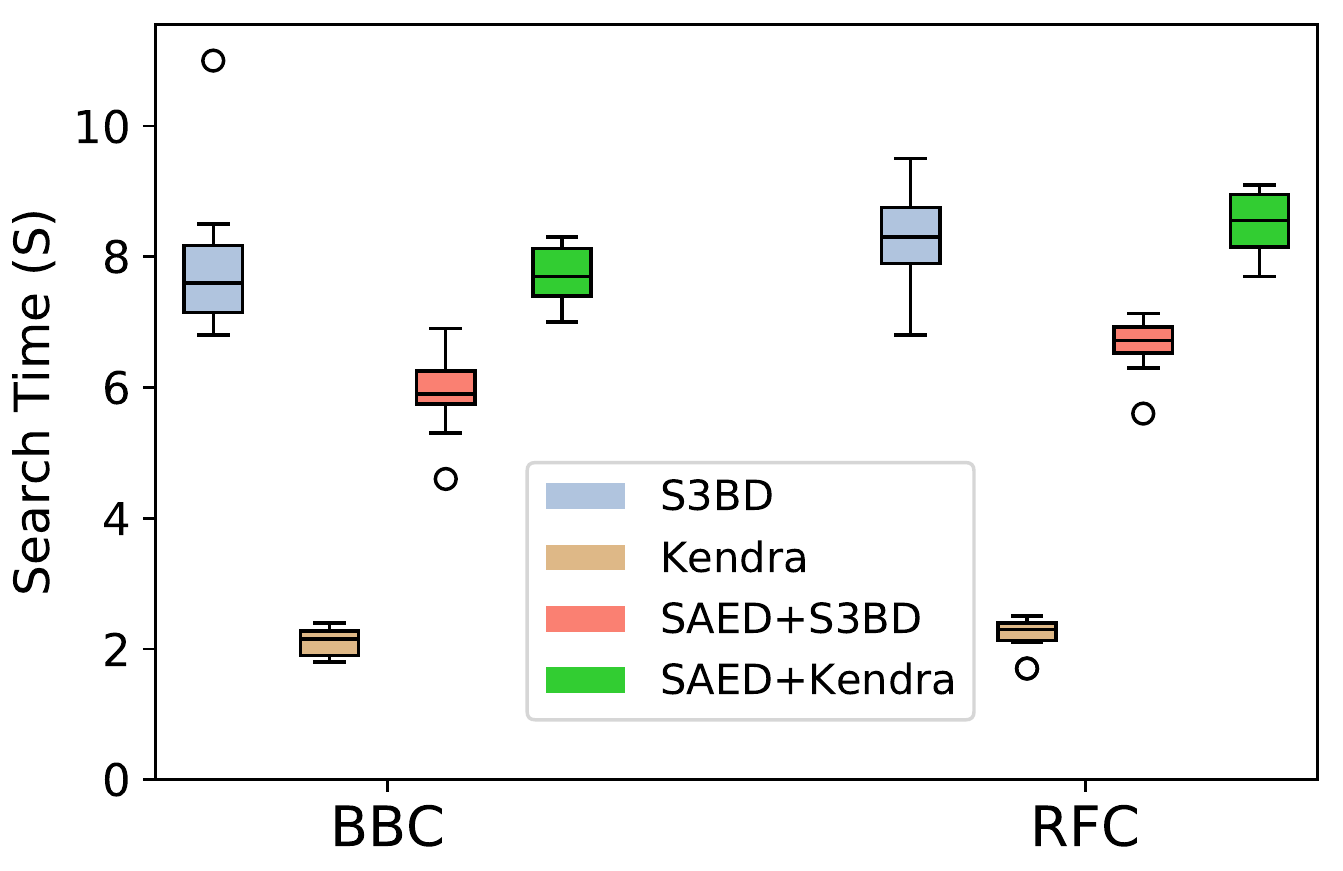}
\caption{Search time comparison among S3BD, Kendra, SAED+S3BD, and SAED+Kendra systems.}
\label{fig: searchtime}
\end{figure}

~\\
Figure~\ref{fig: searchtime} presents the total incurred search time of the experimented queries for each dataset. The search time is calculated as the summation of the elapsed time taken by a query to be processed (\eg expansion, weighting) and turnaround time until the result set is received. To eliminate the impact of any randomness in the computing system, we
searched each set of experimented queries 10 times and reported the results in the form of box plots.
The figure indicates that S3BD system has the highest search time overhead for both datasets which could impact real-time searchability in case of big data. SAED+S3BD incurs less query processing time overhead compared to the original (unmodified) S3BD system.

On the other hand, AWS Kendra causes the lowest time overhead for both datasets compared to SAED+Kendra. SAED+Kendra causes around $4$ times more time overhead compared to original Kendra. However, in the prior set of experiments, we determine that SAED+Kendra achieves a substantially higher search relevancy for most of the queries and, particularly, for datasets with privacy constraints. 

%% file: C4-6-sum.tex
\section{Summary}

A context-aware, personalized, and privacy-preserving  enterprise search service is the need of the hour for data owners who wish to use cloud services. Our approach to address this demand was to separate the search intelligence and privacy aspects from the pattern matching aspect. We developed SAED that achieves privacy and intelligence at the edge tier and leaves the large-scale pattern matching for the cloud tier. SAED is pluggable and can work with any enterprise search solution (\eg AWS Kendra and S3BD) without dictating any change on them. Utilizing edge computing on the user's premises preserves the user's privacy and makes SAED a lightweight solution. Leveraging recurrent neural network-based prediction models, WordNet database, and Word2Vec, SAED proactively expands a search query in a proper contextual direction and weights the expanded query set based on the user's interest. In addition, SAED provides the ability to perform semantic search while the data are stored in the encrypted form on the cloud. In this case, the existing enterprise search solutions just perform the pattern matching without knowing the underlying data. Evaluation results, verified by human users, show that SAED can improve the relevancy of the retrieved results by on average $\approx24\%$ for plain-text and $\approx75\%$ for encrypted generic datasets.

%% file: C5-Memo.tex
\chapter{Multi-Tenancy of Latency-Sensitive Deep Learning Applications on Edge}
\label{chap:concurrent}

\section{Overview}
In the prior chapter, we propose an enterprise search application, namely SAED in the form of a trusted application for enabling secure search over confidential data in the cloud. The SAED application spans across edge-to-cloud continuum and consists of several microservices that run on edge to perform the intelligent aspects of searching (\ie query processing, personalization, and ranking). Our investigation indicates that running a number of microservices on the edge consumes a significant percentage of resource, specifically, edge memory is exhausted and  service(s) can either be killed or failed to execute. Prior studies quantized the NN models to make them lighter but without model management, the system cannot get actual advantage of multi-tenant processing. \name~leverages NN model compression techniques, such as model quantization, and dynamically loads NN models for DL applications to stimulate multi-tenancy on the edge server.  We consider the problem and scale it up in order to come across a unified solution of it. 

\begin{figure}
    \centering
    \includegraphics[width=.8\textwidth]{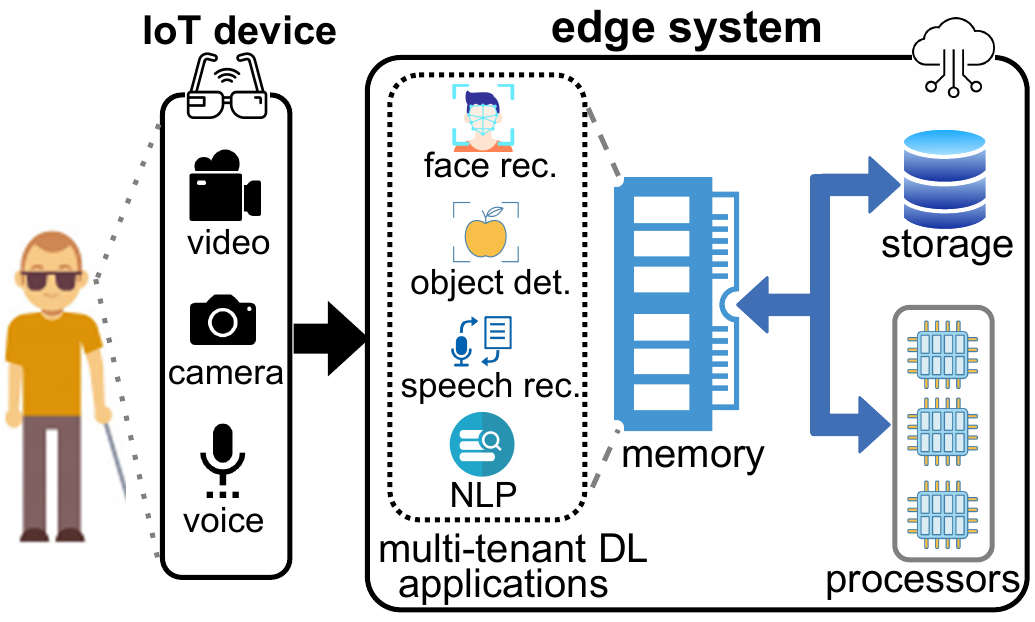}
    \caption{Bird-eye view of SmartSight, an IoT-based system that continuously receives various inputs from the smartglass (IoT device) sensors, and processes them via multi-tenant DL applications running on the edge server.}
\label{fig:high_level}
\vspace{-.3 cm}
\end{figure}

Due to the robust uses of smart IoT-based systems, various application requests (\ie object detection, face recognition, NLP, and motion capture) incoming from users' devices execute on edge tier with low-latency constraint on a daily basis. 
An exemplar use case of such IoT-based systems is SmartSight~\cite{felare2022}, illustrated in Figure~\ref{fig:high_level}, that aims at providing ambient perception for the blind and visually impaired people.
The system operates based on a smartglass (IoT device) and a companion edge server (\eg smartphone). The smartglass continuously captures the inputs via its sensors (\eg camera and microphone) and requests the edge server to process DL-based applications, such as object detection to identify obstacles; face recognition to identify acquainted people; speech recognition, and NLP to understand and react to the user’s commands. To make SmartSight usable, the edge server has to continuously execute multiple (a.k.a. \emph{multi-tenant}) DL application to process incoming requests with low-latency and high accuracy. It is noteworthy that, although cloud datacenters can mitigate the inherent resource limitations of the edge, due to the network latency overhead and data confidentiality~\cite{deng2020edge,zobaed2021saed,hussain2020analyzing}, offloading the latency-sensitive service requests to the cloud is not a tractable approach in many use cases. 

DL applications utilize bulky Neural Network (NN) models at their kernel to infer on the inputs received from the sensors. The NN models have to be kept in memory to enable low-latency (a.k.a. \emph{warm-start}~\cite{dang2017dtrust}) inference operations. Otherwise, because the NN model size is often huge, loading it into the memory in an on-demand manner (a.k.a. \emph{cold-start}) is counterproductive and affects the latency constraint of the DL applications. As the edge servers naturally have a limited memory size (\eg 4 GB in the case of Jetson Nano \cite{nvidia}), multi-tenant execution of DL applications on them leads to a memory contention challenge across the processes~\cite{deng2020edge,gholami2021survey}.

Accordingly, the main challenge of this study is to resolve the memory contention across multi-tenant DL applications without compromising their latency and accuracy constraints.

\begin{table}
\centering
\caption{Load time, inference time, and accuracy of popular NN models individually running
on Samsung Galaxy S20+ as the edge server.}
\resizebox{.95\linewidth}{!}{
\begin{tabular}{|c|c|c|c|c|c|}
\hline

     {\textbf{NN Models}}
                                                & \begin{tabular}[c]{@{}c@{}}\textbf{Bit}\\  \textbf{Width}\end{tabular} & \multicolumn{1}{l|}{\begin{tabular}[c]{@{}l@{}}\textbf{Size}\\ \textbf{(MB)}\end{tabular}} & \begin{tabular}[c]{@{}c@{}}\textbf{Loading}\\\textbf{Time (ms)}\end{tabular} & \begin{tabular}[c]{@{}c@{}}\textbf{Inference}\\\textbf{Time (ms)}\end{tabular} & \begin{tabular}[c]{@{}c@{}}\textbf{Accu-}\\\textbf{racy (\%)}\end{tabular} \\ \hline \hline
\multirow{2}{*}{InceptionV3}                      & FP32                                                 & 105                                                                      & 650                                                       & 100                                                      & 78.50                                                \\ \cline{2-6} 
                                                  & INT8                                                 & 24                                                                       & 380                                                       & 80                                                       & 77.20                                                \\ \thickhline
\multirow{2}{*}{VGG16}                            & FP32                                                 & 528                                                                      & 820                                                       & 52                                                       & 71.30                                                \\ \cline{2-6} 
                                                  & INT8                                                 & 132                                                                      & 185                                                       & 40                                                       & 70.18                                                \\ \thickhline
\multirow{2}{*}{MobileNetV1}                      & FP32                                                 & 89                                                                       & 600                                                       & 15                                                       & 70.56                                                \\ \cline{2-6} 
                                                  & INT8                                                 & 23                                                                       & 192                                                       & 8                                                        & 65.70                                                \\ \thickhline
\multirow{2}{*}{MobileNetV2}                      & FP32                                                 & 26                                                                       & 110                                                       & 10                                                       & 72.08                                                \\ \cline{2-6} 
                                                  & INT8                                                 & 9                                                                        & 65                                                        & 7.5                                                      & 63.70                                                \\ \thickhline
                                                  
\multirow{2}{*}{MobileNetV3}  & FP32                                                 & 14                                                                       & 80.3                                                      & 7.80                                                     & 74.04                                                \\ \cline{2-6} 
\multicolumn{1}{|l|}{}                             & INT8                                                 & 8                                                                        & 47.45                                                     & 6.21                                                     & 71.32                           \\ \thickhline

\multirow{2}{*}{MobileBERT}  & FP32                                                 & 96                                                                       & 1100                                                      & 62                                                       & 81.23                                                \\ \cline{2-6} 
\multicolumn{1}{|l|}{}                            & INT8                                                 & 26                                                                       & 890                                                       & 40                                                       & 77.08                                                \\ \hline
\end{tabular}
}
\label{tab:model_des}
\vspace{ -.5 cm}
\end{table}

In the deep learning context, there are techniques based on the idea of approximate computing, such as quantization~\cite{zhou2018adaptive}, that make the model edge-friendly via compressing its NN model, hence, reducing its inference time and accuracy. To understand the impact of such approximations, we conducted a preliminary experiment using a Samsung Galaxy S20+ as the edge server; and five popular DNN models, namely InceptionV3, VGG16, MobileNetV1, MobileNetV2, MobileNetV3, MobileBERT, each one at two quantization (precision) levels, namely FP32 and INT8 bit widths. In Table~\ref{tab:model_des}, we report the average loading time, inference time, and accuracy for their individual executions. We observe that: (A) for all the models, the loading time is 8---17$\times$ more than its inference time; (B) Loading the high-precision model (FP32 bit width) occupies $\approx$3.5$\times$ more memory than the low-precision (INT8 bit width) one; and (C) Loading a low-precision model can reduce the inference accuracy by around 3---6\%. These results demonstrate that the model compression has a considerable potential to mitigate the memory footprint of the DL applications. Moreover, the model loading time invariably dominates the inference time~\cite{samani2022exploring}. Accordingly, our hypothesis is that \emph{the efficient use of model compression and the edge memory can enhance the multi-tenancy and inference time of DL applications without any major loss on their inference accuracy}. 

We propose each DL application to be equipped with multiple NN models with different precision levels. The low-precision models have a small memory footprint, hence, allowing for a higher multi-tenancy of DL applications with their models loaded into the memory (\ie warm-start inference) that enhances the service latency. However, loading overly low-precision (over-quantized) models to maximize multi-tenancy and warm-start inference is not viable, because it reduces the inference accuracy and renders the multi-tenant DL applications to be futile. On the contrary, loading high-precision (large) NN models on a memory-limited edge system for an indefinite time period unnecessarily occupies an excessive memory space that is detrimental for the multi-tenancy and warm-start inference of other tenants. That is, other tenants face a significant slow down (as noted in Table~\ref{tab:model_des}), because they cannot keep their NN model in memory and have to load it from the storage (\ie cold-start) to perform the inference operation. Therefore, an ideal solution for a multi-tenant edge system should be able to dynamically load a suitable model from the set of models available to the application (a.k.a. model zoo), such that it neither interrupts the execution of other applications, nor causes a cold-start inference for them.

\section{Problem Statement}
The research question that we investigate is: \emph{how to maximize the number of warm-start inferences for multi-tenant DL applications on edge without compromising the inference accuracy?} The question indicates a trade-off between two objectives: fulfilling the latency constraint of DL applications and maintaining their inference accuracy. The former objective entails having the NN models of DL applications loaded into the memory (\ie warm-start inference), whereas, the latter entails retaining high-precision NN models in the memory. 

For application $A_i \in A$ with $M_i = \{ m^{k}_i \ | \ 1 \le k \le q_i\}$ as its model zoo, let $r_i(t)$ be a Boolean function that represents an inference request for $A_i$ at time $t$ with value 1. 
Also, let $m^\ast_i \subseteq M_i$ be an NN model of $A_i$ with size of $s^\ast_i$ that is currently loaded in the memory. This means that, for application $A_j$ that does not have any of its NN models currently in the memory, we have $m^\ast_j = \varnothing$ and $s^\ast_j = 0$. Then, $M^\ast = \bigcup_{i=1}^{n} m^\ast_i$ represents the set of currently loaded NN models that occupy $S^\ast = \sum_{i=1}^{n} s^\ast_i$ of the memory space. 
A cold start event for the request arrives at time $t$ for $A_i$, denoted $C_i(M^\ast,t)$ and shown in Equation (\ref{eq:cold_start}), occurs when there is no NN model in memory for $A_i$ (\ie $M_i \cap M^\ast = \varnothing$).

\begin{equation}
\label{eq:cold_start}
C_i(M^\ast,t) =
\begin{cases}
    r_i(t)  & M_i \cap M^\ast = \varnothing \\
    0 & otherwise
\end{cases} 
\end{equation}

Assume that utilizing $m^\ast_i \in M_i$ results in an inference accuracy that we denote it as $\chi^\ast_i$. Then, based on Equation~(\ref{eq:bi_objective}), for $n$ multi-tenant DL applications, we can formally state the objective function as minimizing the total number of cold-start inferences, while maximizing the accuracy of the inferences. In this case, the total memory size available for the NN models (denoted $S$) serves as the constraint.

\begin{equation}
\label{eq:bi_objective}
\begin{gathered}
    \min \biggl( \int_{t}^{\infty} \sum\limits^{n}_{i=1}   C_i(M^\ast, t) \ dt \biggr) \ , \quad
    \max \biggl( \int_{t}^{\infty} \sum_{i=1}^{n} \chi^\ast_i(t)  \ dt  \biggr)\\
    \textrm{subject to:} \quad \quad \quad
    \forall t ,\ \ \sum\limits_{i=1}^{n} s^\ast_i \le S \quad \quad \quad 
\end{gathered}
\end{equation}

Note that optimal NN model management decisions do not have a greedy nature. That is, minimizing the number of cold-start inferences at a given time $t$ does not necessarily lead to the minimum total number of cold-starts with maximum accuracy during the entire applications' lifetime. In other words, the system may experience a cold-start at time $t$ to prevent multiple ones at a later time. That is why, the objective function of Equation~\ref{eq:bi_objective} includes integrals over $t$ to the $\infty$ to encompass the impacts of the decisions at $t$ on the future cold-starts and accuracy levels. In the objectives, the NN models of application $A_i$ are only chosen from its model zoo ($M_i$), thus, the accuracy ($\mu_i(t)$) and size functions ($s_i$) are discrete functions. It is needless to say that \emph{minimizing the number of cold-start inferences} is equivalent to \emph{maximizing the number of warm-start events}~\cite{manner2018cold}. In the rest of this chapter, we use these two interchangeably.

\section{Solution Statement and Contributions}
To stimulate multi-tenancy on the limited edge memory, we develop a framework, called \name, that takes advantage of a model zoo for each DL application and can dynamically swap the NN models of the applications. To maximize the number of warm-starts with high inference accuracy across multi-tenant DL applications, our approach is to proactively load the high-precision NN models for the applications that are expected to receive inference requests, while loading low-precision models for the others. We utilize the recent memory usage information to predict the memory availability for the next executions while not interrupting other active applications. We develop model management heuristic policies that make use of the expected memory availability and the usage pattern of multi-tenant DL applications to choose a suitable NN model for the requester application right before the inference operation, thereby, both the latency and inference accuracy of the application are fulfilled.

%% file: C5-3-archi.tex


\section{Architectural Overview \& System Design of \name }
\begin{figure}
  \centering
    \includegraphics[width=0.7\textwidth]{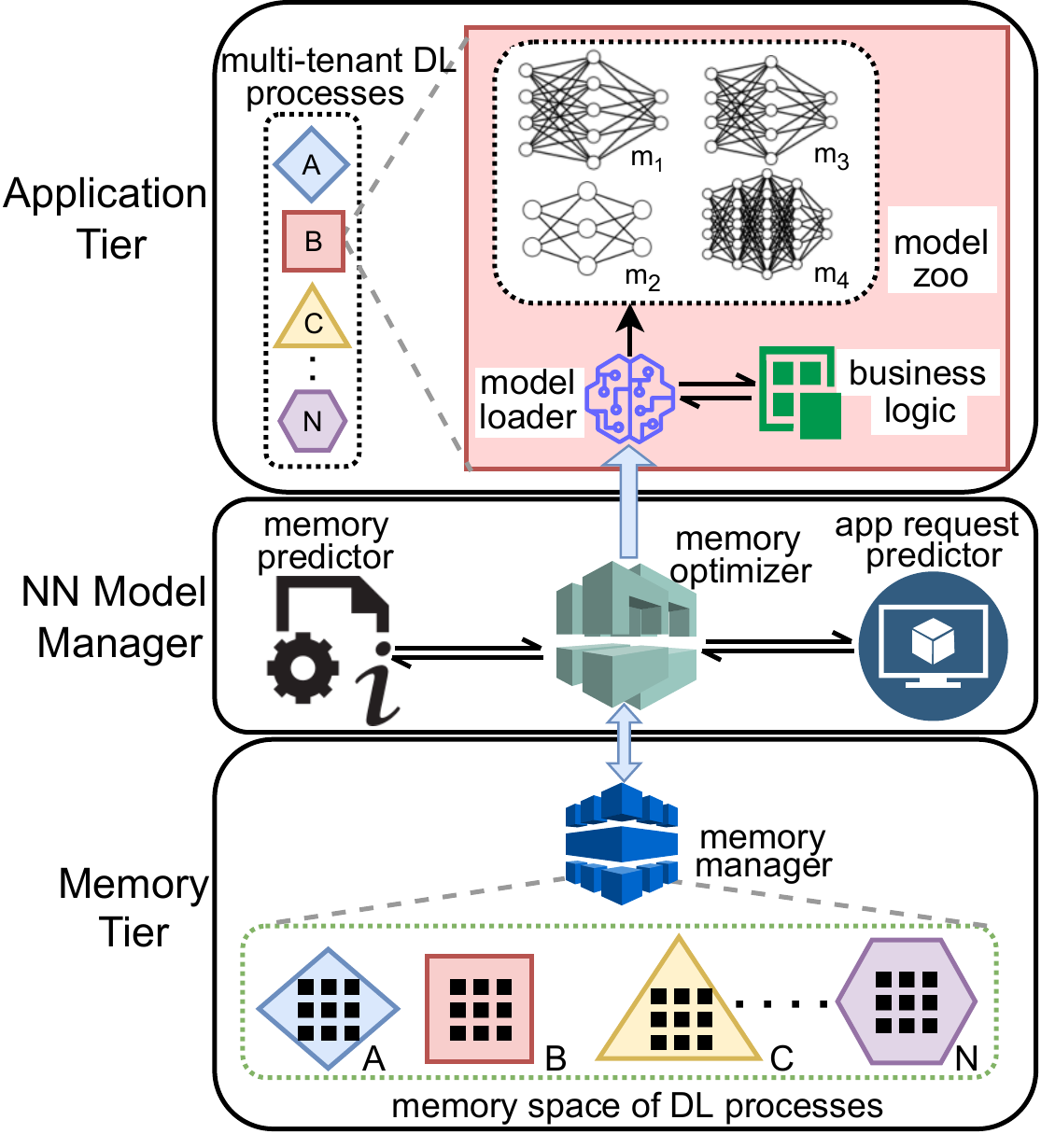}
  \caption{\small{Architectural overview of the \name~framework with three tiers: Application, NN Model Manager, and Memory.}}\label{fig:archi}
  \vspace{-5 mm}
\end{figure}

Figure~\ref{fig:archi} illustrates the architectural overview of \name that facilitates multi-tenancy of DL applications on a resource-limited edge system via enabling the applications to only swap their NN models, instead of the entire application. The framework consists of three tiers: (i) Application tier, (ii) NN model manager, and (iii) Memory tier.

\noindent\textbf{Application Tier.} The incoming multi-modal inputs from the connected IoT devices trigger execution of multi-tenant DL applications in the application tier. The \emph{model zoo} for each DL application acts as a repository that contains NN models with different compression levels (sizes) and inference accuracy (a.k.a. various \emph{precision levels}). The \emph{model loader} is responsible for loading the chosen NN model from the model zoo into the edge memory. 


\noindent\textbf{NN Model Manager.} NN model manager comprises of three components: (i) \emph{application request predictor}, (ii) \emph{memory predictor}, and (iii) \emph{memory optimizer}. ``Application request predictor'' collects historical requests to each application and trains a lightweight (edge-friendly) many-to-one vanilla recurrent neural network (RNN) time series prediction model, similar to the one in \cite{pang2020innovative}, to periodically foresee the inference request arrivals for each application. 
Upon arrival of each request, ``memory predictor'' is in charge of predicting the memory availability based on the recent memory allocations in the entire edge system. We leverage the historical memory allocation data and train another many-to-one vanilla RNN time-series prediction model to predict the available memory.

Memory optimizer interacts with the 
application ``request predictor'' and ``memory predictor'' to receive: (A) the request arrival time for different applications plus the information of their model zoo; and (B) the memory availability information. Then, the memory optimizer feeds the received information to an NN model management policy that determines the highest possible precision NN model that can be loaded to serve the inference request of a DL application with the minimum impact (in terms of the prediction accuracy or latency) on the execution of other applications. Upon facing memory shortage for an arriving inference request, the memory optimizer scavenges the memory allocated to the NN models of other applications via either loading a lower-precision model or forcing them to cold-start. After procuring adequate memory, the memory optimizer informs the ``model loader'' to load the appropriate NN model of the  requested application.

 \noindent\textbf{Memory Tier.} The tier includes the ``memory spaces'' allocated to the applications; and a ``memory manager'' that keeps track of the currently loaded models, the available memory spaces, and the current status of the applications. The memory manager communicates these information to the NN Model Manager to efficiently allocates them to the arriving requests.

\section{Heuristics to Manage Models of Multi-tenant Applications}
\subsection{\textit{Overview }}~\\
Recall that the aim of NN model management policy is to minimize the number of cold-start inferences and maximize the inference accuracy for multi-tenant DL applications on the edge servers. To that end, the memory optimizer strives to maximize the time to retain the loaded models in the edge memory. However, due to limitations in the available memory space, it is not possible to retain the highest precision NN model of all applications in the memory. To resolve this memory contention, the NN models of the applications that are 
unlikely to be requested in the near future should be assigned a lower priority to remain in the memory. Furthermore, \name makes it possible to dynamically load NN models for the applications. This means that, upon predicting time $t$ as the inference request time for a given DL application, \name can be instructed to load the high-precision NN model of that application immediately before performing the inference. Similarly, in the face of a memory shortage, for the application(s) that are unlikely to be requested at time $t$, \name can be instructed to unload their NN models or, more interestingly, replace them with a lower precision one. 

\begin{figure}
    \centering
    \includegraphics[width=.8\textwidth]{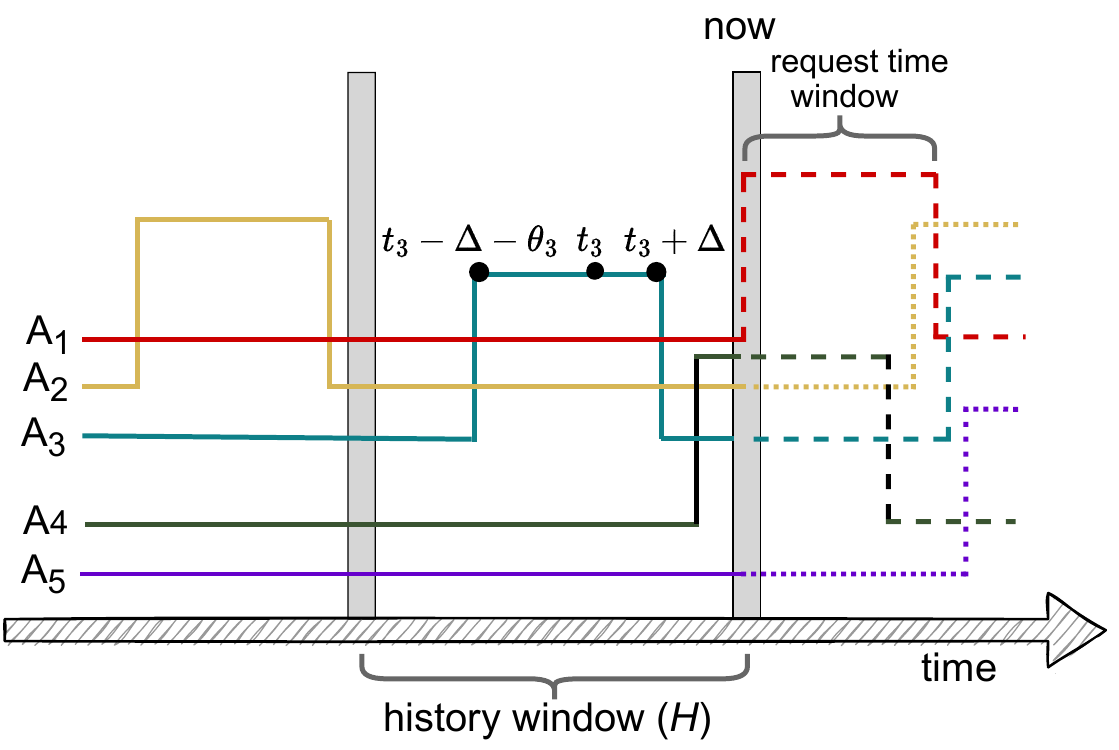}
    \caption{A sample scenario of inference requests for five multi-tenant applications, namely $A_1$ to $A_5$. Each pulse represents the time window within which an inference request is expected. Solid lines expresses the event that has already happened and dashed lines after ``now'' are the request predictions.}
\label{fig:sim}
\vspace{-5 mm}
\end{figure}

However, we know that the request arrivals are inherently uncertain \cite{felare2022} and no prediction model can precisely capture the exact request time for an application. To capture the uncertainty, we consider a request time window, denoted as $\Delta$, around each predicted request time. 
The value of $\Delta$ is obtained from profiling past request predictions and calculating the mean difference of actual arrival time and the predicted ones across all applications.  
In addition, there is a time overhead, denoted as $\theta_i$, to load the chosen NN model of an application $A_i$ into the memory. In sum, to prevent a cold-start for $A_i$ that is predicted to perform inference at time $t$, as shown in Figure~\ref{fig:sim}, the NN model  has to be loaded at time $(t_i-\Delta-\theta_i)$ and kept in memory until $(t_i+\Delta)$.    
Furthermore, there is uncertainty in predictions of ``no request'' for an application at a given time. That is, at time $t$, there can be an inference request for an application that was predicted not to have an request at that time. To make the system robust against this type of uncertainty and to avoid cold-start inferences in these circumstances, an ideal policy should load low-precision NN models for these applications. Hence, an unpredicted inference request can be still served as a warm-start by the low-precision model and the latency constraint is maintained. 

In this work, the set of applications whose NN models are retained in memory outside of their predicted request time window are called the \emph{minimalist} set, and denoted as $A^\prime$. Similarly, the set of applications that are in their request time window and we load a high-precision model for them are called \emph{maximalist}, and denoted as $A^\ast$. 
To resolve the memory contention, the policy can be based on scavenging memory from the minimalist applications to procure the required memory space for the maximalist ones. That is, in the event that application $A_i$ is predicted to have an inference at time $t_i$, it becomes a member of $A^\ast$ set at time $t_i-\Delta-\theta_i$, and then becomes a member of $A^\prime$ set after $t_i+\Delta$; thus, its model can be evicted from the memory in the event the memory space is needed for another maximalist application. The NN model eviction is only permitted from $A^\prime$ set and we aim at retaining a low-precision model for the applications in this set. However, due to high inference demand, $A^\prime$ have to unload their models (\ie switch to cold-start) to free space for the model of the applications that are in the maximalist set. In an extreme situation, if $A^\prime$ is empty, or the scavenged memory from $A^\prime$ cannot procure sufficient space to load the suitable model for application $A_i$, the next (smaller) model for $A_i$ is considered, and the aforementioned steps are repeated. Ultimately, if the scavenged memory space is inadequate for the lowest precision model of $A_i$, an inference failure occurs.

The memory contention problem can be reduced to the classic binary Knapsack optimization problem \cite{chen2021fepim} where from a collection of items, each one with a weight and a value, we need to select items such that the total value is maximized, while the total weight is bounded to a limit. This problem is known to be NP-Complete,
hence, we can rely on the heuristic-based solutions for it \cite{christensenheuristic}. In the next part, we discuss four NN model management (a.k.a. \emph{NN model eviction}) policies to manage the memory for multi-tenant DL applications such that the number of warm-start inferences is maximized without any major impact on the inference accuracy.
\subsection{\textit{Policy 1: Largest-First Model Eviction (LFE) }}~\\
In this policy, to allocate memory for the NN model of a maximalist process, we first evict NN models from set ($A^\prime$ that occupy the highest memory space, until there is enough space to allocate the high-precision NN model of $A^\ast$. For that purpose, members of $A^\prime$ are sorted based on the size of their currently loaded NN model in the descending order. In the event that evicting all the NN models of $A_\prime$ does not free enough memory space to allocate the NN model of the request, a lower precision NN model (smaller in size) is tried for allocation. This procedure continues until a model from the model zoo can be allocated in the memory; otherwise, the edge system is not able to serve that request at that time. 

\subsection{\textit{Policy 2: Best-Fit Model Eviction (BFE) }}~\\
The limitation of LFE is to evict the largest NN models of the minimalist applications, irrespective of the exact memory requirement. This means that adopting LFE  can free more memory space than the actual requirement. To tackle the issue, we implement the BFE policy where applications in the minimalist set are sorted based on the difference between their model sizes and the actual memory requirement. Then, the NN model with a minimum difference is chosen for eviction. 
The memory requirement for a maximalist application is first calculated based on its highest precision (largest) NN model to gain the highest inference accuracy. However, in the event that evicting the NN models of all the minimalist applications do not free enough memory space to allocate the desired NN model, BFE iteratively selects the next high-precision model from the model zoo of the requested application. 

\subsection{\textit{Policy 3: Warm-Start-aware Best-Fit Model Eviction (WS-BFE) }}~\\
Let $A_i\in A^\ast$ an application that is currently in the maximalist set, and $A_j\in A^\prime$ an application that is currently in the minimalist set. It is technically possible that the predicted request time window of $A_i$ overlaps with the one for $A_j$. In this case, LFE and BFE policies potentially choose to evict the NN model of $A_j$ in favor of the $A_i$ model. This is because both of these policies are backward-looking and ignore the fact that $A_j$ can be requested soon after evicting its NN model. Such an eviction decision increases the likelihood of a cold-start inference and to avoid that, we develop WS-BFE that assigns the lowest eviction priority to those applications in $A^\prime$ that have overlapping time window with $A_i$. 

In our early experiments, we realized that another reason for cold-start inferences is due to uncertain nature of request arrivals. That is, a minimalist application is unexpectedly requested. To minimize the likelihood of cold-start inference in these circumstances, we implement WS-BFE to \emph{replace} the evicted NN model with the lowest-precision (\ie smallest) NN model of that application. As such, in the event of an unpredicted request the minimalist applications, there is a low-precision model available to carry out a warm-start inference. 

\subsection{\textit{Policy 4: Intelligent Warm-Start-aware Best-Fit Eviction (iWS-BFE) }}~\\
To make WS-BFE robust against uncertainties in the application request time prediction, we enhance it by applying the Bayesian theory and proposing a new policy, called iWS-BFE. This policy is inspired from the widely-adopted LRU-K cache management policy~\cite{o1993lru} that considers \emph{the least recently used (\ie requested) applications are not likely to be requested in the near future}. 
Similarly, iWS-BFE only considers members of $A^\prime$ as eviction candidates, denoted by $E^\prime$, that are not recently requested.
Figure~\ref{fig:sim}, shows a scenario of predicted request times for $A_1$---$A_5$. To procure memory for $A_1$, we have $A^\prime=\{A_2, A_3, A_5\}$. Because $A_3$ was requested during the ``history window'' ($H$), it is likely to be requested in the near future. Hence, iWS-BFE, chooses $E^\prime=\{A_2, A_5\}$ for eviction. The value of $H$ is determined based on the mean request inter-arrival time of all applications.  

In addition to considering LRU, iWS-BFE also makes use of the request prediction, provided by \name. That is, it considers the most appropriate application for eviction as the one that has not been recently requested, and is predicted to be requested the latest in future. However, the request time predictions are uncertain, and the system can receive an unexpected request from members of $E^\prime$ in the current request window. To make iWS-BFE robust against such uncertainty, we calculate the probability of an unexpected request. For application $A_j\in E^\prime$, let $r_j$ denote an unexpected request. Then, the probability of $r_j$ occurring during the current request window (\ie [$t, t+\Delta$]) is defined as $P(r_j | A_i \in A^\ast)$. The application that is likely to be requested unexpectedly is not an optimal choice for eviction. Therefore, in Equation~\ref{eq:scoreiws}, to calculate the fitness score of $A_j$ for eviction (denoted $Score(A_j)$), we consider $1 - P(r_j | A_i \in A^\ast)$. To take the predicted request time of $A_j$ into consideration, we calculate the distance between its predicted request time and the current time (\ie $t_j-t_i$). To confine the value between [0,1], we normalize the distance based on the latest predicted distance across all $k$ applications. 


\vspace{-3 mm}
\begin{equation}
\label{eq:scoreiws}
\begin{gathered}
Score(A_j) = \frac{t_j-t_i}{\displaystyle \max_{k\in E^\prime} (t_k - t_i)} \cdotp 
\bigl[ 1 - P(r_j | A_i \in A^\ast)\bigr] 
\end{gathered}
 \end{equation}

The pseudo-code of the iWS-BFE policy is provided in Algorithm~\ref{alg:memory_man}. It begins with an initial set of eviction candidates, called $\tau \subseteq A^\prime$, that is formed based on the applications that were not requested during the history window ($H$). From $\tau$, in Step $3$, a list of eviction candidates (denoted $E$) whose elements do not overlap with the request window of active application ($A_i$) is derived. Next, in Step 4, we use Equation~\ref{eq:scoreiws} to calculate the fitness score for each $E_k \in E$ and then, build a max-heap tree of $E$ based on the fitness scores (Step 5). In Steps 6---10, the policy iteratively retrieves the application with the highest fitness score (\ie the max-heap root, denoted $w$) and foresees the amount of memory that can be scavenged upon replacing its loaded model with the lowest-precision one. Once the policy finds enough memory to be scavenged such that the NN model of $A_i$ (denoted $m_i$) can be loaded, in Step 13, it enacts all the NN model replacement decisions and then loads $m_i$ in Step 14. In the event that the scavenged memory is insufficient, the policy switches to the next NN model for $A_i$ that has a lower size and accuracy (Step 17). In the worst case that even the smallest NN model of $A_i$ cannot fit in the memory, the inference request fails (Step 17)~\cite{mokhtari2020autonomous}.
%



\begin{algorithm}
	\SetAlgoLined\DontPrintSemicolon
	\SetKwInOut{Input}{Input}
	\SetKwInOut{Output}{Output}
	\SetKwFunction{algo}{algo}
	\SetKwFunction{proc}{Procedure}{}{}
	\SetKwFunction{main}{\textbf{ChooseCenter}}
		
	\SetKwBlock{Function}{Function \texttt{iWS-BFE($A^\prime$, $A^\ast$, $A_i$, $H$)}}{end}
	
	\Function{
	   $\tau \gets$ Select $\forall A^\prime_j \in A^\prime$ not requested during $H$\;
	   $E \gets$  Determine $\forall A^\prime_j \in \tau$ non-overlapping with request window of $A_i$ \;
	   
	$\forall E_k\in E$ calculate fitness score using Equation~\ref{eq:scoreiws}\;
       Build max-heap tree of $E$ based on fitness score\;

        \While {$size(m_i) >$ available memory}  
        {  
        $w \gets$ Extract root of the max-heap tree \;
        \textbf{If}{ $w=\emptyset$} \textbf{then} break the loop \;
        Measure memory scavenged by replacing model of $w$ with its lowest-precision one \; 
        Add scavenged amount to \emph{available memory} \;
        }

	    \If {$size(m_i) \leq$ available memory} 
	    {
            Enact NN model replacement(s) decisions \;
	    Scavenge the leftover memory to load $m_i$ \;

	    }
	    \Else 
	    {

            \textbf{If} there is no model left to check \textbf{then} the inference request fails \;
             Repeat Step 6---10 with the next (smaller) model \;
	    }
	     
	}
	\caption{Pseudo-code for iWS-BFE NN model eviction policy}
	\label{alg:memory_man}
	
\end{algorithm}

%% file: C5-4-perf.tex
\section{Performance Evaluation}
\label{sec:edgeevltn}

\subsection{\textit{Experimental Setup and Evaluation Metrics }}
To evaluate the efficacy of \name and its NN model eviction policies, we benchmarked five different DL applications, namely face recognition, speech recognition, image classification, next sentence prediction, and text classification, and recorded their real characteristics, including the model size, and the inference accuracy (shown in Table~\ref{tab:app_model_size}).
We have developed the E2C simulator that enables modeling the IoT-based systems with different characteristics and configurations, and is available publicly for the community access through our Github page\footnote{Github page of the E2C simulator: \url{ https://github.com/hpcclab/E2C-Sim.git}}. The simulator has implemented all of the NN model eviction policies, and the user can quickly deploy and examine any one of them. 

\begin{table}[]
\centering
\resizebox{.7\linewidth}{!}{
\begin{tabular}{|c|c|c|c|c|}
\hline

\textbf{Application}                  & \textbf{NN Model}                                                                    & \textbf{\begin{tabular}[c]{@{}c@{}}Bit\\ Width\end{tabular}} & \textbf{\begin{tabular}[c]{@{}c@{}}Size\\ (MB)\end{tabular}} & \textbf{\begin{tabular}[c]{@{}c@{}}Accuracy\\ (\%)\end{tabular}} \\ \hline \hline
\multirow{3}{*}{Face recognition}     & \multirow{3}{*}{VGG-Face}                                                            & FP32                                                         & 535.1                                                        & 90.2                                                             \\ \cline{3-5} 
                                      &                                                                                      & FP16                                                         & 378.8                                                        & 82.5                                                             \\ \cline{3-5} 
                                      &                                                                                      & INT8                                                         & 144.2                                                        & 71.8                                                             \\ \thickhline
\multirow{3}{*}{Image classification} & \multirow{3}{*}{VIT-base-patch16}                                                    & FP32                                                         & 346.4                                                        & 94.5                                                             \\ \cline{3-5} 
                                      &                                                                                      & FP16                                                         & 242.2                                                        & 81.3                                                             \\ \cline{3-5} 
                                      &                                                                                      & INT8                                                         & 106.7                                                        & 72.2                                                             \\ \thickhline
\multirow{3}{*}{Speech recognition}   & \multirow{3}{*}{S2T-librisspeech}                                                    & FP32                                                         & 285.2                                                        & 89.7                                                             \\ \cline{3-5} 
                                      &                                                                                      & FP16                                                         & 228.0                                                          & 77.2                                                             \\ \cline{3-5} 
                                      &                                                                                      & INT8                                                         & 78.4                                                         & 68.0                                                               \\ \thickhline
\multirow{3}{*}{Sentence prediction}  & \multirow{3}{*}{\begin{tabular}[c]{@{}c@{}}Paraphrase-Mini\\ LM-L12-v2\end{tabular}} & FP32                                                         & 471.3                                                        & 88.2                                                             \\ \cline{3-5} 
                                      &                                                                                      & FP16                                                         & 377.6                                                        & 81.7                                                             \\ \cline{3-5} 
                                      &                                                                                      & INT8                                                         & 98.9                                                         & 76.2                                                             \\ \thickhline
\multirow{3}{*}{Text classification}  & \multirow{3}{*}{Roberta-base}                                                        & FP32                                                         & 499.0                                                          & 91.1                                                             \\ \cline{3-5} 
                                      &                                                                                      & FP16                                                         & 392.2                                                        & 82.4                                                             \\ \cline{3-5} 
                                      &                                                                                      & INT8                                                         & 132.3                                                        & 76.6                                                             \\ \thickhline
\end{tabular}
}
\caption{Application-specific models with different precision variants that are experimented.}
\label{tab:app_model_size}
\end{table}

\begin{figure*} 
    \centering
    \includegraphics[width=.95\textwidth]{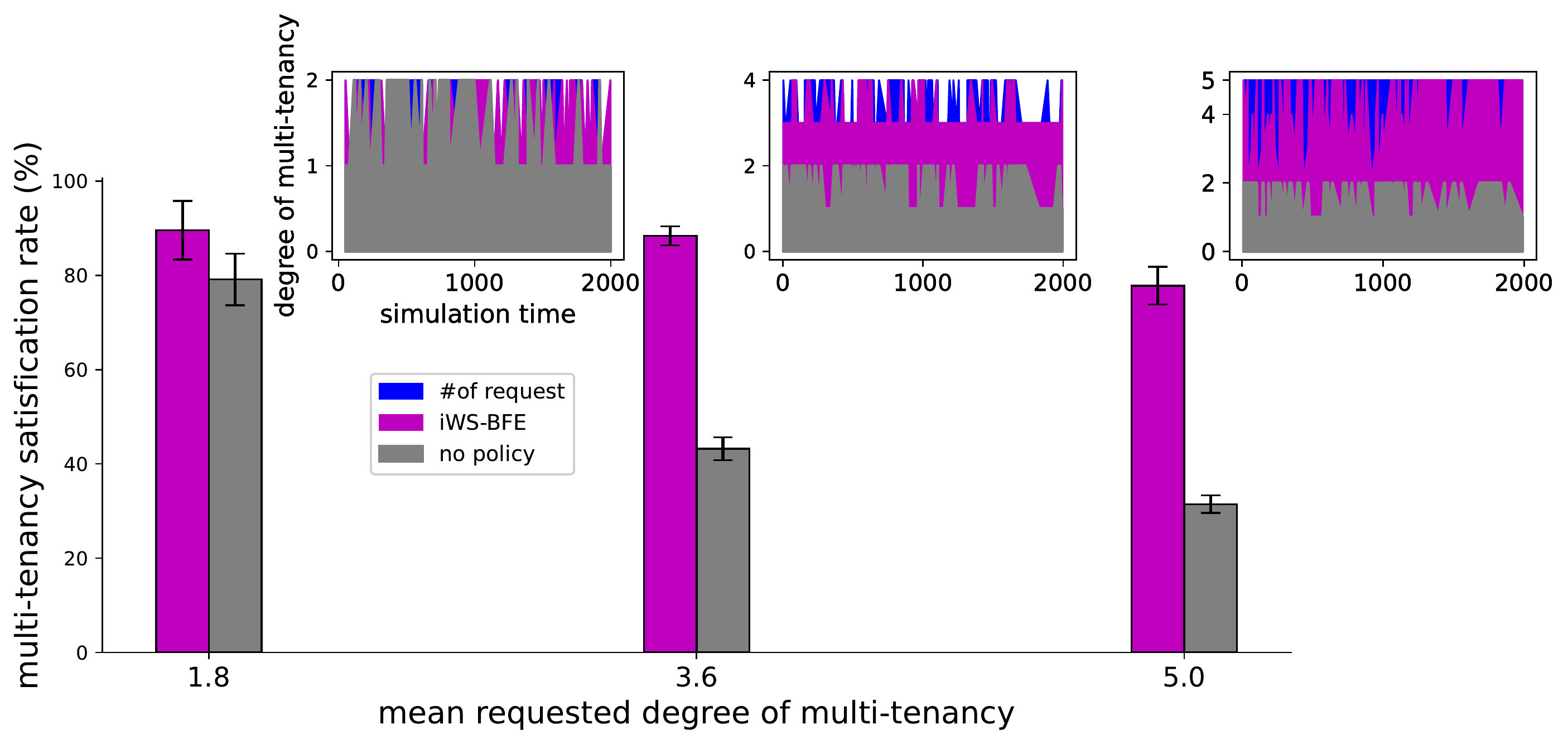}
    \caption{The impact of \name and its iWS-BFE eviction policy on satisfying the requested multi-tenancy. The large graph represents the summative analysis via increasing the mean of multi-tenancy requested in the horizontal axis, and showing the percentage of requests that were satisfied in the vertical axis. For each case, the smaller graph more granularly represents the number of concurrent requests issued and fulfilled during the simulation time.}
\label{fig:concE}
\end{figure*}
The simulator also enables us to generate workload traces that include the request arrival times for each application during the simulation time. We configure the actual workload to include an equal number of requests for the five applications, and the inter-arrival times between requests for each application are distributed exponentially within the workload.
To study the uncertainty exists in the inference request predictions, in the evaluations, we generate two sets of workloads, one includes the predicted arrival times for the multi-tenant applications, and the other one includes the actual arrival times of the applications. The distribution of request arrivals in the actual workload deviates from the distribution of requests in the predicted workload. The degree of deviation between the two is measured based on the Kullback-Leibler (KL)~\cite{van2014renyi} divergence. We explore the impact of this deviation in the experiments of next subsections.

Our evaluation metrics are: (A) The \emph{degree of multi-tenancy} under different request arrival intensity; (B) The \emph{inference latency}; (C) the \emph{inference accuracy}; and (D) The \emph{robustness} metric to measure the tolerance of different eviction policies against the uncertainty exists in the request predictions.

\subsection{\textit{Impact of \name on the Degree of Multi-tenancy }}~\\

This experiment is to examine the efficacy of \name~in satisfying the incoming requests to the edge server. To that end, as shown in Figure~\ref{fig:concE}, we increased the workload intensity, via the mean number of concurrent requests issued, and in each case measured the \emph{multi-tenancy satisfaction rate}, which is the percentage of warm-start inferences out of the total incoming requests during the simulation time. We examined two cases: (A) without any solution to stimulate multi-tenancy (called, no policy); and (B) with \name and its iWS-BFE policy in place. The experiment was repeated 10 times and the average rate and 95\% confidence intervals for each data point is reported.

The experiment shows that the degree of multi-tenancy achieved by adopting \name and its iWS-BFE is remarkably higher than the situation where \name is not in place. The smaller graphs show that this superiority occurs consistently during the simulation time. We also notice that the impact of employing \name is more effective for higher degrees of multi-tenancy. In particular, we can see that with the mean degree of multi-tenancy is 5, using \name and its iWS-BFE policy achieves $\approx$130\% higher satisfaction rate than no policy when mean requested degree of multi-tenancy is larger than 2. This experiment justifies the efficacy of \name and the NN model management in stimulating multi-tenancy of DL applications. 

\subsection{\textit{Impact of the Eviction Policies on the Cold-Start Inference }}~\\
The purpose of this experiment is to 
   \begin{figure}
    \centering
    \includegraphics[width=.75\textwidth]{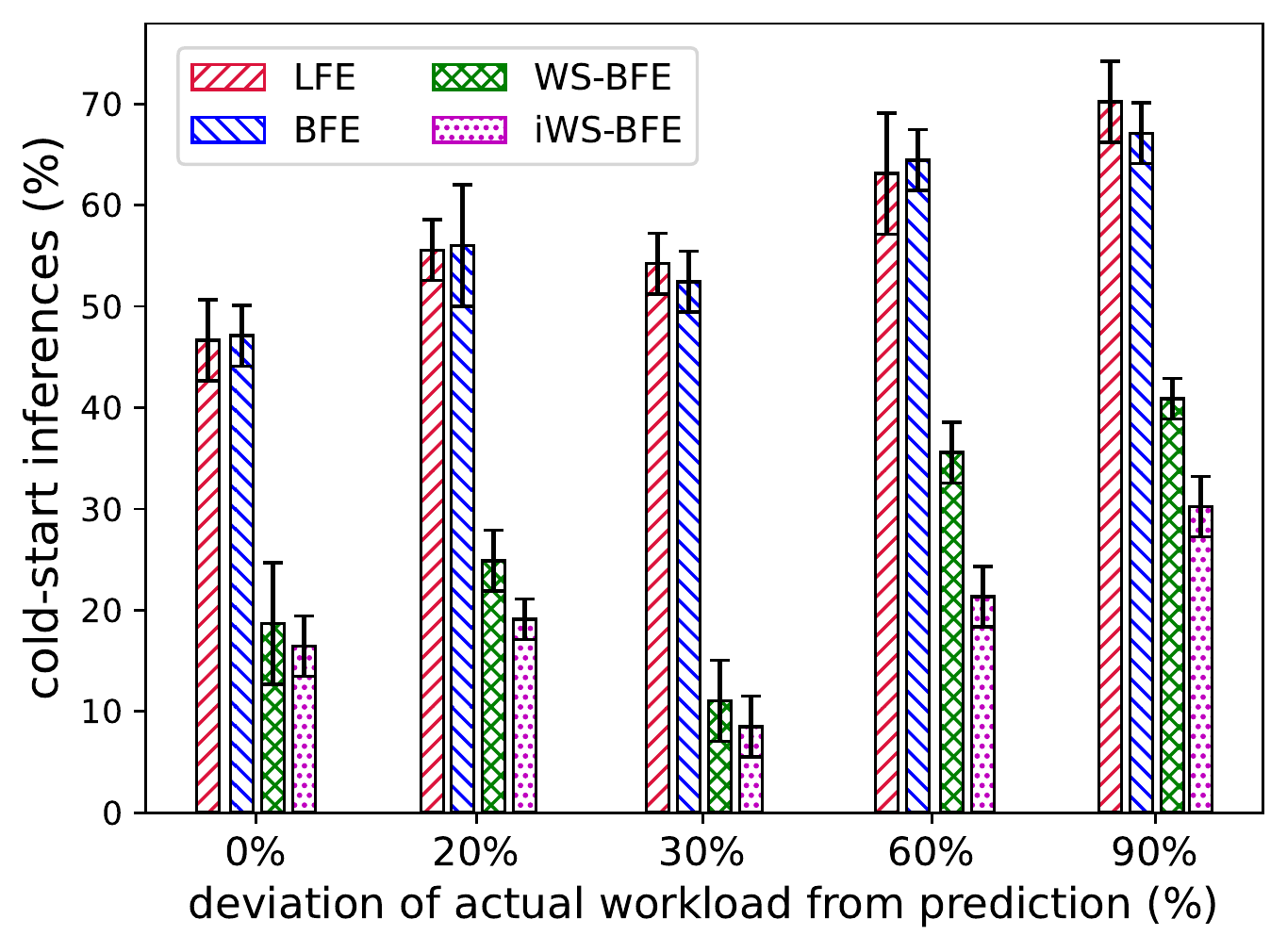}
    \caption{Measuring the percentage of cold-start inferences of multi-tenant applications resulted from 
    the proposed eviction policies. The horizontal axis shows the deviation between predicted and actual inference request times.}
\label{fig:ncoldsVd}
\end{figure}
analyze the impact of different NN model eviction policies on the number of cold-start inferences. For that purpose, we measure  percentage of cold-start inferences caused by employing different eviction policies, particularly, upon varying the deviation of request prediction from the actual requests. 

The results, illustrated in  Figure~\ref{fig:ncoldsVd}, show that LFE and BFE perform poorly and cause a remarkable number of cold-start inferences, whereas, WS-BFE and iWS-BFE mitigate the cold-start inferences by at least 65\%. This is because, in LFE and BFE, upon evicting an NN model, its corresponding application suffers from a cold-start inference in the event of an unpredicted request. In contrast, in WS-BFE and iWS-BFE, the evicted model is replaced with a low-precision one, hence, unpredicted calls to the corresponding application do not lead to cold-start inferences. It is noteworthy that, regardless of the employed policy, the percentage of cold-start inferences rises upon increasing the deviation between predicted and actual request times. Nonetheless, we see that even under 90\% deviation, iWS-BFE still substantially outperforms other policies. On average, it yields 102\% less cold-start in compare to LFE and BFE, and 40\% less than WS-BFE. 

\begin{figure}    
    \centering
    \includegraphics[width=.75\textwidth]{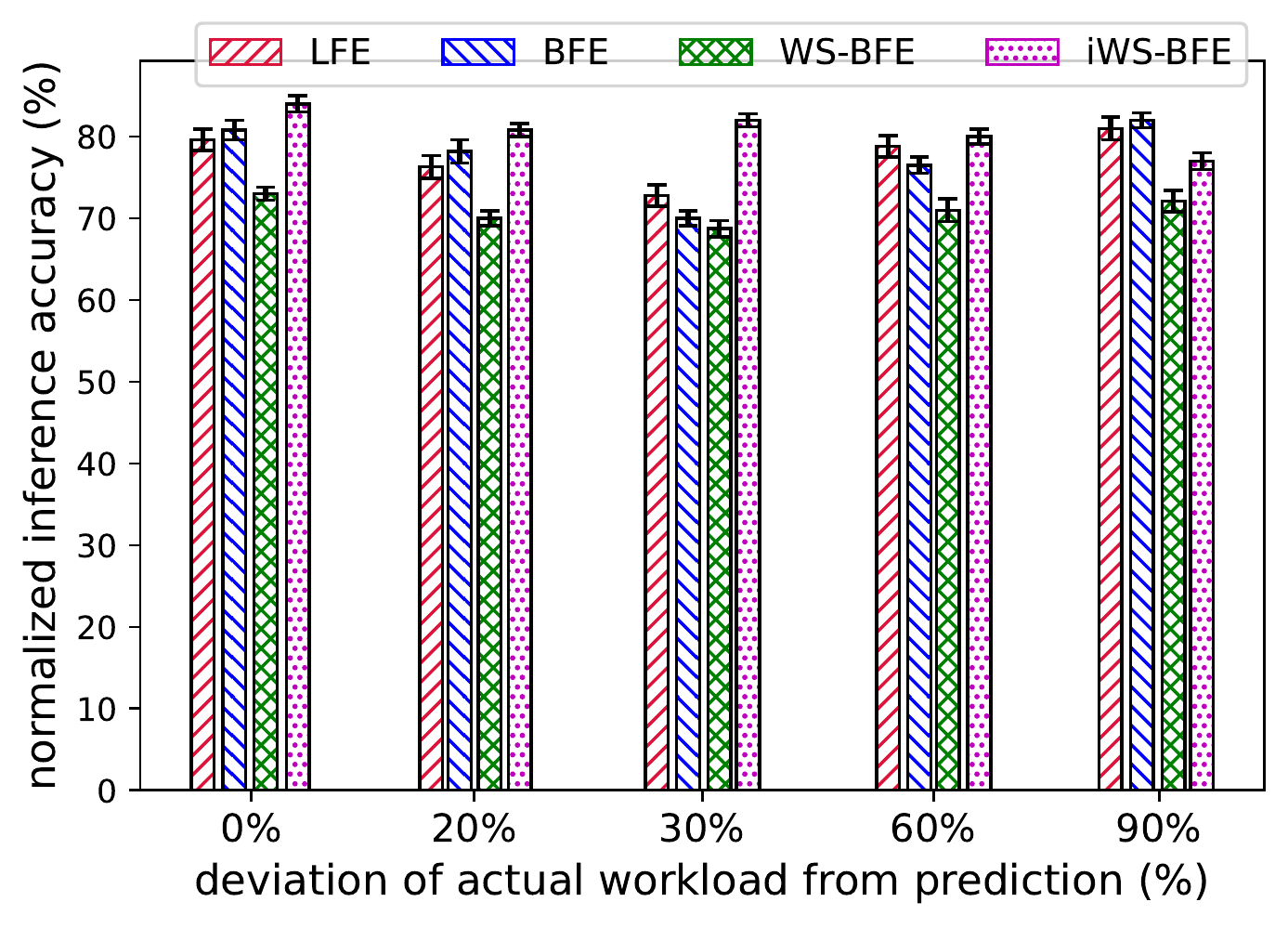}
    \caption{Measuring the normalized inference accuracy of applications resulted from employing the different eviction policies.}
\label{fig:accVd}
\end{figure}

\subsection{\textit{Impact of the Eviction Policies on the Inference Accuracy }}~\\
In this experiment, we analyze the average inference accuracy caused by employing different model eviction policies. Because the accuracy largely varies across different applications, we perform min-max normalization on the accuracy values. 
Also, for the cold-start inferences, in the accuracy measurements, we  consider the accuracy provided by the NN model after it is loaded into the memory.  


Figure~\ref{fig:accVd} shows the normalized mean inference accuracy obtained from employing different NN model eviction policies upon changing the deviation between predicted and actual request times. According to the figure, LFE and BFE policies outperform WS-BFE. This is because, these two policies do not retain the low-precision models in the memory. Therefore, their inference requests either lead to a cold-start (that was explored in the previous experiment), or they load high-precision models that provide a high inference accuracy. Nonetheless, we observe that iWS-BFE outperforms LFE and BFE in most of the cases, except the one with 90\% deviation. The reason for the higher inference accuracy of iWS-BFE is that, it nominates cold-start candidates intelligently, based on their probability of future invocations. This results indicate the importance of the scoring (described in Equation~\ref{eq:scoreiws}) on efficiently nominating cold-start candidates. It is noteworthy that the higher inference accuracy of LFE and BFE at 90\% deviation comes with the cost of substantially higher cold-start inferences that are detrimental to the ``usability'' of the IoT-based systems.

\subsection{\textit{Bi-Objective Analysis of NN Model Eviction Policies }}
\begin{figure} 
    \centering
    \includegraphics[width=.75\textwidth]{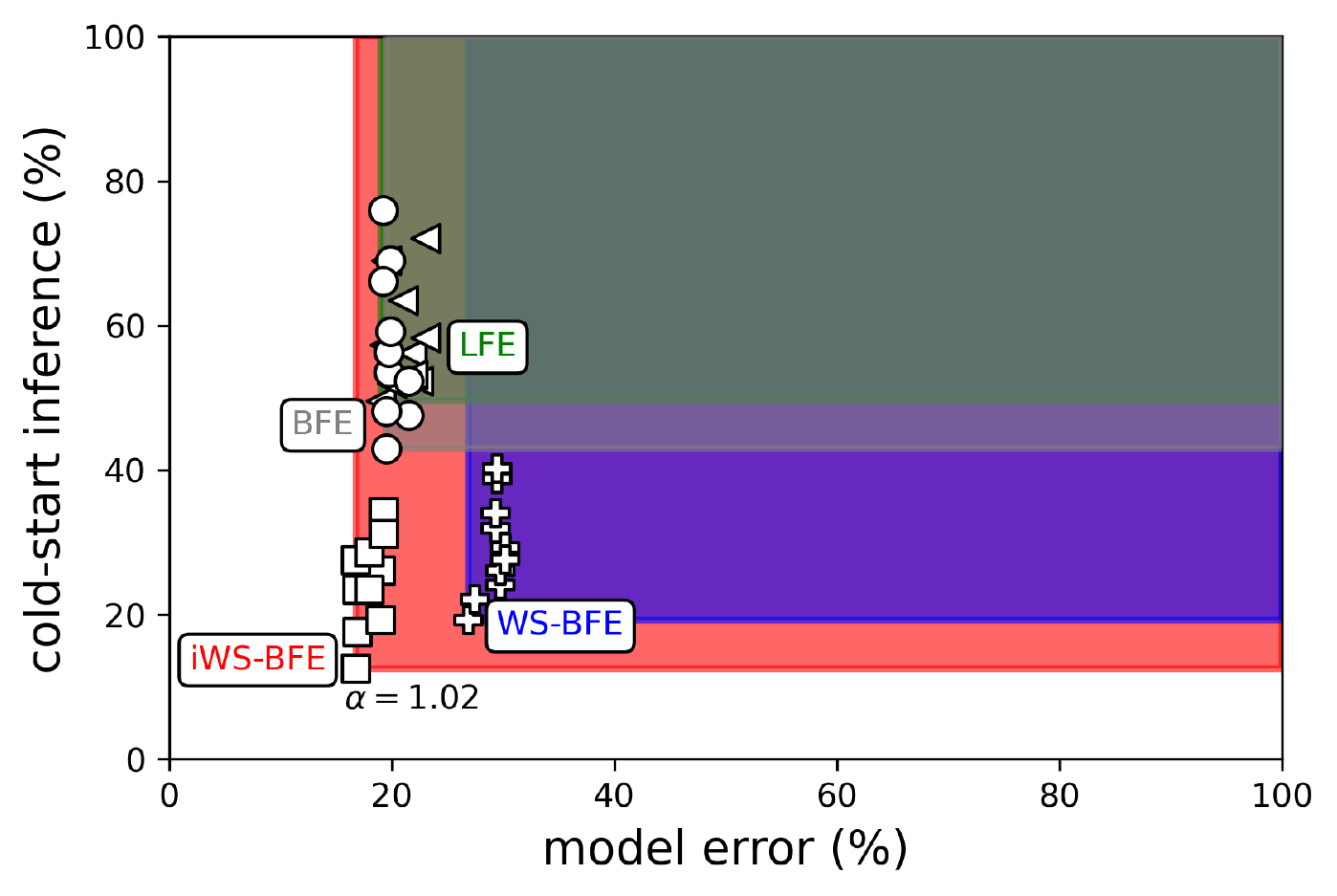}
    \caption{\small{Bi-objective analysis of the different model selection policies.}}
\label{fig:pareto}
\end{figure}

~\\
Recall that the NN model management for multi-tenant applications in a resource-limited edge system is a bi-objective optimization problem that aims at minimizing the number of cold-start inferences and maximizing the inference accuracy. However, these two are generally conflicting objectives and there is not a single optimal solution that can satisfy both objectives. Instead, there could be a range of solutions that dominate other solutions. To analyze which one of the studied policies dominate others, in Figure~\ref{fig:pareto}, we plot the percentage of cold-start inferences versus the model error (defined as 100-accuracy) for different policies  
and $\Delta$ values. Let $D$ and $\sigma$ be the mean and standard deviation of residuals of predicted versus actual request times. Then, $\Delta=D\pm\alpha\cdotp\sigma$ ranges by changing the value of $0\leq\alpha\leq2$. The deviation of actual versus predicted workload in this experiment is 30\%.

For each policy, the colored area shows the cold-start inferences and model error rate that are dominated by that policy. An ideal policy should approach the graph origin (\ie resulting in zero cold-start and zero model error). In Figure~\ref{fig:pareto}, we observe that \name dominates other policies and form the Pareto-front, particularly with $\alpha=1.02$. We can conclude that the iWS-BFE policy can significantly improve the usability of the systems via causing fewer cold-start inferences and offering a higher inference accuracy.


\subsection{\textit{Analyzing Robustness against Uncertainties }}~\\
The goal of this experiment is to study how the eviction policies of \name~make the IoT-based system robust against the uncertainty exists between the predicted and actual application request predictor. We define the \emph{robustness metric}, shown in Equation~\ref{eq:robust}, to encompass the ratio of warm-start inferences (denoted $\varpi_{i}$) to the total number of requests (denoted $\gamma_i$), and the mean prediction accuracy ($\psi_i$) of each application $i$ throughout the simulation period. 


\begin{equation}
\label{eq:robust}
R= \frac{1}{n} \cdotp \displaystyle\sum^n_{i=1}\bigg[{ \frac{\varpi_{i}}{\gamma_{i}}  \cdotp\psi_i } \bigg] 
\end{equation}
 \begin{figure}
    \centering
    \includegraphics[width=.75\textwidth]{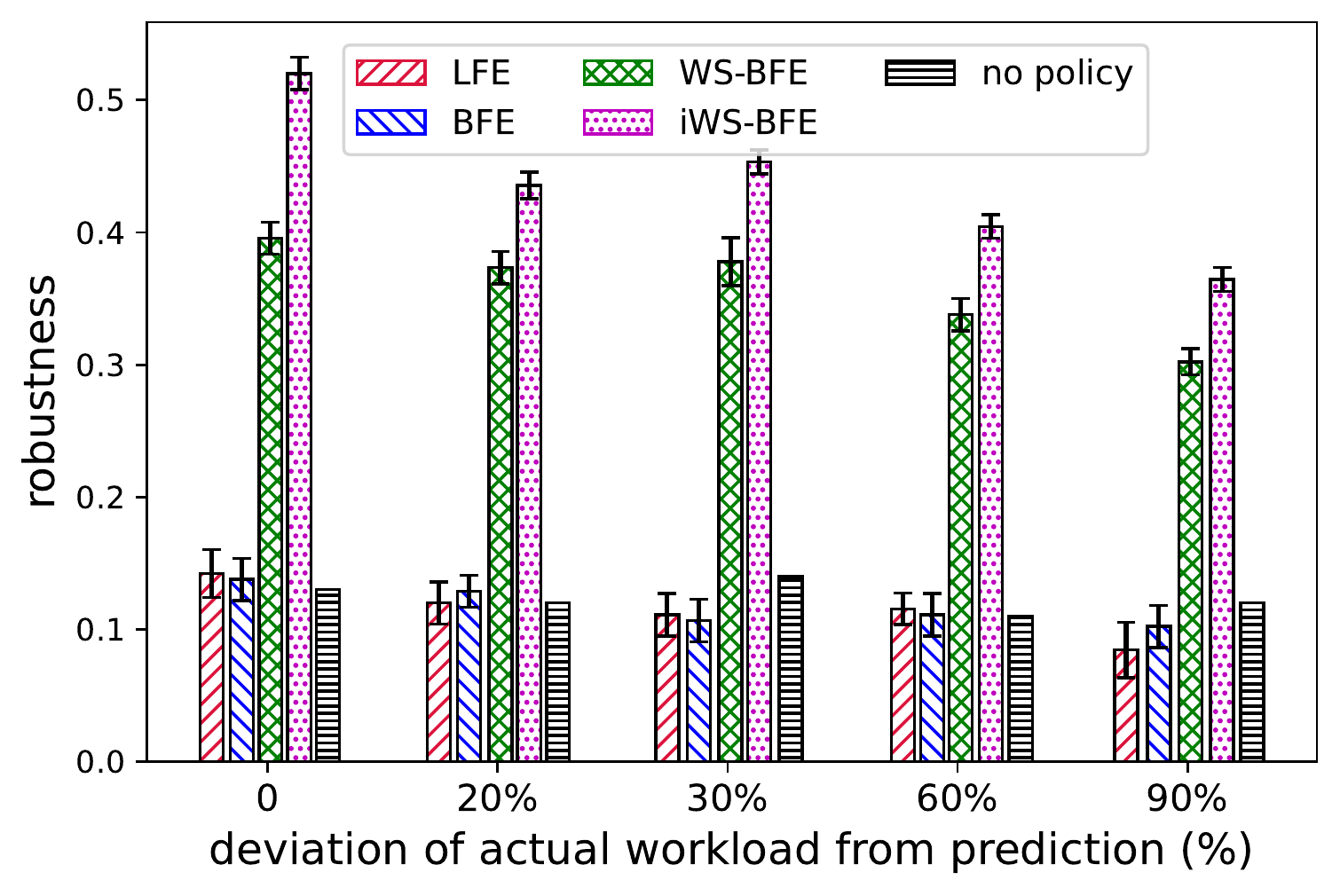}
    \caption{Robustness of the system against uncertainty in the prediction of inference requests. }
\label{fig:robust}
\end{figure}

Figure~\ref{fig:robust} represents the robustness score achieved by adopting the proposed policies and no policy (a.k.a. baseline) against uncertainties in the inference request prediction. We observe that deploying \name with any policy provides more robustness than the circumstance where \name is not in place (no policy). We also notice that the robustness value consistently drops 
because the rate of inference failure and cold-starts rise for higher deviations.
We observe that WS-BFE and iWS-BFE are more robust against deviation than the LFE and BFE. This is because, LFE and BFE do not replace their NN models with a lower-precision one upon eviction, which leads to cold-start inferences for the applications.




\subsection{\textit{Evaluating the Fairness of NN Model Eviction Policies }}
\begin{figure} 
    \centering
    \includegraphics[width=.8\textwidth]{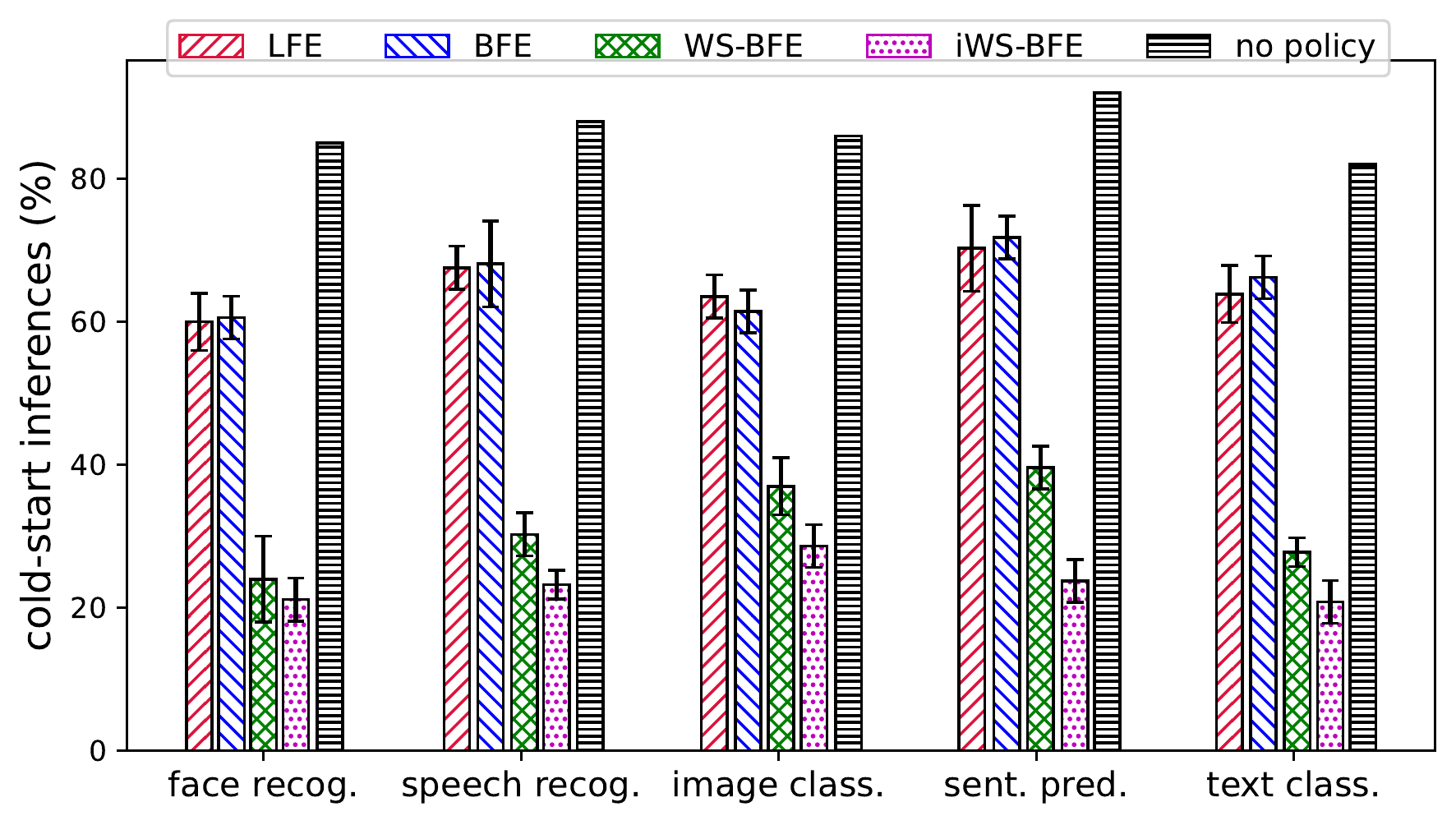}
    \caption{The percentage of cold-start inferences using different NN model eviction policies versus no policy.}
\label{fig:appcold}
\end{figure}

\begin{figure}
    \centering
    \includegraphics[width=.75\textwidth]{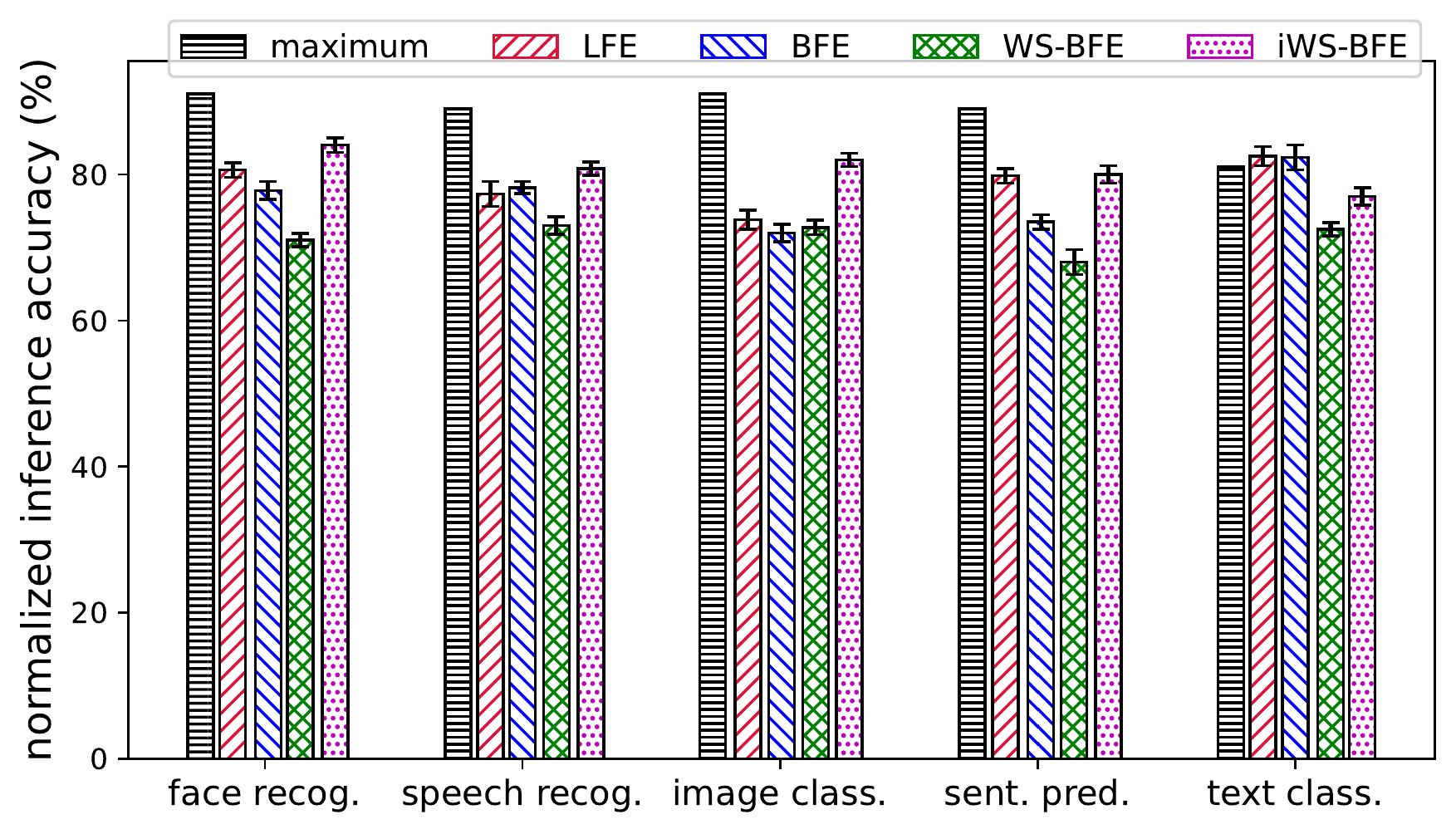}
    \caption{The inference accuracy obtained from the different policies. The ``maximum'' is the benchmark, showing the accuracy of the highest-precision model for each application.}
\label{fig:appaccuracy}
\end{figure}

~\\
In this experiment, our goal is to examine whether the achievements of \name and its policies, explored in the previous experiments,  is fairly distributed across all applications, or some applications benefit more than the others. To that end, we analyze the distribution of cold-start inference and accuracy across different DL applications. The name and the NN model characteristics of the examined DL applications are listed in Table~\ref{tab:app_model_size}. Figures~\ref{fig:appcold} and~\ref{fig:appaccuracy}, respectively, express the percentage of cold-start inferences and inference accuracy for each application upon using various NN model eviction policies. It is noteworthy that in Figure~\ref{fig:appcold}, ``no policy'' indicates the situation where \name is not in place, and in Figure~\ref{fig:appaccuracy}, ``maximum'' serve as the benchmark, by showing the use of highest-precision NN model for each application. While Figure~\ref{fig:appcold} shows that WS-BFE and iWS-BFE remarkably outperform the other policies across all the applications, Figure~\ref{fig:appaccuracy} illustrates that, particularly for iWS-BFE, the outperformance does not come with the cost of lower inference accuracy for the applications. More importantly, in both figures, we observe that, for each policy, the percentage of cold-start inferences and accuracy do not fluctuate significantly from one application to the other. This shows that policies are not biased to any particular DL application. Specifically, the rate of cold-start inferences and the accuracy are fairly distributed across different applications.


%% file: C5-5-sum.tex
\section{Summary}
    Smart IoT-based systems often desire continuous execution of multiple latency-sensitive Deep Learning (DL) applications. The edge servers serve as the cornerstone of such IoT-based systems, however, their resource limitations hamper the continuous execution of multiple (multi-tenant) DL applications.  The research aims to stimulate the degree of multi-tenancy of such applications without compromising their latency and accuracy objectives.

We developed a framework, called \name, to facilitate multi-tenancy of DL applications via enabling swapping only their NN models. The framework was also equipped with model management policies, particularly iWS-BFE, to choose suitable models for eviction and loading to edge memory, such that the percentage of warm-start inferences is maximized without any major loss in the inference accuracy of the applications. Evaluation results indicate that \name can improve the degree of multi-tenancy by $2\times$, and iWS-BFE can increase warm-start inferences by 60\%. They also show how different policies are robust against uncertainty in the inference request predictions. Last but not the least, the experiments show that the policies are not biased to a certain application in their decisions.

%% file: C6-Conc.tex
\chapter{Conclusion and Future Research Directions } \label{chap:conc}
This chapter summarizes the research and major findings of this dissertation.
Additionally, research topics that have surfaced during this research but have not been covered in this dissertation are brought up and discussed. These potential pathways for the future can be investigated further by other researchers working in this field.

\section{Discussion }
In this dissertation, our main objective was to enable confidential computing across edge-to-cloud continuum by maintaining data integrity and confidentiality during executions that span across the continuum. We provide three trusted applications to perform secure clustering and semantic searching over confidential data without revealing any meaningful information to any off-premise tiers. In addition, for model management of DL applications, we develop a framework that can effectively facilitate multi-tenancy of DL applications via enabling swapping only their NN models.  

In Chapter~\ref{chap:clus}, we developed solutions for topic-based clustering of both static (ClustCrypt and S-ClusPr) and dynamic unstructured encrypted big datasets (SD-ClusPr and FD-ClusPr). The proposed solutions approximate the number of clusters for a dataset within a feasible time complexity. For that purpose, they leverage the tokens' co-occurrences to measures the tendency of each token to stay with or segregate from other tokens and use that to estimate
the number of clusters. Next, we develop a probabilistic approach
to determine the center of each cluster and disseminate encrypted tokens to the most topically related cluster.
Experimental evaluations reveal that for static datasets, S-ClusPr can improve the clustering coherency on average by $65\%$. Similarly, for semi-dynamic and dynamic datasets, SD-ClusPr and FD-ClusPr can improve the coherency by $55\%$. By incorporating ClustCrypt and ClusPr within the context of a secure semantic search system, we learned that the more coherent and accurate topic-based clustering can improve the relevancy of search results. 

In Chapter~\ref{chap:saed}, we propose an open-source generic pluggable module, namely SAED into existing search services (\eg AWS kendra, S3BD) to perform context-aware, personalized, and secure search without dictating any change on them. SAED can search over the data that is either plain-text or encrypted using client side encryption
before outsourcing to the cloud (\ie AWS S3). Upon verified by human users, experimental evaluations indicate SAED can improve the relevancy of the retrieved results by on average $\approx24\%$ for plain-text and $\approx75\%$ for encrypted datasets.

Our solution in Chapter~\ref{chap:saed} entailed continuously and simultaneously maintaining multiple DL models that process confidential user data on the trusted edge tier. This was challenging considering the memory limitations on the edge tier. Moreover, such ML models could not be outsourced to Clouds because of the user's privacy. As such in Chapter~\ref{chap:concurrent}, we propose a framework, namely Edge-MultiAI that that operates based on the idea of  approximate computing and ushers the NN models of the DL applications into the edge memory such that the degree of multi-tenancy and the number of warm-starts are maximized. \name~leverages NN model compression techniques, such as model quantization, and dynamically loads NN models for DL applications to stimulate multi-tenancy on the edge server. 
    We also devise a model management heuristic for \name, called \emph{iWS-BFE}, that functions based on the Bayesian theory to predict the inference requests for multi-tenant applications, and uses it to choose the appropriate NN models for loading, hence, increasing the number of warm-start inferences. We evaluate the efficacy and robustness of \name under various configurations.  Evaluation results indicate that \name can improve the degree of multi-tenancy by $2\times$, and iWS-BFE can increase warm-start inferences by 60\%. They also show how different policies are robust against uncertainty in the inference request predictions. Last but not the least, the experiments show that the policies are not biased to a certain application in their decisions.



\section{Future Research Directions}
Based on our findings during the exploration of AI-driven confidential computing paradigm across the edge-to-cloud continuum,  there are several points where the work could be expanded upon that were not covered in this dissertation.

\subsection{\textit{Hierarchical Clustering of unstructured Data }}~\\

We can employ \emph{active learning} to enable the automatic hierarchical clustering of tokens with similar topics~\cite{dasgupta2008hierarchical,kumar2020active,min2020three}. 
The active learning paradigm was inspired by situations in which it is simple to collect enormous quantities of unlabeled data (\ie pictures and videos downloaded from the internet, speech signals obtained from recordings made with microphones, and so on), but it is difficult or expensive to gain their labels.

 We can exploit the meaning of the deciphered tokens and their distributions in the available clusters. With  this information, we incorporate Wikipedia knowledge to formulate hierarchical relationships among the tokens across the confidential dataset. Later, for new a token, we measure the relatedness between the token and the representation of each topic to propagate the hierarchy.

\subsection{\textit{Building Classifier from the Encrypted Clusters }}~\\

Clustering is a classic unsupervised learning that groups a massive amount of unlabeled data. We can employ \emph{active learning} on the clusters to build a classifier that potentially increases the use-cases in trusted computing for unstructured data paradigm. 
 Active learning can leverage the knowledge while querying on cluster to measure the relatedness with the new token to form a decision boundary of a classifier. The resultant classifier offers substantially lower cost than traditional supervised learning~\cite{dasgupta2008hierarchical}.

\subsection{\textit{Introducing Elasticity in Confidential Search }}
The current implementation of SAED framework needs dependency of a connected edge server to facilitate the searching. One idea is to including flexibility in the framework which will reduce the burden of edge communication with acceptable performance degradation.  For instance, when the user is on the move and does not have access to the edge, SAED should shrink to the bare minimum search intelligence and vice versa. 

\subsection{\textit{Adding Energy in Model Management Schemes }}~\\
In Edge-MultiAI, NN model management for multi-tenant applications in a resource-limited edge system is a bi-objective optimization problem that aims at minimizing the number of cold-start inferences and maximizing the inference accuracy. Since the edge servers have limited energy, often use battery, considering energy is crucial to assign a job on an edge. Otherwise, due to dead battery, the system could be abruptly switched off that leads to execution failure. 
Subsequently, we can add the energy as a third objective  into the problem. In this way, we frame the problem as maximizing the warm-starts and total accuracy with memory size and energy budget as constraints. 

\subsection{\textit{Cloud Offloading for Latency-tolerant Applications }}~\\
We only consider low-latency applications in Edge-MultiAI. We can also add offloading option to the cloud tier in the framework. In this way, the system could decide whether processing a task locally (probably with a few cold-starts) is more beneficial (in terms of latency and energy consumption) or offloading it to the cloud server.





%% file: F-biosketch.tex
\begin{biosketch}
   \indent \Author{} received his
   Bachelor of  Science in computer science and engineering in the fall of 2015 from Islamic University of Technology (IUT), Bangladesh. He started his professional career in December of 2015, as a System Engineer in one of the top tech giants named ``Huawei Technologies Ltd''. After around a year and half, he planned to enrich his academic knowledge by pursuing higher education. Hence, \Author~started his Ph.D. journey in computer science in the fall of 2017 at the University of Louisiana at Lafayette. \Author~ received his M.Sc. degree in computer science in the spring of 2019 during his Ph.D. journey. His research interests are: Natural language processing, Data analytics, and Cloud computing.
\end{biosketch}